\documentclass[fleqn,usenatbib,useAMS]{mnras}

\usepackage{graphicx}	
\usepackage{amsmath}	
\usepackage{amssymb}	
\usepackage{multicol}        
\usepackage{bm}		
\usepackage{pdflscape}	
\usepackage{rotating}
\usepackage{geometry}
\usepackage{longtable}

\newcommand{\ngal}{\,17\,}

\usepackage[T1]{fontenc}
\usepackage{ae,aecompl}

\usepackage{newtxtext,newtxmath}

\title[PHANGS--HST Aperture Corrections \& Cluster Morphologies]{Bright, Relatively Isolated Star Clusters in PHANGS-HST Galaxies: Aperture Corrections, Quantitative Morphologies, and Comparison with Synthetic Stellar Population Models}

\author[Deger et al.]
{Sinan Deger$^{1}$,\thanks{Contact e-mail:\href{mailto:sdeger@caltech.edu}{sdeger@caltech.edu}}
Janice~C.~Lee$^{2,3}$,
Bradley~C.~Whitmore$^{4}$,
David~A.~Thilker$^{5}$,
\newauthor
Mederic Boquien$^{6}$, 
Rupali Chandar$^{7}$,
Daniel~A.~Dale$^{8}$,
Leonardo Ubeda$^{4}$,
\newauthor
Rick White$^{4}$,
Kathryn Grasha$^{9}$,
Simon~C.~O.~Glover$^{10}$,
Andreas Schruba$^{11}$,
\newauthor
Ashley~T.~Barnes$^{12}$,
Ralf Klessen$^{10, 13}$, 
J.~M.~Diederik Kruijssen$^{14}$,
Erik Rosolowsky$^{15}$,
\newauthor
Thomas G. Williams$^{16}$
\\
$^{1}$TAPIR, California Institute of Technology, Pasadena, CA 91125\\
$^{2}$IPAC, California Institute of Technology, Pasadena, CA, USA\\
$^{3}$Gemini Observatory/NSF’s NOIRLab, 950 N. Cherry Avenue, Tucson, AZ, USA\\
$^{4}$Space Telescope Science Institute, 3700 San Martin Drive, Baltimore, MD, USA\\
$^{5}$Department of Physics and Astronomy, The Johns Hopkins University, Baltimore, MD, USA\\
$^{6}$Centro de Astronomía (CITEVA), Universidad de Antofagasta, Avenida Angamos 601, Antofagasta, Chile\\
$^{7}$University of Toledo, 2801 W. Bancroft St., Mail Stop 111, Toledo, OH, 43606\\
$^{8}$Department of Physics and Astronomy, University of Wyoming, Laramie, WY 82071, USA\\
$^{9}$Research School of Astronomy and Astrophysics, Australian National University, Canberra, ACT 2611, Australia\\
$^{10}$Universit\"{a}t Heidelberg, Zentrum f\"{u}r Astronomie, Institut f\"{u}r Theoretische Astrophysik, Albert-Ueberle-Str 2, D-69120 Heidelberg, Germany\\
$^{11}$Max-Planck-Institut f\"{u}r extraterrestrische Physik, Giessenbachstra{\ss}e 1, D-85748 Garching, Germany\\
$^{12}$Argelander-Institut f\"{u}r Astronomie, Universit\"{a}t Bonn, Auf dem H\"{u}gel 71, 53121, Bonn, Germany\\
$^{13}$Universit\"{a}t Heidelberg, Interdisziplin\"{a}res Zentrum f\"{u}r Wissenschaftliches Rechnen, Im Neuenheimer Feld 205, D-69120 Heidelberg, Germany\\
$^{14}$Astronomisches Rechen-Institut, Zentrum f\"{u}r Astronomie der Universit\"{a}t Heidelberg, M\"{o}nchhofstra\ss e 12-14, D-69120 Heidelberg, Germany\\
$^{15}$Department of Physics, University of Alberta, Edmonton, AB T6G 2E1, Canada\\
$^{16}$Max-Planck-Institut f\"{u}r Astronomie, K\"{o}nigstuhl 17, D-69117, Heidelberg, Germany\\
}

\date{Last updated 2015 May 22; in original form 2013 September 5}

\pubyear{2021}

\begin{document}
\label{firstpage}
\pagerange{\pageref{firstpage}--\pageref{lastpage}}
\maketitle


\begin{abstract}
Using PHANGS--HST NUV-U-B-V-I imaging of 17 nearby spiral galaxies, we study samples of star clusters and stellar associations, visually selected to be bright and relatively isolated, for three purposes: to compute aperture corrections for star cluster photometry, to explore the utility of quantitative morphologies in the analysis of clusters and associations, and to compare to synthetic stellar population models. We provide a technical summary of our procedures to determine aperture corrections, a standard step in the production of star cluster candidate catalogs, and compare to prior work. We also use this specialized sample to launch an analysis into the measurement of star cluster light profiles. We focus on one measure, $M_{20}$ (normalized second order moment of the brightest 20\% of pixels), applied previously to study the morphologies of galaxies. We find that $M_{20}$ in combination with UB-VI colors, yields a parameter space where distinct loci are formed by single-peaked symmetric clusters, single-peaked asymmetric clusters, and multi-peaked associations. We discuss the potential applications for using $M_{20}$ to gain insight into the formation and evolution of clusters and associations. Finally, we compare the color distributions of this sample with various synthetic stellar population models. One finding is that the standard procedure of using a single-metallicity SSP track to fit the entire population of clusters in a given galaxy should be revisited, as the oldest globular clusters will be more metal-poor compared to clusters formed recently.
\end{abstract}

\begin{keywords}
galaxies: star formation -- galaxies: star clusters: general
\end{keywords}

\section{Introduction}
Understanding the star formation histories and properties of galaxies has strong ties with understanding the evolution of star clusters. Stars are formed in clustered environments within giant molecular clouds following a hierarchy \citep{Lada03}, with the dense parts of the hierarchy converting the accreted molecular gas into stars. There are many factors that moderate this process, and that cause the star formation cycle to fade. Different factors also determine whether clusters of stars remain in a bound state following this clustered formation. Many clusters will dissolve due to internal \citep[e.g. stellar feedback,][]{Matzner02, Fall10} and external processes \citep[tidal disturbances within the interstellar medium, gravitational shocks, orbital decay due to dynamical friction, e.g.][]{Fall09}. Yet a subset of these structures remain bound clusters of stars of a singular origin, and evolve to old ages. The mechanics that determine long-term boundedness are not as yet fully understood, and its investigation requires observing the star formation paradigm in multiple scales and stages. 

The \textit{Hubble Space Telescope} (HST) has enabled the systematic study of star clusters throughout the nearby galaxy population \citep[e.g.,][]{Holtzman92,Meurer95,Whitmore95}. HST increased the number of star clusters detected per galaxy from hundreds to thousands, and unlocked observational access to the wealth of clusters and associations in nearby galaxies. The high angular resolution of HST (FWHM $\sim 0\farcs08$, pixel scale = $0\farcs04$ for WFC3/UVIS) is essential for the study of star clusters at the parsec scale in galaxies at distances larger than 1 Mpc, where the typical half-light radius of compact clusters is a few parsecs \citep{pz10, Ryon17, Krumholz19}. One program that utilized this observational strength of HST is The Legacy ExtraGalactic UV Survey \citep[LEGUS,][]{legus}, which observed 50 galaxies ranging from dwarves to massive spirals, within 12 Mpc.

In this paper, we analyze data obtained by the PHANGS--HST treasury survey \citep{lee21}. The PHANGS--HST Treasury survey has obtained 5-band (NUBVI, mainly HST WFC3) imaging of 38 nearby spiral galaxies. Observations started in April 2019, and the program completed observations in May 2021. The distances of the galaxies range from 4.5 Mpc to 23 Mpc \citep{Anand21, lee21}, with galaxy stellar masses in the range $10^{9.5}-10^{11} \mathrm{M_{\odot}}$. The program will ultimately yield catalogs of tens of thousands of star clusters and associations. 

To enable the study of star clusters, they must first be identified and their observed and physical properties measured. Standard steps in the development of star cluster catalogs are the morphological classification of candidates, and the identification of a bright isolated sample of objects to compute aperture corrections. The large number of star cluster candidates that can be identified thanks to the angular resolution of HST has heightened the need for uniform automated classification frameworks. Recognition-based neural networks (popular examples include VGG \citealt{vgg}, ResNet \citealt{resnet}) are impressive in tackling star cluster classification tasks, attaining accuracy comparable to humans at speeds far surpassing the average human identifier \citep[][]{Messa18, Grasha19, wei20, Perez20, whitmore21}. This is evidence that many-layered feed-forward convolutional neural networks are able, through repeat exposure to a training set of manually classified star clusters, to approximate a functional form for the complex range of star cluster morphologies. However, this approximate representation is both difficult to access, and is also likely to be overly complex. Gaining insight especially into star cluster evolution would benefit highly from tracking this evolution in widely used observed parameters (e.g. observed colors), and well-defined and tested characterization of the cluster morphology.

Morphological analyses of star clusters are also important in that they guide the inferences we make on broad cluster populations. In the current paradigm of star cluster analysis, morphology is a key factor in determining which structures are included, and which are excluded from study. As discussed by \cite{Krumholz19}, an inclusive approach would consider multiply peaked structures, such as stellar associations, in the determination of ensemble population properties such as the mass function. Whereas an exclusive approach would discard those on grounds that these are unlikely to be relaxed systems, and keep only those sources that are singly-peaked and compact. The exclusive approach tends to rely on qualitative descriptions of cluster morphology, such as compactness and visual roundness, in hopes of selecting the structures with the highest likelihood of being bound star clusters. The exclusive selection favors selection of clusters of older ages. This results in a disparity in cluster age functions derived using exclusive and inclusive star cluster catalogs \citep[for a detailed overview, see][]{Krumholz19}. One goal of this paper is to identify a quantitative morphological parameter which would be useful for guiding construction of star cluster catalogs.

Similar morphological investigations have been carried out in the extragalactic community, and some of the techniques developed there can be adapted for use with star clusters. A multitude of parameters have been employed to chart the morphological properties of galaxies. Some popular parametric measures of galaxy morphology that assume an underlying light profile for galaxies (e.g. \citealt{deV48} and \citealt{Sersic63} profiles) are not immediately applicable to star cluster morphologies.  Measures that are nonparametric, however, do not make this assumption and therefore have potential applications outside of galaxy morphologies. We test one such parameter here, the $M_{20}$ parameter, which has been used in conjunction with the Gini coefficient ($G$) in \cite{Lotz04} for galaxy merger classification. The $M_{20}$ parameter (summarized in Section~\ref{sec:morphology} and discussed in detail in \citealt{Lotz04}), or the normalized second order moment of the brightest 20\% pixels, is especially sensitive to light profiles containing multiple bright peaks. We investigate the utility of the application of $M_{20}$ in the classification of star cluster light profiles, in hopes of defining a framework for a quantitative parametrization of star cluster morphologies, and jointly, gaining insight into their evolution. $M_{20}$ also has many features (preserving effectiveness at low signal-to-noise ratios, getting less affected by diffuse background light) that make it a promising starting point, paving the way to investigating other quantitative morphology measures.

Another key difference from the study of morphologies of galaxies that makes the process more difficult for star clusters is that the underlying galactic backgrounds on which star clusters are identified and classified can be far more complex.  To simplify the problem for a first study, a sample of bright, fairly isolated clusters can be used.  Such samples are identified as a standard part of the production of star cluster catalogs for the purposes of computing aperture corrections \citep[][]{Adamo17, Cook19}.  Here, we describe the construction of such samples for the PHANGS-HST galaxies, and also document our process of computing aperture corrections, before turning our attention to using the sample for testing the utility of $M_{20}$.

A standard approach in computing the ages and masses of star clusters is through the analysis of their spectral energy distributions (SED's). SED's of individual clusters can be obtained by integrating the flux encapsulated within an annulus for a set of bands where observations have been made. Ages and masses are then derived via an SED fitting procedure. Here we only provide a brief overview of this process, and direct the interested reader to works such as \cite{dacunha08, franzetti08, han12, moustakas13, chevallard16}. A common first step in SED fitting is the formulation of a grid of synthetic template models. To avoid overly complicated fitting procedures with potentially unreliable outcomes, certain parameters forming this grid can be fixed a priori. An example to this is choosing a preset metallicity for the synthetic models. The best fitting model from within this grid can then be determined by using a goodness fit measure, often by picking the model that returns the minimum $\chi^{2}$ value for the fit. Frequently, the stellar populations forming these template SED's are chosen to be that of the simple (i.e. single-aged) stellar population (SSP), named as such because the population is formed as a result of a single and instantaneous burst of star formation. This choice follows from the hypothesis that clusters we observe are congregations of stars born at the same time, or are at least most closely approximated by such models. This assumption has been put to test by various past work in the literature \citep[e.g.][]{koleva08, kudryavtseva12, fukui14, kuncarayakti16}. Furthermore, especially in studies of star clusters in nearby main-sequence galaxies with $M_{*}>10^{10}~\mathrm{M_{\odot}}$, solar metallicity is often chosen to represent the population, driven by studies into the metallicity content of nearby galaxies \citep[e.g.][]{gallazzi05}. The SED fitting procedure for the PHANGS-HST program has been detailed in \cite{turner21}.

In this paper we compare the solar metallicity SSP models with various other synthetic stellar population models, as a test to the use of SSP models in star cluster analysis. To this effect, we generate models that have exponentially declining star formation histories for various metallicity values. In tandem, we also test the effects of nebular emission from the gas surrounding the stellar population. Our investigation in this paper focuses on comparing the color evolution of this set of synthetic stellar population models with the observed colors of our sample of bright, fairly isolated clusters. How closely the color evolution of models track the colors of observed clusters provides insight into the efficacy of the models. Though more detailed future work is needed for improved clarity, the test we detail in this paper provides an invaluable first look.

The remainder of this paper is organized as follows. We provide a summary of the PHANGS--HST survey, and overview of the 5-band HST imaging in Section~\ref{sec:PHANGS--HST}. In Section~\ref{sec:tr_sample}, we discuss the details of how the bright, isolated cluster and association sample is constructed.  We also describe the ensemble observed and physical properties of this specialized sample, and compare with the properties of the full cluster population.  We then describe how we use clusters from this sample to derive the aperture corrections for the PHANGS--HST data processing pipeline in Section~\ref{sec:ap_corr}. We discuss the quality assurance process applied to the clusters prior to their use in the aperture correction computation in the same section. Next, we provide the details of quantitative morphology analysis in Section~\ref{sec:morphology}.We compare the colors of our specialized sample with the color evolution of various synthetic stellar population models in Section~\ref{sec:stellarpop_models}. Finally, we provide our conclusions and plans for future work in Section~\ref{sec:conclusions}.

\section{PHANGS--HST Imaging}
\label{sec:PHANGS--HST}
The analyses in this paper are based on HST UV-optical imaging for \ngal galaxies for which samples of bright, isolated star clusters (as described in Section~\ref{sec:tr_sample}) were constructed during the first year of the survey (Table~\ref{tab:sample_overview}). A complete description of the PHANGS--HST Treasury program (GO-15654) is given in \citet{lee21}; we provide a brief summary of the imaging observations here.  Three exposures with sub-pixel dithering were taken with the WFC3 UVIS camera in each of five filters: F275W (NUV), F336W~(U), F438W~(B), F555W~(V), F814W~(I).  The exposures for each HST pointing were obtained within a three-orbit visit, and yielded total exposure times of ${\sim}2200$s (NUV), ${\sim}1100$s~(U), ${\sim}1100$s~(B), ${\sim}670$s~(V), ${\sim}830$s~(I).  For 8 of these galaxies, suitable WFC3 or ACS data in one or more of these filters were taken by prior programs.  The available archival data were obtained from MAST and processed in a consistent manner with our new observations.  The three exposures in each filter were first  "drizzled" to improve sampling of the PSF. The five drizzled images were then all aligned, and placed onto a common grid with pixel scale of $0\farcs04$ (the native WFC3 pixel scale). Astrometry was calibrated using stars from the GAIA DR2 catalog. These aligned images were used to identify the sample used to compute aperture corrections, and measure quantitative morphological parameters, as described below.  Public releases of these PHANGS--HST processed images are available at \url{https://archive.stsci.edu/hlsp/phangs-hst}.

\section{Samples of Bright, Isolated Stars and Star Clusters}
\label{sec:tr_sample}  

\subsection{Sample Construction}
\label{subsec:sample_construction}

The samples of star clusters, associations, and stars studied here were originally assembled to investigate aperture corrections, but they can find broader application as they comprise a set of particularly high signal-to-noise (S/N) and relatively isolated objects.  Here we describe the construction of the sample.

This specialized sample is manually identified by human visual inspection of the HST imaging by coauthor BCW.  About 20~single-peaked clusters, 10 associations, and 10~point sources (stars) are selected in each galaxy. The relatively small size of these specialized samples is well-suited for selection via visual inspection to ensure that only the best examples of each class morphological class are included. The single identifier approach allows for uniformity across all our galaxies. Even though the multi-identifier approach in principle may benefit from statistical averaging, it also introduces systematic biases in classifications due to subtle differences in the classification definitions assumed by different workers. We favored the uniformity inherent in the single identifier approach for the construction of our specialized sample, which was also the strategy followed by LEGUS to form their ``training samples'' \citep{Adamo17}. The strengths and weaknesses of the single versus multi-identifier approach has been discussed in \cite{wei20}, and \cite{whitmore21}, and the reader is referred to those papers for further details. 

The objects studied here fall into the standard four classes used to characterize sources in star cluster candidate catalogs \citep[e.g.,][]{Grasha15, Adamo17, Cook19, wei20}: 
\begin{itemize}
\item Class~1: star cluster -- single peak, circularly symmetric, but with radial profile more extended than point source
\item Class~2: star cluster -- same as Class~1, but elongated or asymmetric 
\item Class~3: stellar association -- asymmetric, multiple peaks 
\item Class~4: not a star cluster or stellar association.  A broad range of sources are included in this class~(e.g., image artifacts, background galaxies, individual stars), but for the purposes of the sample used in this analysis, only point sources (stars) are identified.
\end{itemize}
A total of 199 class~1, 163 class~2, 159 class~3 objects, and 157 stars are selected across the \ngal galaxies studied here (Table~\ref{tab:sample_overview}).

\begin{table}
\centering
\centerline{Sample Overview}
\begin{tabular}{|c|c|c|c|c|c|}
\hline
  \multicolumn{1}{|c|}{Galaxy} &
  \multicolumn{1}{c|}{$d \pm \delta d$ [Mpc]} &
  \multicolumn{1}{c|}{Class 1} &
  \multicolumn{1}{c|}{Class 2} &
  \multicolumn{1}{c|}{Class 3} &
  \multicolumn{1}{c|}{Stars} \\
\hline
  NGC 4826 & ~4.41$\pm$0.19 & 11 & 10 & 11 & 10\\
  NGC 3351 & ~9.96$\pm$0.33 & 12 & 11 & 10 & 10\\
  NGC 3627 & 11.32$\pm$0.48 & 10 & 09 & 09 & 08\\
  NGC 2835 & 12.22$\pm$0.94 & 10 & 11 & 09 & 08\\
  NGC 5248 & 14.87$\pm$1.34 & 10 & 09 & 08 & 06\\
  NGC 4571 & 14.9~$\pm$1.1 & 10 & 10 & 10 & 10\\
  NGC 4689 & 15.00~$\pm$2.25 & 10 & 10 & 08 & 10\\
  NGC 4569 & 15.76$\pm$2.36 & 10 & 10 & 10 & 10\\
  NGC 4535 & 15.77$\pm$0.37 & 12 & 10 & 10 & 10\\
  NGC 1792 & 16.20~$\pm$2.43 & 10 & 09 & 09 & 10\\
  NGC 4548 & 16.22$\pm$0.38 & 11 & 09 & 10 & 10\\
  NGC 1566 & 17.69$\pm$2.00 & 16 & 10 & 08 & 10\\
  NGC 1433 & 18.63$\pm$1.86 & 11 & 10 & 07 & 07\\
  NGC 1559 & 19.44$\pm$0.45 & 18 & 08 & 09 & 09\\
  NGC 1365 & 19.57$\pm$0.78 & 16 & 10 & 10 & 08\\
  NGC 4654 & 21.98$\pm$1.16 & 10 & 10 & 10 & 10\\
  NGC 2775 & 23.15$\pm$3.47 & 12 & 07 & 10 & 11\\
\hline
  \textit{Total} & - & 199 & 163 & 159 & 157 \\
\hline\end{tabular}
\caption{\label{tab:sample_overview} Overview of the sample analyzed in the paper. The first column shows the galaxy ID. The second column shows the distance to the galaxy together with its uncertainty, from \citet{Anand21}. Columns three to six show the number of class 1, 2, 3 objects and stars visually identified for the galaxy, respectively. We display the total number in our sample for each class in the bottom row.}
\end{table}

The details of the visual selection process are as follows. A F555W~(V) image is used alongside a color image (F814W, F555W, F336W) produced using the Hubble Legacy Archive (HLA; \citealt{Whitmore16}) algorithm, as provided in a website interface developed by coauthor RLW, and also provided for each galaxy at \url{https://archive.stsci.edu/hlsp/phangs-hst} as an interactive display . The first step is to identify class~1 star clusters that have high S/N, and are unsaturated, and isolated (i.e., the source dominates the flux up to 20~pixels from its center, or within $0.8\arcsec$).  We generally look for these objects in the outskirts of galaxies, but when possible, several class~1 clusters in the inner bulge are also included to provide a crosscheck on the degree of variation of radial profile with environment. An effort is also made at this point to look for bluer class~1 clusters if they exist, using the color image, to allow for the inspection of cluster light profiles in the bluest filters (F275W, F336W). 

A set of stars (point sources) is also identified simultaneously with the identification of class~1 clusters, since these generally can also be found in the outer parts of a galaxy. The 20~pixel isolation criterion is again required. An important selection criterion for the stars is a significantly steeper light profile compared to the class~1 clusters described above, with a distinctly smaller FWHM. 

An analogous procedure is performed for class~2 clusters (i.e., asymmetric or elongated, single central peak), with an important difference. The 20~pixel isolation criterion is relaxed, since there are often few if any class~2 clusters that would meet this criterion. Instead, a 10~pixel isolation criterion is used. As will be discussed in the next section, this led to a strategy for computing the aperture correction using the combination of inner profiles of class~1 and~2 clusters with the outer profiles for only class~1 clusters ($10{-}20$~pixels). Many class~2 clusters are in crowded regions so even this 10~pixel criterion may be relaxed to provide a sufficient sample, with the requirement that the cluster at least dominates the region out to 10~pixels. In general, the aperture corrections for class~2 clusters are less well determined compared to those for class~1 due to the more crowded environments in which they are often found. 

Finally, a set of class~3 associations is selected. These are not used in the derivation of aperture corrections, which would be problematic given that these objects often represent the densest portions of much larger associations of stars (i.e., spanning tens or even hundreds of pixels; \citealt{larson21, lee21}). Rather, we identify them for inclusion in this sample as archetypes of stellar associations for other applications, such as the exploration of quantitative morphologies presented later.

On rare occasions, it is challenging to identify enough class~1 and class~2 bright, isolated clusters. Though not included in this paper, the PHANGS--HST galaxy NGC~1317 suffers from such a shortcoming. In these instances, we plan to derive the aperture correction using median aperture corrections of galaxies close in distance to it.

It is important to note that this sample is not intended to be complete, and its properties may not necessarily be representative of the properties of the overall star cluster population. By construction, objects in this sample are brighter and more isolated than the general population of clusters and associations (see next section), reflecting the primary goal of using them to determine aperture corrections. Another important note is that the construction of this sample is distinct from that of the PHANGS--HST star cluster catalogs as described in \cite{lee21} and \cite{thilker21}. The construction of the specialized sample we analyze in this paper relies on independent human visual inspection and targeted examination, while the full star cluster catalogs are generated by an extensive automated selection and processing pipeline, with subsequent candidate inspection and morphological classification performed by both BCW and deep neural networks \citep{wei20, whitmore21}. 

\begin{figure*}
    \centering
    \includegraphics[scale=0.36]{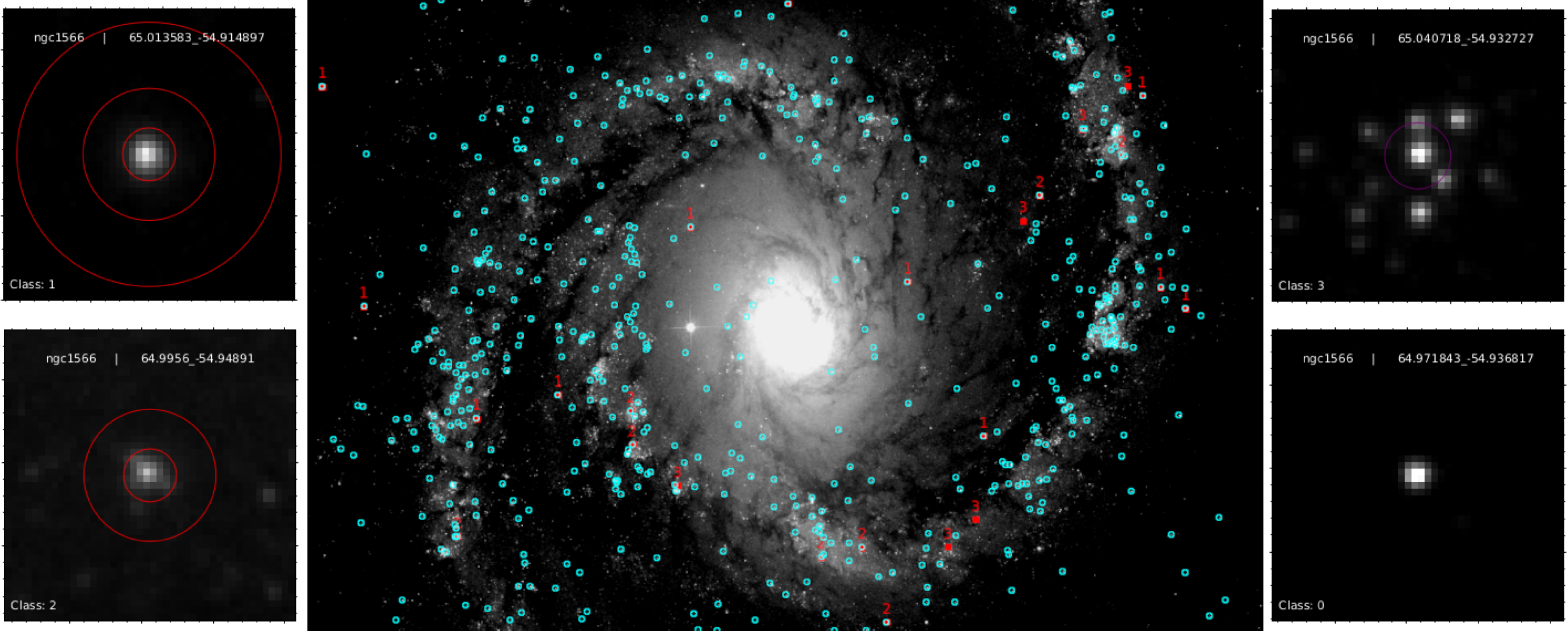}
    \caption{A cutout from NGC~1566 V-band image. The red markers denote objects residing in the bright, isolated sample for NGC~1566 (description in Section~\ref{sec:tr_sample}), with their morphology classes noted next to the marker, whereas cyan markers indicate visually classified PHANGS--HST star clusters and associations from the full catalogs. The postage stamps to the left and right of the figure display a class~1 (top left) cluster, class~2 (bottom left) cluster, class~3 (top right) association, and a star (bottom right) from this galaxy. The postage stamps and the central cutout are all shown with logarithmic stretch.}
    \label{fig:ngc1566_cutout}
\end{figure*}

For one of the PHANGS--HST galaxies, NGC~1566, we illustrate the spatial distribution of the selected sources on the V-band image in the central panel of Figure~\ref{fig:ngc1566_cutout}, where the red numbers indicate the morphological classification. The cutouts surrounding the central panel are examples of each class of object in the sample. All cutout panels in this paper are produced using \textit{aplpy} \citep{aplpy2012, aplpy2019}. In the central panel we also mark the locations of the greater population of star clusters from the PHANGS--HST pipeline.

\subsection{Sample Properties}

In this section, we describe the general properties of the sample. Key characteristics are provided in Table~\ref{tab:sample_prop}. As mentioned earlier, about 10~objects in each class~are selected in each galaxy, and the total sample contains 199 class~1, 163 class~2, 159 class~3 objects, and 158 stars across \ngal galaxies.

\begin{figure*}
    \centering
    \includegraphics[scale=0.7]{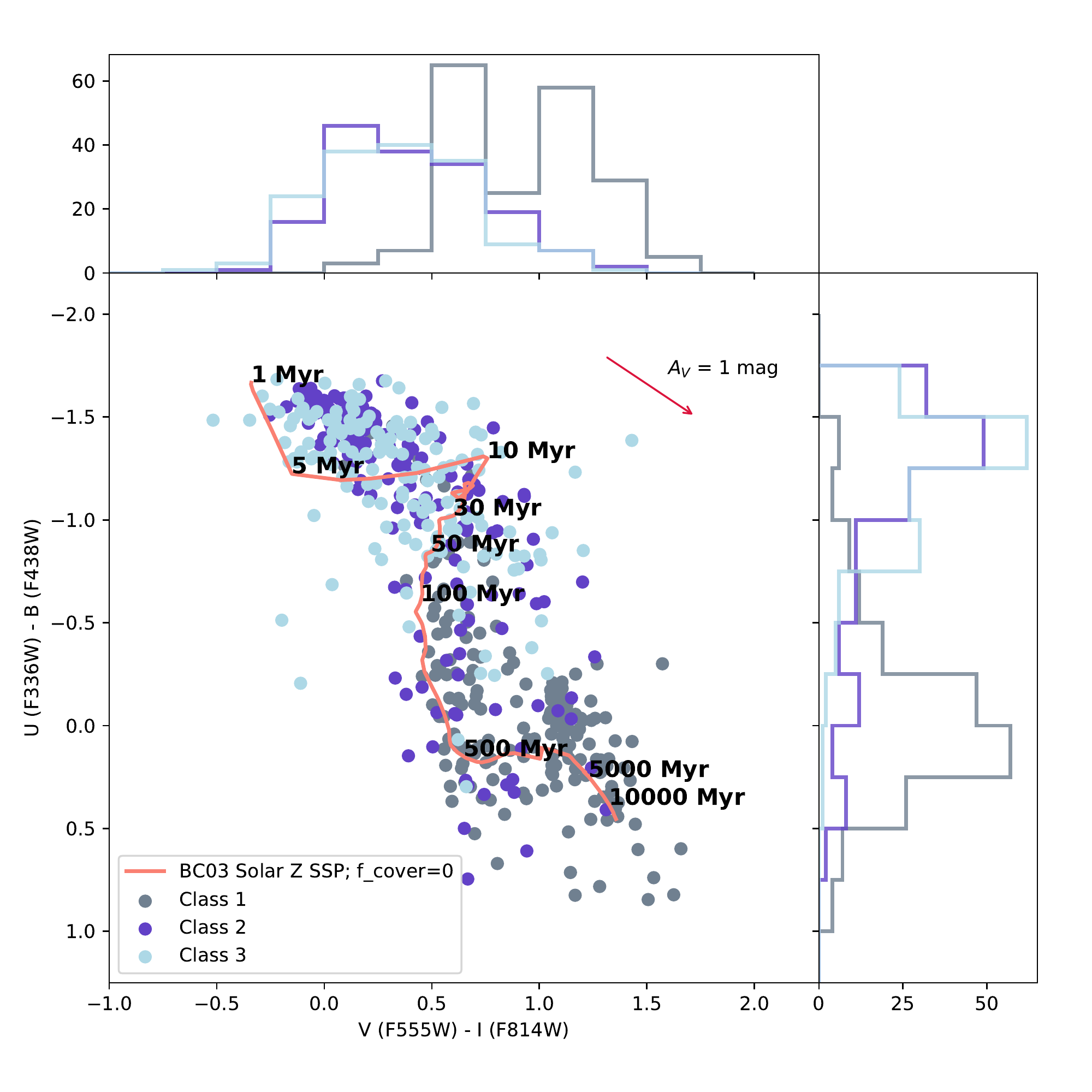}
    \caption{The U-B vs V-I color--color plot of the sample of bright, relatively isolated clusters identified in 17 PHANGS-HST galaxies, color-coded by cluster morphological type. The gray, purple, and blue color markers represent classes~1, 2, and~3, respectively. Attached to the x- and y-axes are histograms depicting the distribution of these clusters in the V-I, and U-B colors, respectively. Aperture and foreground extinction corrections have been applied to magnitudes.  We also plot the curve tracking the evolution of a BC03 single stellar population (SSP) model on the UB-VI space, for the solar metallicity case with a covering fraction of zero (no nebular emission). The red arrow in the panel shows the reddening vector for $A_{V} = 1$~mag assuming a Milky Way extinction law and dust type. We indicate the positions of a few select ages on the SSP track.}
    \label{fig:ub-vi}
\end{figure*}

\begin{figure*}
    \centering
    \includegraphics[scale=0.75]{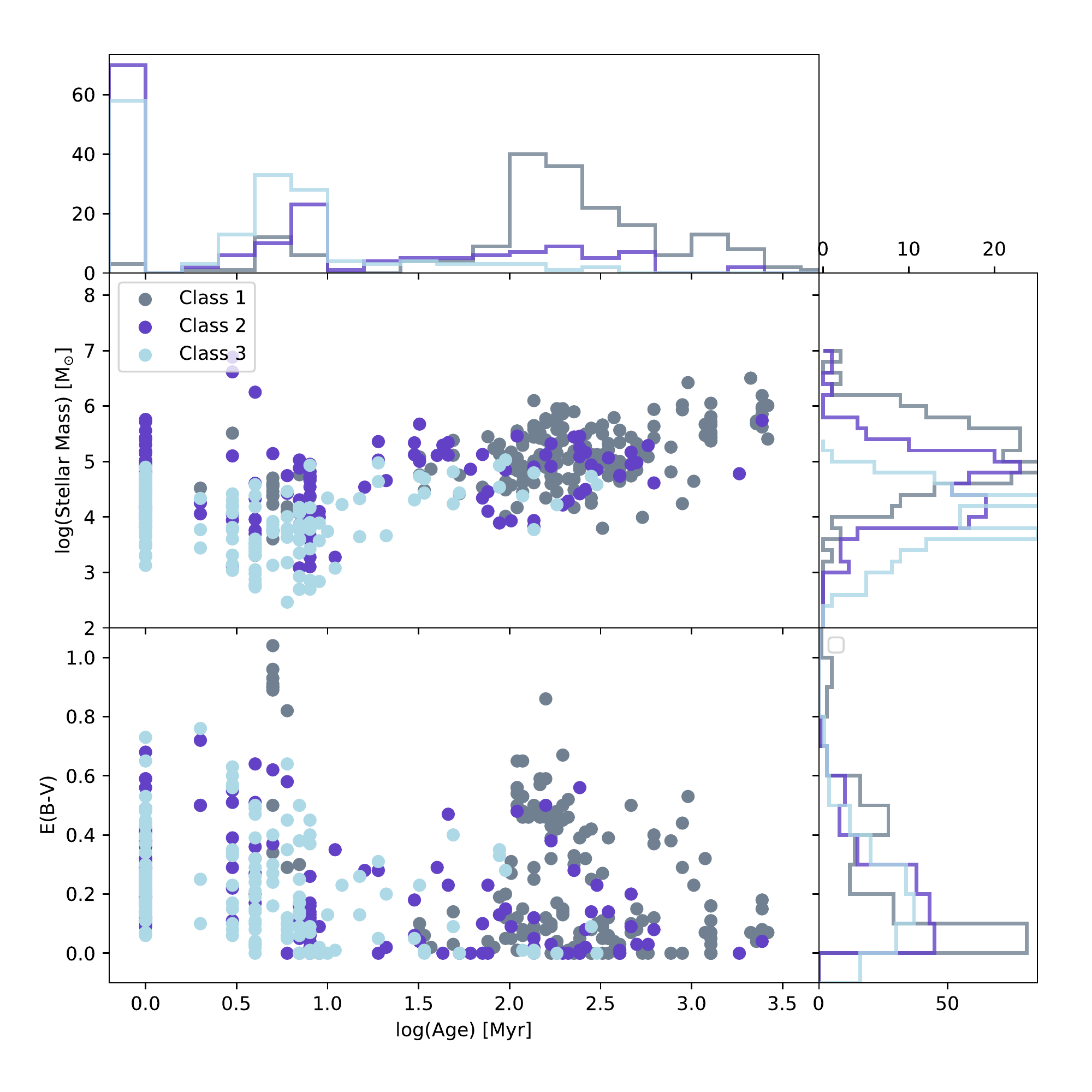}
    \caption{ A joint panel showing the distributions of our clusters in terms of their ages, stellar masses, and reddening values, determined via $\chi^{2}$ minimization fit to their SEDs. The plots are color coded according to cluster morphological type, where gray, purple, and gray markers denote classes~1, 2, and~3, respectively. The two main panels show the age versus stellar mass, and age versus reddening plots of our sample, from top to bottom. Attached to each of these panels are histograms showing the distributions of the morphological type in the respective quantities represented by the panel axes.}
    \label{fig:age-mass-red}
\end{figure*}

We start by visualizing the sample on two diagrams commonly used to characterize the properties of star clusters and associations: the color--color diagram in Figure~\ref{fig:ub-vi}, and the age--stellar mass--reddening diagram in Figure~\ref{fig:age-mass-red}. Ages and masses are derived via SED fitting of the 5-band HST photometry with CIGALE \citep[][]{Boquien19} as described in \cite{turner21}. It is important to note that such figures are usually shown for complete cluster populations in a single galaxy to provide insight into cluster formation history and the physics of cluster evolution. However, here we use these diagrams as an aid to illustrate the properties of the composite population of our specialized bright, isolated sample gathered from \ngal galaxies, so they should not be interpreted in the customary manner.

Figure~\ref{fig:ub-vi} shows our composite sample of class~1, 2, and~3 structures from across \ngal galaxies. We find that class~2 and~3 objects preferentially reside towards the upper left, or the bluest area of the UB-VI diagram. This area is also populated by the single stellar population (SSP) model at its youth with ages between $1{-}30$~Myr. On the other hand, we find that the class~1 clusters dominantly reside on the red portion of the UB-VI diagram, in the area occupied by the oldest ages of the SSP evolutionary track. Almost all our class~1 clusters are by the part of the SSP track that has evolved for at least $100$~Myr. The SSP track in the figure was generated using stellar population synthesis models from \cite[][BC03]{BC03}, and the particular SSP we plot is one of solar metallicity with no nebular emission. Even though the majority of the data is consistent with this track plus reasonable reddening (median E(B-V) of the sample is 0.18~mag, reddening for $A_{V} = 1$~mag shown in panel), there are deviations that motivate further investigation. One such sub-population is the group of class 1 clusters offset from the track by roughly 0.3 mags in U-B color between the ages of 500 Myr - 5 Gyr. We investigate this, and other differences between the observations and the SSP track in Section~\ref{sec:stellarpop_models}, where we explore other synthetic stellar population models beyond the solar metallicity SSP, which is commonly adopted for SED fitting of star clusters in nearby galaxies \citep[][]{Adamo17, turner21}.

We find that these age trends are supported by the ages derived from the 5-band SED fitting. Figure~\ref{fig:age-mass-red} shows that the class~2 and~3 objects are dominantly young (median ages for either class~is 4~Myr), whereas class~1 clusters are old (median age 196~Myr). The top panel shows that the lowest stellar mass range is preferentially occupied by class~3's, the density of class~1's peaking at the high stellar mass end, with class~2's residing in between. The bottom panel shows the reddening due to dust as computed by our SED fitting. The class~1 clusters concentrated at around log(Age/Myr) of $2.2$ in this panel are the old globular clusters, as mentioned above, for which the ages are underestimated and reddenings overestimated. We find that class~2 and class~3 objects are reddened by dust more than class~1 objects, especially if the population described in the previous sentence is overlooked. This follows naturally from the isolation requirements we employ for each morphology type. Our class~2 and~3 objects are much younger, and more likely to reside in crowded areas. They therefore have a much higher likelihood to be in regions with dust. Our class~1 clusters, on the other hand, are far less likely to come from areas containing significant amounts of dust. 

\begin{figure*}
    \centering
    \includegraphics[width=.48\textwidth]{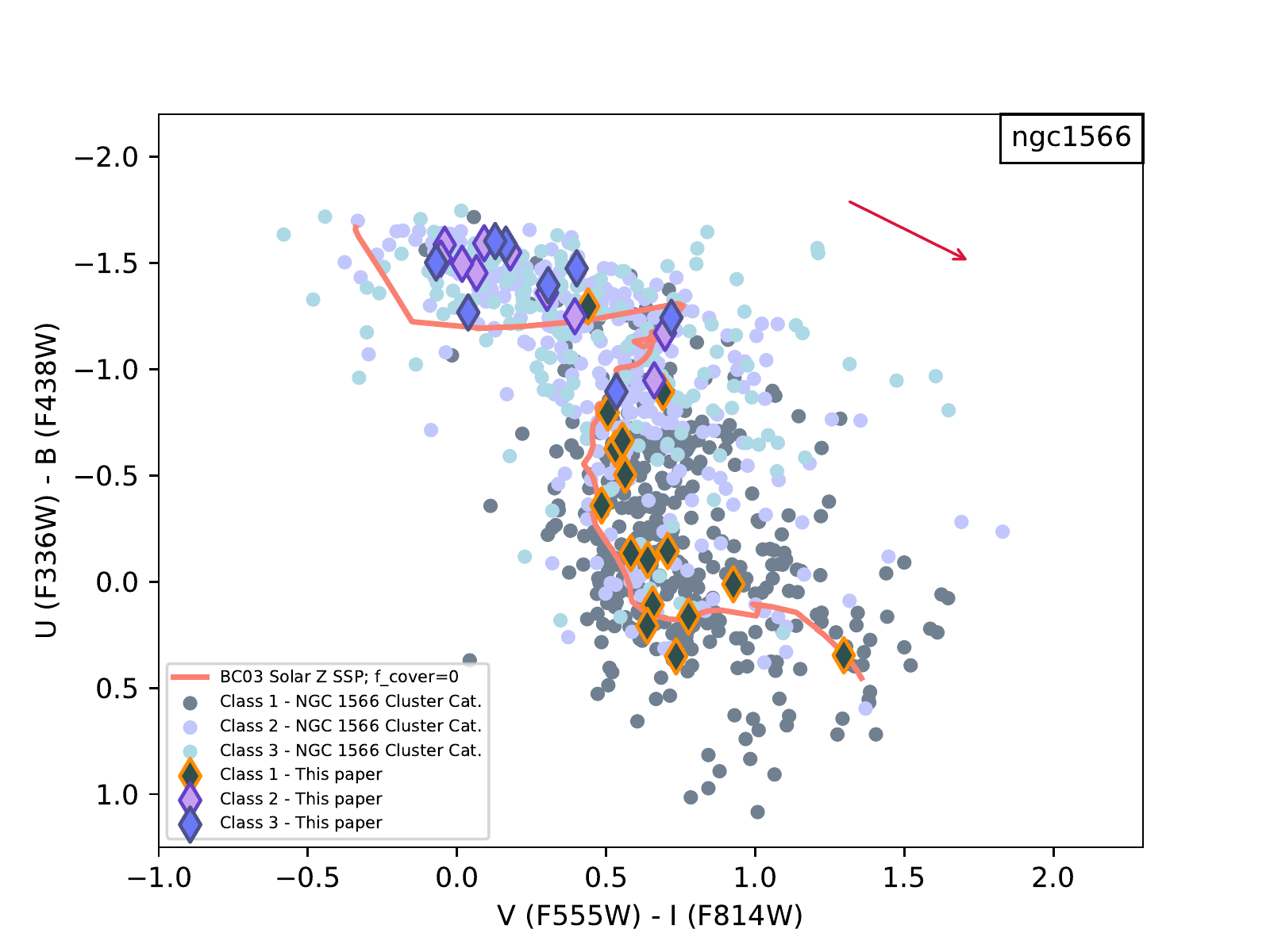}
    \includegraphics[width=.48\textwidth]{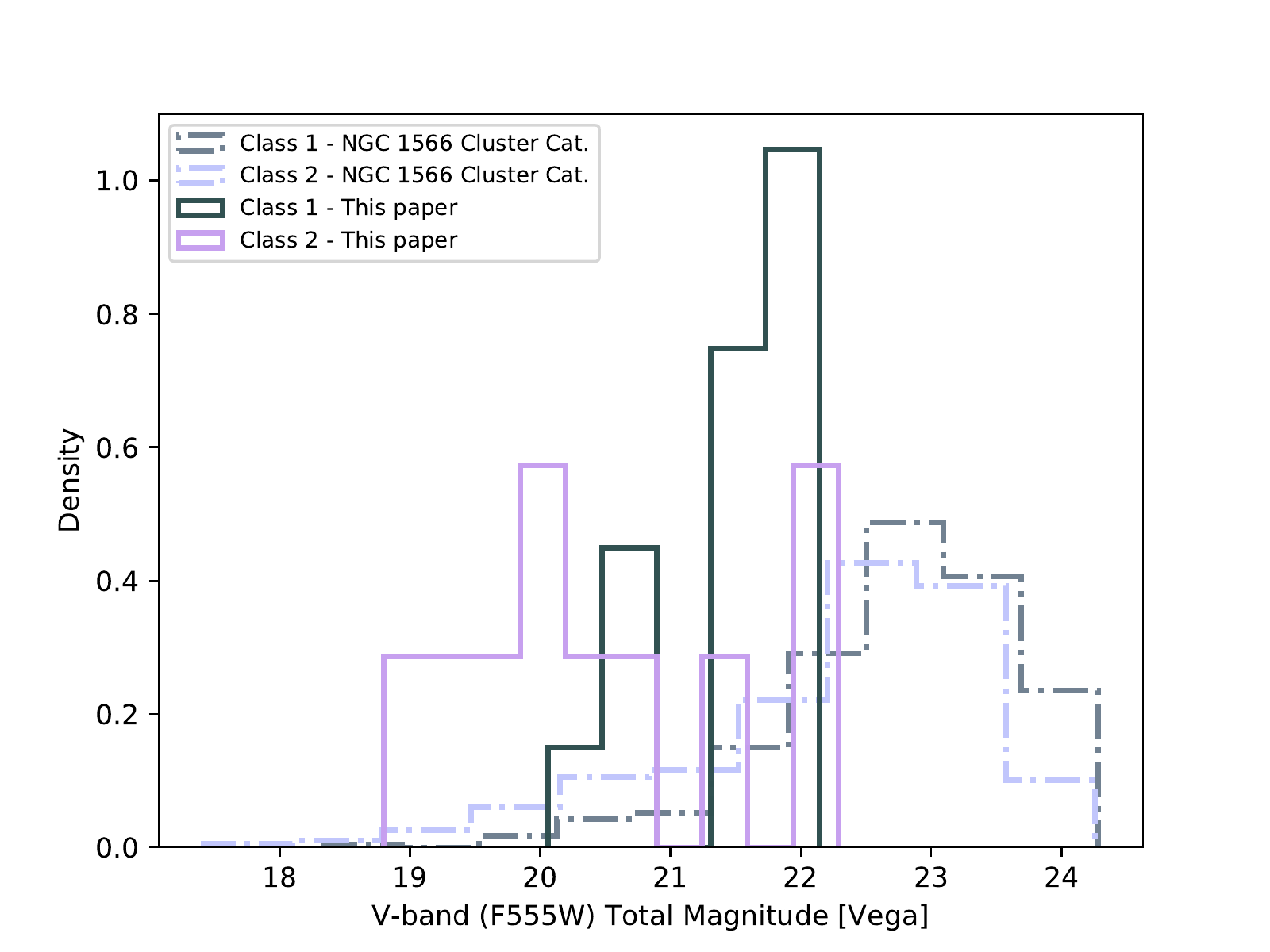}
    \caption{\textit{Left panel --} The UB-VI color--color diagram comparing the distributions of the bright, isolated sample we analyze in this paper (diamond-shaped markers) and the PHANGS--HST cluster catalog (circular markers) for NGC~1566. The gray markers denote class~1, purple markers class~2, and blue markers class~3 objects. The reddening vector corresponding to $A_{V} = 1$~mag is shown as the red arrow. \textit{Right panel --} The V-band total magnitude normalized histograms for the sample in this paper (full lines), and the full cluster catalog (dashed \& dotted lines) of NGC~1566. The gray lines represent class~1 clusters, and purple lines represent class~2 clusters.}
    \label{fig:ngc1566_comparison}
\end{figure*}

Having shown the entire sample across \ngal galaxies on the UB-VI space, we now compare the UB-VI distributions of the isolated, bright sample and the full cluster catalog for one of our fields, NGC~1566. We have shown the spatial distribution of these two samples on the same galaxy in Figure~\ref{fig:ngc1566_cutout}. In Figure~\ref{fig:ngc1566_comparison}, we show the comparison of the UB-VI color--color distributions in the left panel, and the V-band total (i.e., having had the aperture corrections applied, computation of which we discuss below) magnitude comparison in the right panel. The left panel reveals that the different morphology types from the full cluster sample do follow a similar trend in the UB-VI space as our sample here, shown in Figure~\ref{fig:ub-vi}. The full sample shows more spread around the SSP track, as is expected. In the absence of the isolation requirement, the full sample is more susceptible to reddening from dust. Furthermore, the full sample includes objects with lower masses, and stochastic sampling of the stellar inital mass function also contributes to increased scatter \citep[e.g.,][and references therein]{hannon19}. 

Despite the increased scatter of the full sample around the SSP track, the isolated, bright clusters of each morphological type in NGC~1566 reside where the density of the same class of objects from the full sample peaks. The left panel compares the distributions of the V-band apparent magnitudes of these two samples. The two panels of Figure~\ref{fig:ngc1566_comparison} demonstrates that even though the sample used in this paper is not representative of the full cluster sample, it is solely because they are drawn from the brightest end of the $m_{V}$ distribution. The morphological types in this sample still show the key characteristics of clusters from a more complete set. The lack of generalizability to complete cluster samples therefore does not translate to an inability to make meaningful inferences.

Finally, we present the distribution of Multiple Concentration Index (MCI) values for the sample, which are used to select star cluster candidates in the PHANGS--HST cluster catalog pipeline \citep{Whitmore99, lee21, thilker21}. We briefly summarize the parameters forming this space here, and encourage the interested reader to refer to \cite{thilker21} for details. 

Previous work has widely utilized the concentration index (CI) as a key metric to distinguish candidate clusters of stars from point sources \citep{Adamo17, Cook19}. This metric is a measure of the difference in photometry between circular apertures of different radii (the measure can also be represented as CI$_{ij}$, where~$i$ and~$j$ denote the radius in pixels of circular apertures to be compared), and the choice of 1~pixel and 3~pixel radii circular apertures has been standard practice. The optimal CI threshold that results with maximal distinguishing power is then determined using sets of bright and isolated star clusters and point sources, similar in construction to the current sample. 

The MCI metrics aim to provide a better informed selection that is easily calibrated to varying distances of the host galaxy, using multiple CI's in contrast to the single CI measure. 

Two MCI parameters are defined: MCI$_{\rm in}$ (based on aperture photometry with radii of $1.0, 1.5, 2.0, 2.5$ pixels) and MCI$_{\rm out}$ (based on aperture photometry with radii of $2.5, 3.0, 4.0, 5.0$ pixels). MCI$_{\rm in}$ therefore carries information about the inner parts of a given source, which is the span of the core of an average star cluster at the distances of the PHANGS--HST roster of galaxies, whereas MCI$_{\rm out}$ traces the faint outer extent of the source.  The exact definition of these parameters are given in \cite{lee21} and \cite{thilker21}.  

\begin{figure}
    \centering
    \includegraphics[scale=0.55]{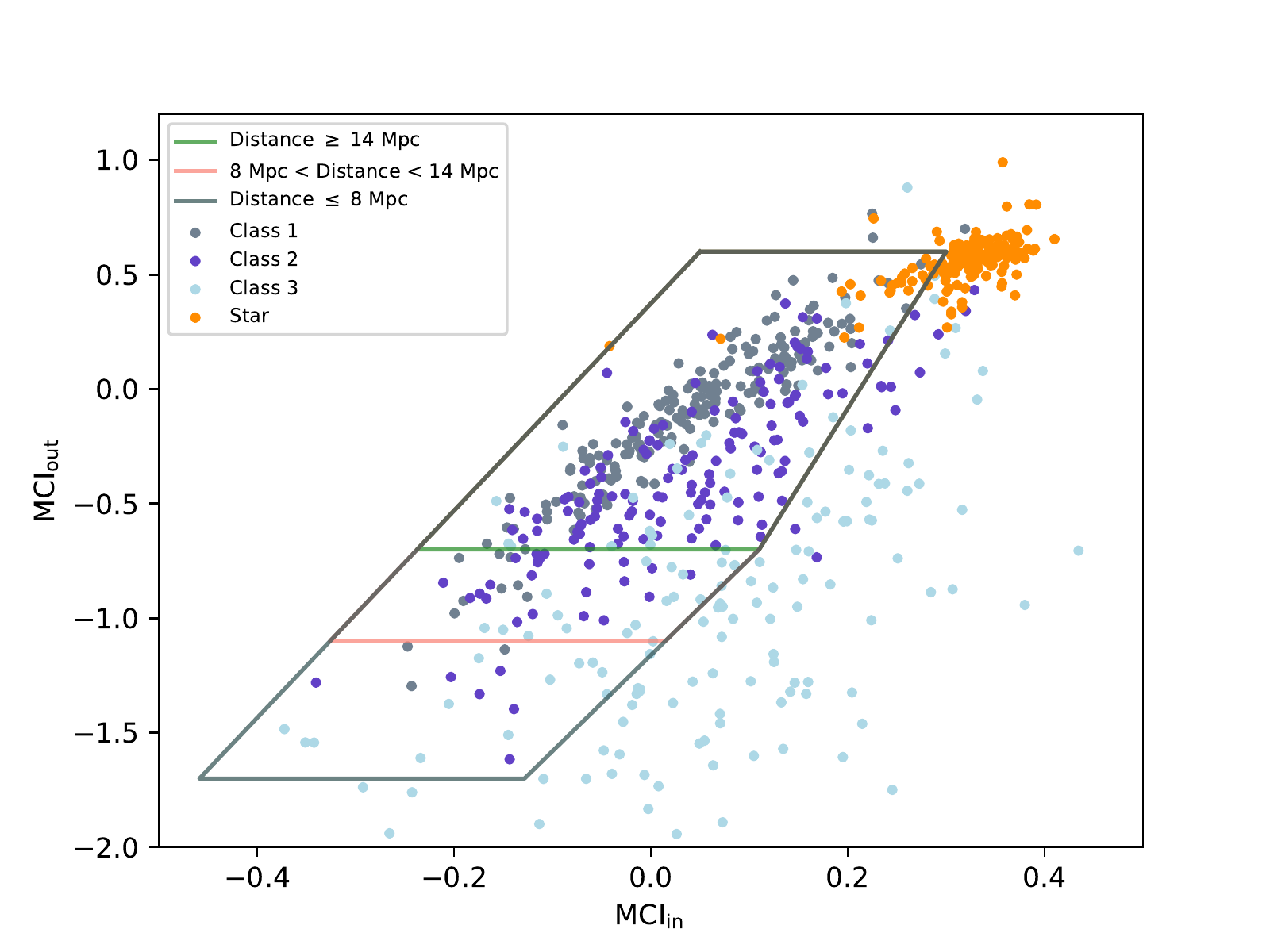}
    \caption{The inner and outer multiple concentration indices (MCI) of specialized sample of bright, relatively isolated class~1, 2, 3 objects and stars, which are used to select star cluster candidates in the PHANGS-HST pipeline. Orange markers denote stars, gray markers class~1, purple markers class~2, and blue markers class~3 objects collected from \ngal galaxies. The polygons provide the candidate selection regions, used in PHANGS--HST data products. The different lower edges of the polygons is dependent on the distance of the parent galaxy. The lower boundary of MCI$_{\rm out}$ is closer to zero for galaxies that are further away (i.e., the selection area shrinks as the clusters become less resolved).}
    \label{fig:mci_fig}
\end{figure}

Figure~\ref{fig:mci_fig} shows the bright, isolated sample in the MCI$_{\rm in}$ versus MCI$_{\rm out}$ space. The polygon defines the region within which candidate clusters are selected for human visual inspection and morphological classification in the PHANGS--HST pipeline from the initial list of source detections.  The four classes of objects are coded by color. 

Three separate polygons are employed by the PHANGS--HST star cluster detection pipeline, all three of which share the same top, left, and right edges. They only differ in their bottom edges, which vary according to three distance bins, as displayed in the legend of the figure. The most distant bin (galaxies with distances greater than $14$~Mpc) has the shallowest edge in MCI$_{\rm out}$, the intermediate distance bin (distance between $8$ and $14$~Mpc) is deeper, and the lowest distance bin (distance below $8$~Mpc) has the deepest polygon edge in MCI$_{\rm out}$. 

We find that almost all class~1 and~2 clusters in our visually selected bright, isolated sample reside within these polygons. Class~2 clusters are positioned similar to class~1's, being a class of centrally concentrated clusters themselves, albeit with a wider spread.  Class~3 associations display the highest amount of spread on this space among the star clusters in our sample, as expected. A significant fraction of Class 3 associations also resides outside of the polygon, as the polygon is not constructed to select for these structures. Rather, we rely on a separate pipeline, based on a watershed algorithm, to identify stellar associations over multiple scales \citep{lee21, larson21}. Finally, the stars are concentrated at the top right of the polygon, with mild overlap with the area surrounded by the polygon. We note that the PHANGS--HST cluster detection pipeline utilizes stellar exclusion regions to account for this overlap, and increase the success rate of cluster detection.  Given that the construction of this specialized bright, isolated sample is independent of the automated PHANGS--HST pipeline selection of cluster candidates, the distribution of this specialized sample in the MCI plane provides a useful check on the robustness of the selection region.

\section{Aperture Corrections}
\label{sec:ap_corr}

In this section, we detail the methodology employed to obtain the aperture corrections for photometry of the PHANGS--HST star clusters. In summary, an average aperture correction in the V-band is derived for each galaxy using the bright, isolated sources described above.  Constant offsets from the V-band aperture correction are applied to compute the corrections for the other four bands (NUV, U,~B,~I).  Photometry for all PHANGS--HST clusters is performed using a circular aperture with a radius of 4~pixels.  The size of the aperture is chosen to roughly correspond to the half-light radius, so the corrections are expected to be ${\sim}0.75$~mag.

In Section~\ref{subsec:comp_ap_corr}, we introduce our three-step aperture correction methodology. In Section~\ref{subsec:quality_cuts}, we provide the quality assurance criteria clusters need to satisfy to be eligible to partake in the aperture correction computation for their respective galaxies. Next, in Section~\ref{subsec:other_bands}, we provide the rationale for choosing to use a constant offset to derive aperture corrections in bands other than the V-band, rather than deriving corrections directly in each filter. In Section~\ref{subsec:apcorr_results} we show the results following the description of our aperture correction methodology. The three step procedure and application of constant offsets represent attempted improvements upon previous practice \citep[e.g.,][]{Adamo17}. We provide comparisons between the techniques in Section~\ref{subsec:apcorr_comparison}. Finally, we investigate the trend between our computed aperture corrections and galaxy distance in Section~\ref{subsec:apcorr_distance}.

The derived aperture corrections and uncertainties are presented in ~\ref{tab:apcor_components} and Tables~\ref{tab:ap_corr_prop}.

\subsection{Computation of the Aperture Corrections}
\label{subsec:comp_ap_corr}

\begin{figure*}
    \centering
    \includegraphics[scale=0.7]{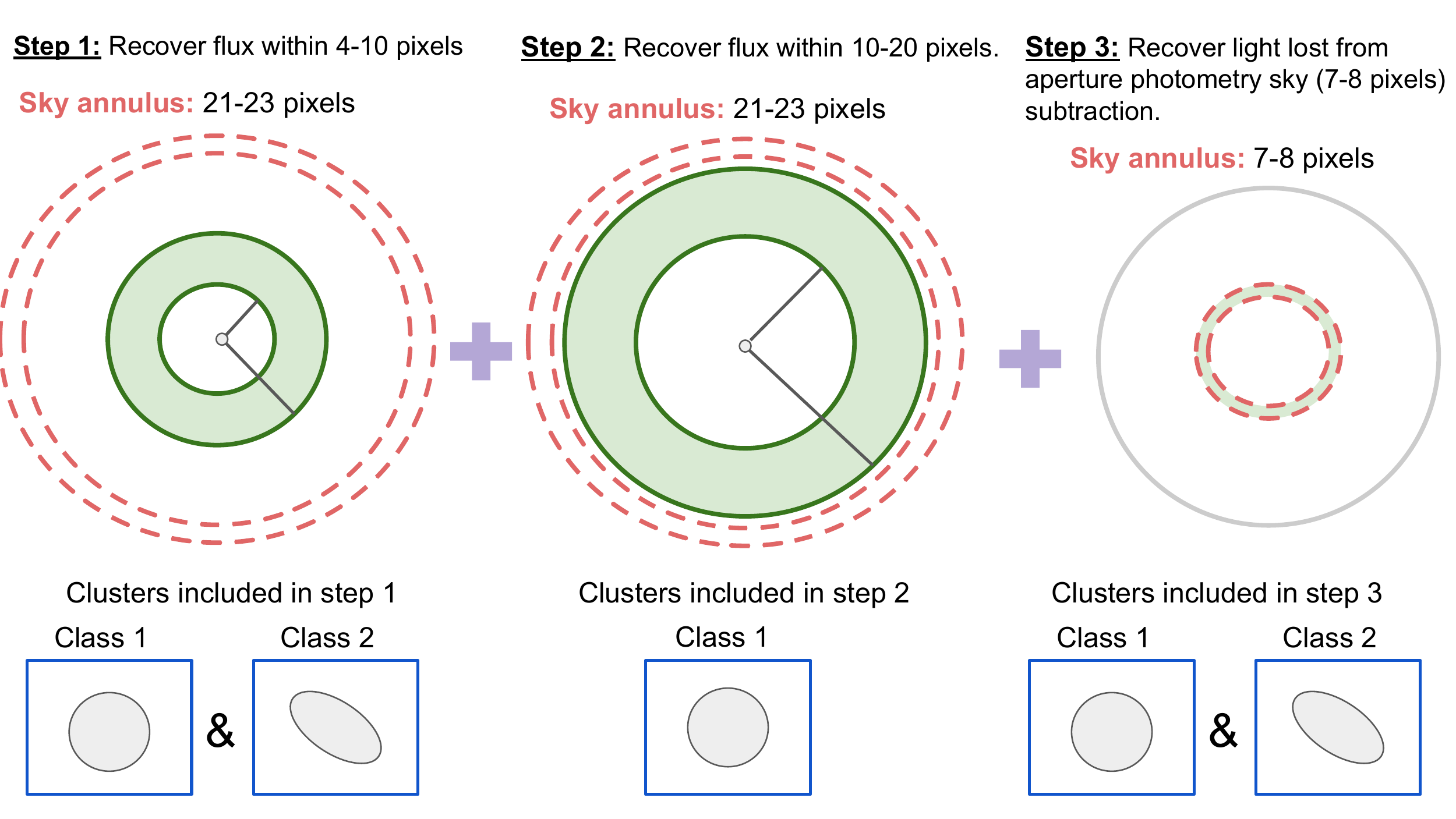}
    \caption{The three-step aperture correction methodology employed by PHANGS--HST, devised to recover the average cluster flux left out by the four-pixel aperture photometry used in the PHANGS--HST data products. The steps illustrated here correspond to the same steps detailed in Section~\ref{sec:ap_corr}.}
    \label{fig:apcorr_strategy}
\end{figure*}

The motivation behind the application of aperture corrections to star cluster photometry is that the modest size of apertures used to perform the photometry does not contain the entire flux of the typical star cluster, and using a larger aperture to encompass a higher fraction of light leads to contamination by other sources due to the crowded nature of many of the regions. A circular aperture containing $\sim$50\% of the cluster flux is normally used. Our aperture correction methodology therefore revolves around recovering an estimate of the light missed by the circular aperture 4~pixels in radius employed by PHANGS--HST data processing, and the 7 to 8~pixel definition for the sky subtraction. Our steps, as illustrated in Figure~\ref{fig:apcorr_strategy}, are:

\begin{itemize}
    \item \textbf{Step~1:} The goal of step~1 is to recover the cluster light between 4 and 10~pixels (Figure~\ref{fig:apcorr_strategy}, left section). Both class~1 and class~2 objects can be used for this step, as our selection scheme for these objects ensures that they do not have neighboring objects closer than 10~pixels.
    
    For each class~1 and~2 cluster in the isolated, bright sample, we first perform 
    aperture photometry in a series of circular apertures with radii increasing from 4~pixels to 10~pixels in single pixel increments. In each case, the sky is set to be between 21 and 23~pixels. We next normalize the flux recovered from each aperture by the value at 10~pixels, doing so for each object individually. Next, we find the mean value of the normalized flux for each aperture, averaging over all objects, subject to the viability conditions described in the next section. Finally, we obtain the correction from the mean growth curve using,
    \begin{equation}
        AC_{\rm s1} = -2.5\times \log(1/f_{4}),
    \end{equation}
    where $AC_{\rm s1}$ denotes the contribution to the aperture correction from step~1, and $f_{4}$ indicates the mean normalized flux within an aperture having a radius of 4~pixels.
    
    \item \textbf{Step~2:} The goal of step~2 is to recover the cluster light between 10 and 20~pixels (Figure~\ref{fig:apcorr_strategy}, middle section). Generally, only class~1 clusters are eligible for a robust estimation of the contribution from this step, since class~2 clusters are typically found in more crowded regions, as discussed above. For this step we therefore limit our sample to clusters of this class~only.
    
    We obtain the contribution to the total correction from this step as follows. For each class~1 cluster that satisfies the viability criteria, we perform aperture photometry in a series of circular apertures with radii between 10 and 20 pixels, with the sky again set to be between 21 and 23~pixels. Then, we normalize the flux at each aperture by the value at 20~pixels for each object individually. Next, find the mean value for each aperture, again averaging over all objects. Finally, we obtain the correction from the mean growth curve using,
    \begin{equation}
        AC_{\rm s2} = -2.5\times \log(1/f_{10}),
    \end{equation}
    where $AC_{\rm s2}$ denotes the contribution to the aperture correction from step~2, and $f_{10}$ indicates the mean normalized flux within an aperture having a radius of 10~pixels.
    
    \item \textbf{Step~3:} The goal of step~3 is to recover the cluster light subtracted away by the definition of the sky annulus utilized by the PHANGS--HST photometry (Figure~\ref{fig:apcorr_strategy}, right section). The PHANGS--HST photometry estimates the median sky using a circular annulus between 7 to 8~pixels from the center of the cluster candidate. Step~3 uses both class~1 and~2 clusters to determine an estimate of cluster light left in this annulus.  
    
    For all class~1 and~2 clusters in a galaxy that satisfy the viability criteria, we perform aperture photometry up to 20~pixels, with sky set to be between 7 and 8~pixels. Then, we calculate the correction for step~3 using,
    \begin{equation}
        AC_{\rm s3} = -2.5\times \log(f_{8}/f_{7}),
    \end{equation}
    where $AC_{\rm s3}$ denotes the contribution to the aperture correction from step~3, and $f_{8}$ and $f_{7}$ indicate the mean normalized flux within apertures having radii of 8 and 7~pixels, respectively. 

\end{itemize}

The total aperture correction is then:
\begin{equation}
AC_{\rm tot} = AC_{\rm s1} + AC_{\rm s2} + AC_{\rm s3}~.
\end{equation}

\begin{table*}
\centering
\centerline{Aperture Corrections of PHANGS-HST Galaxies}
\begin{tabular}{|c|c|c|c|c|c|c|c|c|c|}
\hline
  \multicolumn{1}{|c|}{Galaxy ID} &
  \multicolumn{1}{c|}{$d \pm \delta d$} &
  \multicolumn{1}{c|}{Avg. Ap. Corr.} &
  \multicolumn{1}{c|}{Step 1 Cont.}  &
  \multicolumn{1}{c|}{Step 2 Cont.} &
  \multicolumn{1}{c|}{Step 3 Cont.} &
  \multicolumn{1}{c|}{Total Cat. 1} &
  \multicolumn{1}{c|}{Viable Cat. 1} &
  \multicolumn{1}{c|}{Total Cat. 2} &
  \multicolumn{1}{c|}{Viable Cat. 2} \\
\hline
  NGC 4826 & ~4.41$\pm$0.19 & -0.72 & -0.573 & -0.127 & -0.023 & 11 & 10 & 10 & 10\\
  NGC 3351 & ~9.96$\pm$0.33 & -0.68 & -0.532 & -0.134 & -0.018 & 12 & 10 & 11 & 10\\
  NGC 3627 & 11.32$\pm$0.48 & -0.81 & -0.613 & -0.191 & -0.004 & 10 & 10 & 9 & 9\\
  NGC 2835 & 12.22$\pm$0.94 & -0.81 & -0.618 & -0.178 & -0.019 & 10 & 9 & 11 & 11\\
  NGC 5248 & 14.87$\pm$1.34 & -0.65 & -0.488 & -0.157 & -0.009 & 10 & 9 & 9 & 9\\
  NGC 4571 & 14.9~$\pm$1.1 & -0.61 & -0.416 & -0.172 & -0.017 & 10 & 9 & 10 & 8\\
  NGC 4689 & 15.00~$\pm$2.25 & -0.63 & -0.497 & -0.12 & -0.018 & 10 & 9 & 10 & 10\\
  NGC 4569 & 15.76$\pm$2.36 & -0.76 & -0.541 & -0.212 & -0.004 & 10 & 10 & 10 & 9\\
  NGC 4535 & 15.77$\pm$0.37 & -0.71 & -0.533 & -0.166 & -0.013 & 12 & 10 & 10 & 8\\
  NGC 1792 & 16.20~$\pm$2.43 & -0.8 & -0.614 & -0.185 & -0.005 & 10 & 9 & 10 & 10\\
  NGC 4548 & 16.22$\pm$0.38 & -0.64 & -0.456 & -0.162 & -0.025 & 11 & 11 & 9 & 8\\
  NGC 1566 & 17.69$\pm$2.00 & -0.62 & -0.496 & -0.114 & -0.015 & 16 & 16 & 10 & 10\\
  NGC 1433 & 18.63$\pm$1.86 & -0.62 & -0.51 & -0.096 & -0.019 & 11 & 8 & 10 & 10\\
  NGC 1559 & 19.44$\pm$0.45 & -0.67 & -0.52 & -0.138 & -0.011 & 18 & 17 & 8 & 7\\
  NGC 1365 & 19.57$\pm$0.78 & -0.61 & -0.505 & -0.099 & -0.009 & 16 & 14 & 10 & 10\\
  NGC 4654 & 21.98$\pm$1.16 & -0.83 & -0.576 & -0.247 & -0.007 & 10 & 8 & 10 & 8\\
  NGC 2775 & 23.15$\pm$3.47 & -0.45 & -0.371 & -0.071 & -0.01 & 12 & 10 & 7 & 5\\
\hline\end{tabular}
\caption{\label{tab:apcor_components} Table showing the final aperture corrections per galaxy, contributions from individual steps detailed in Section~\ref{sec:ap_corr}, and numbers of clusters from the bright, isolated sample used for the measurement. The first column is the galaxy ID. Second column shows the distance to the galaxy with its uncertainty, as compiled by \citet{Anand21}. Third column shows the final average aperture correction derived for the galaxy. Columns through four to six display the contributions from the steps detailed in Section~\ref{sec:ap_corr}. Columns~7 and~8 show the total number of class~1 objects in the sample, and those that satisfy the viability conditions of Section~\ref{subsec:quality_cuts}, respectively. Columns~9 and~10 display the same quantities but for class 2 clusters.}
\end{table*}

\begin{table*}
\centering
\centerline{Aperture Correction Scatter}
\begin{tabular}{|c|c|c|c|c|c|}
\hline
  \multicolumn{1}{|c|}{Galaxy ID} &
  \multicolumn{1}{c|}{Avg. Ap. Corr} &
  \multicolumn{1}{c|}{Mean Ind. Corr.} &
  \multicolumn{1}{c|}{Median Ind. Corr.} &
  \multicolumn{1}{c|}{Ind. Corr. St. Dev.} &
  \multicolumn{1}{c|}{Range (max - min)} \\ 
\hline
  NGC 4826 & -0.72 & -0.64 & -0.62 & 0.26 & 1.26\\
  NGC 3351 & -0.68 & -0.57 & -0.55 & 0.26 & 1.0\\
  NGC 3627 & -0.81 & -0.67 & -0.63 & 0.21 & 0.91\\
  NGC 2835 & -0.81 & -0.67 & -0.61 & 0.28 & 1.02\\
  NGC 5248 & -0.65 & -0.56 & -0.55 & 0.19 & 0.74\\
  NGC 4571 & -0.61 & -0.52 & -0.49 & 0.19 & 0.71\\
  NGC 4689 & -0.63 & -0.53 & -0.56 & 0.14 & 0.53\\
  NGC 4569 & -0.76 & -0.66 & -0.66 & 0.22 & 0.88\\
  NGC 4535 & -0.71 & -0.58 & -0.57 & 0.15 & 0.67\\
  NGC 1792 & -0.8 & -0.65 & -0.69 & 0.18 & 0.62\\
  NGC 4548 & -0.64 & -0.53 & -0.5 & 0.28 & 1.01\\
  NGC 1566 & -0.62 & -0.55 & -0.51 & 0.2 & 0.84\\
  NGC 1433 & -0.62 & -0.54 & -0.5 & 0.23 & 0.8\\
  NGC 1559 & -0.67 & -0.57 & -0.56 & 0.18 & 0.67\\
  NGC 1365 & -0.61 & -0.52 & -0.49 & 0.24 & 1.16\\
  NGC 4654 & -0.83 & -0.64 & -0.63 & 0.18 & 0.73\\
  NGC 2775 & -0.45 & -0.45 & -0.41 & 0.21 & 0.79\\
\hline\end{tabular}
\caption{\label{tab:ap_corr_prop} This table provides further details on the aperture correction results presented in Table \ref{tab:apcor_components}. The first and second columns are repeated from Table \ref{tab:apcor_components} and show the galaxy ID, and average aperture correction, respectively. The third and fourth columns are the mean and the median of the distribution of individual cluster contributions to the aperture correction. The fifth column is the standard deviation of the same distribution. Finally, the sixth column is the minimum value of this distribution, subtracted from the maximum.}
\end{table*}

\subsection{Quality Cuts}
\label{subsec:quality_cuts}

In this subsection, we detail the viability conditions demanded of clusters to be eligible for participation in the derivation of the aperture correction. These conditions are as follows.

\begin{itemize}
    \item Any object to be used in the aperture correction calculation needs to have a V-band Vega magnitude brighter than 24 in an aperture with a radius of 4~pixels.
    
    \item The normalized growth curves of individual clusters are required to not decrease by more than 2\% between consecutive aperture radius values.
\end{itemize}

\begin{figure}
    \centering
    \includegraphics[scale=0.55]{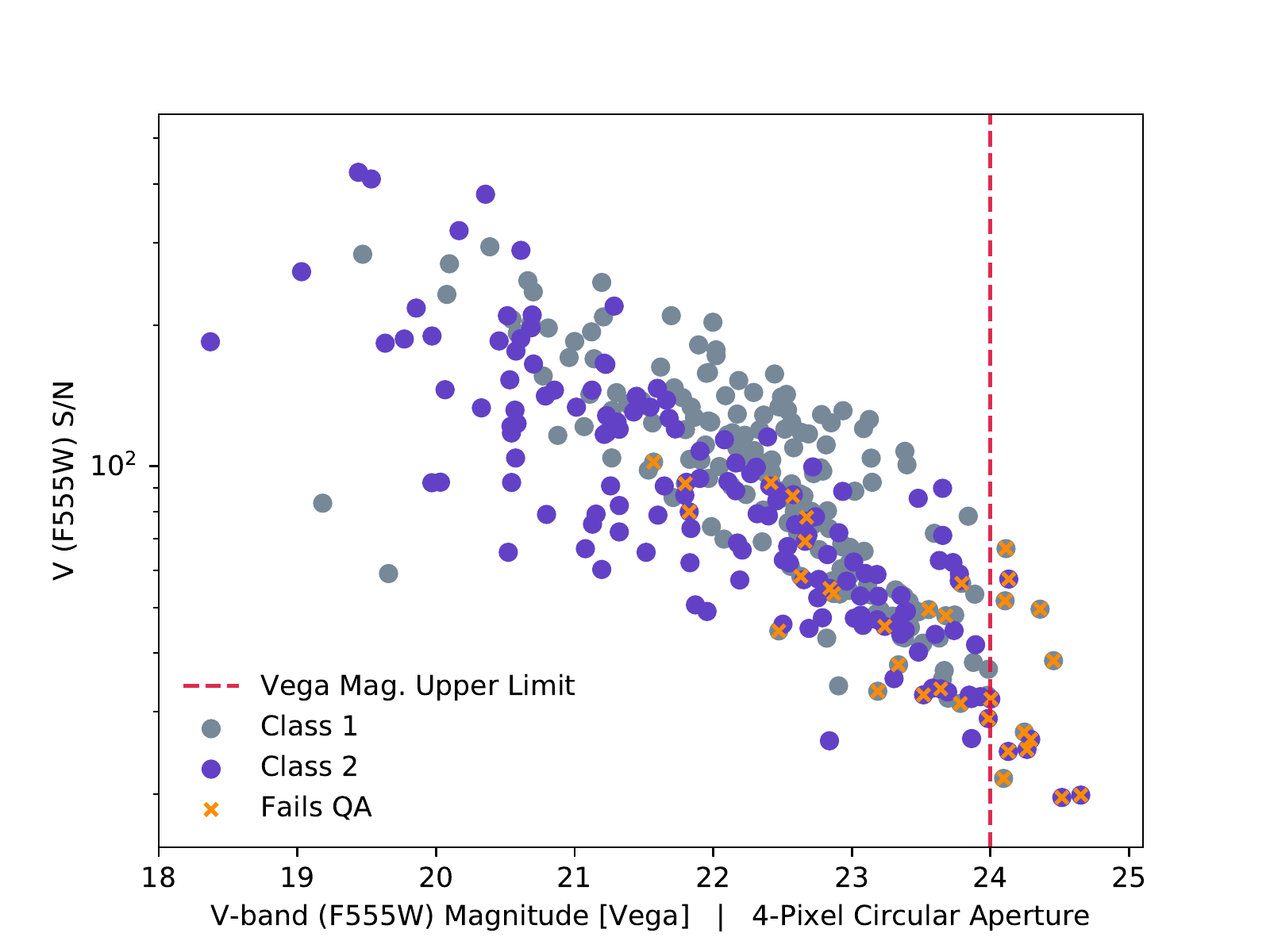}
    \caption{The V-band signal-to-noise (S/N) ratio in log scale versus the F555W Vega magnitude plot of all class~1 and~2 clusters. Both quantities are calculated for a circular aperture 4~pixels in radius. Markers are color-coded by the visually-identified morphological class of the objects; gray markers denote class~1, and purple markers class~2 clusters. Objects that fail to pass the viability conditions described in Section~\ref{subsec:quality_cuts} are shown with an orange cross across the data point, meaning those objects were not used in the derivation of the aperture correction of their respective fields. The V-band lower limit of 24~mags we request as a viability criterion is shown via the red dashed line.}
    \label{fig:v-band_s-to-n}
\end{figure}

Class~1 and~2 clusters in the bright, isolated sample are subjected to these conditions in their V-band imaging, and only those that satisfy both are used in the derivation of the aperture correction. We demonstrate the results of this test in Figure~\ref{fig:v-band_s-to-n}. The clusters that fail either viability condition are denoted with an orange cross. 

\subsection{Aperture Corrections in Bands Other Than the \textit{V}-band}
\label{subsec:other_bands}
In principle, we can apply the above procedures to derive aperture corrections for the clusters in each of the five HST bands used in the PHANGS--HST survey.  However, we instead opt to apply fixed offsets to the measured V-band aperture correction to compute the corrections in the NUV, U, B, and I bands. This effectively assumes the absence of any color gradients. The offsets we apply to the measured V-band aperture correction for NUV, U, B, and I bands are $-0.19$, $-0.12$, $-0.03$, and $-0.12$~mags, respectively. These corrections are determined on the basis of the FWHM of the WFC3 point spread function at their corresponding wavelengths. 

\begin{figure*}
    \centering
    \includegraphics[scale=0.4]{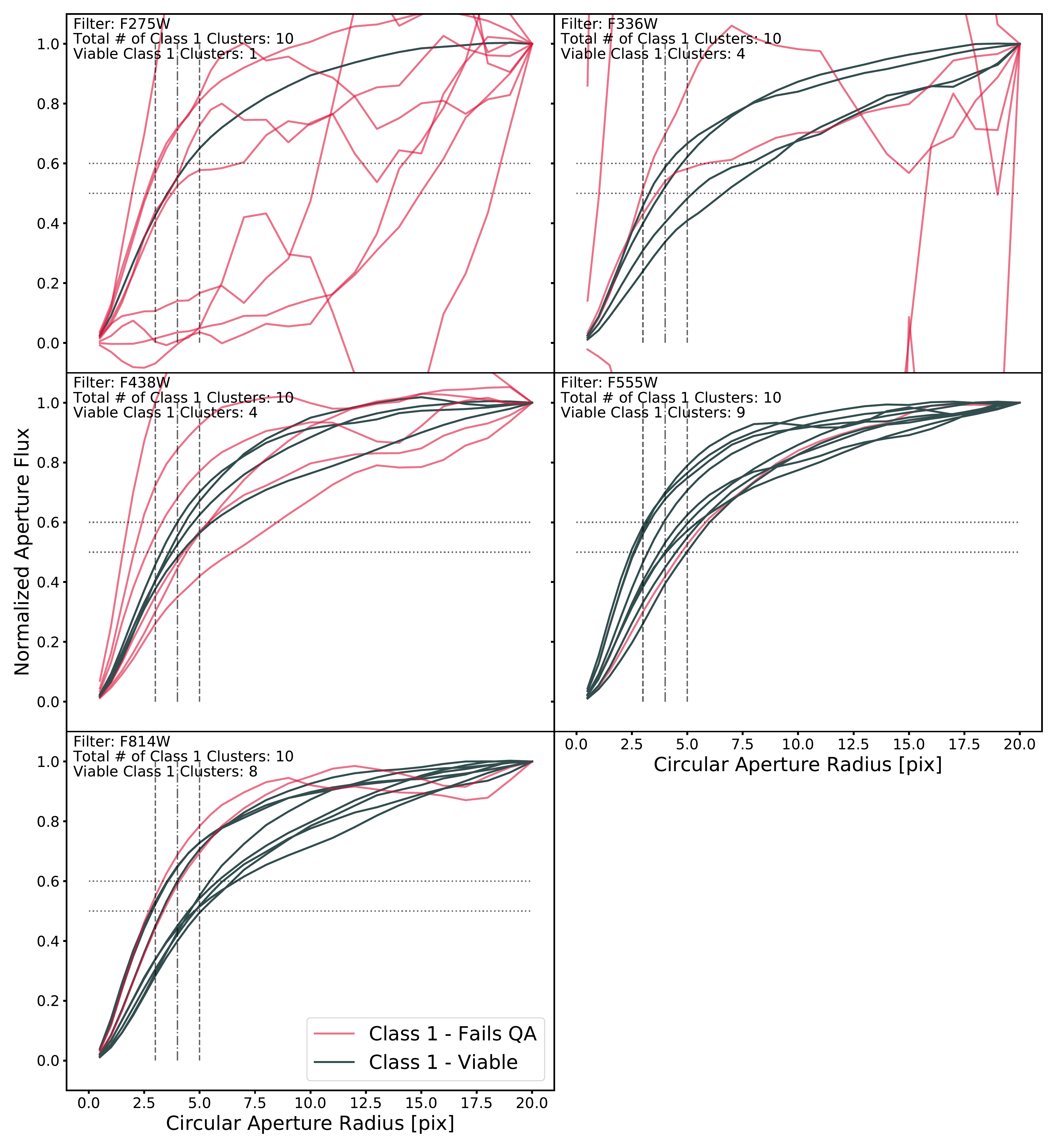}
    \caption{Five panel figure showing the growth curves of class~1 clusters from NGC1792. Each panel shows the growth curves up to 20~pixels for a different filter, as denoted at the top left corner of individual panels. The red curves represent cases that fail to satisfy the viability criteria as detailed in Section~\ref{subsec:quality_cuts}, and gray curves cases that satisfy them. The horizontal dotted lines depict the half, and sixty percent of the total flux, from bottom to top. The vertical dotted lines depict 3, 4 and 5~pixel radius, from left to right.}
    \label{fig:5band_gc}
\end{figure*}

The main reason motivating the use of fixed offsets is the difficulty in maintaining a statistically robust set of clusters in bands bluer than the V-band. We find that many of our fields have either too few, or no clusters that satisfied the viability criteria detailed in Section~\ref{subsec:quality_cuts}, especially in NUV. We demonstrate the difficulty in deriving a robust aperture correction in all five PHANGS--HST bands in Figure~\ref{fig:5band_gc} for NGC~1792. Even though only one cluster was eliminated from use in the derivation of the aperture correction in the V-band, all clusters but one fail the viability criteria in the NUV. Attempting to include clusters to the sample that satisfies the viability conditions in NUV forces them to be in crowded regions, resulting in their light profiles being affected by neighboring objects, thereby providing less reliable corrections. We decided to proceed with applying the constant offsets to a robust measure of the V-band correction, instead.

\begin{figure*}
    \centering
    \includegraphics[width=.57\textwidth]{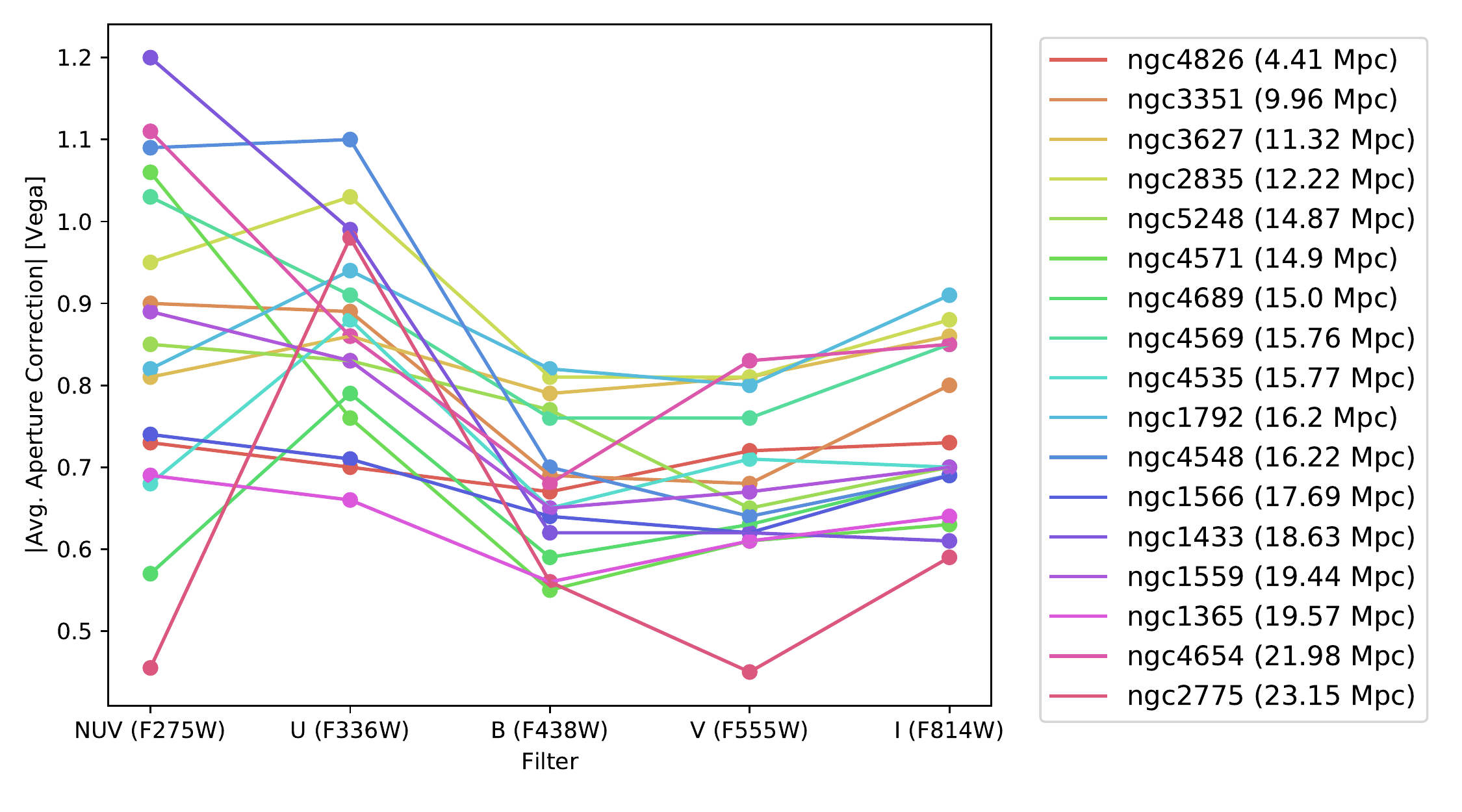}
    \includegraphics[width=.42\textwidth]{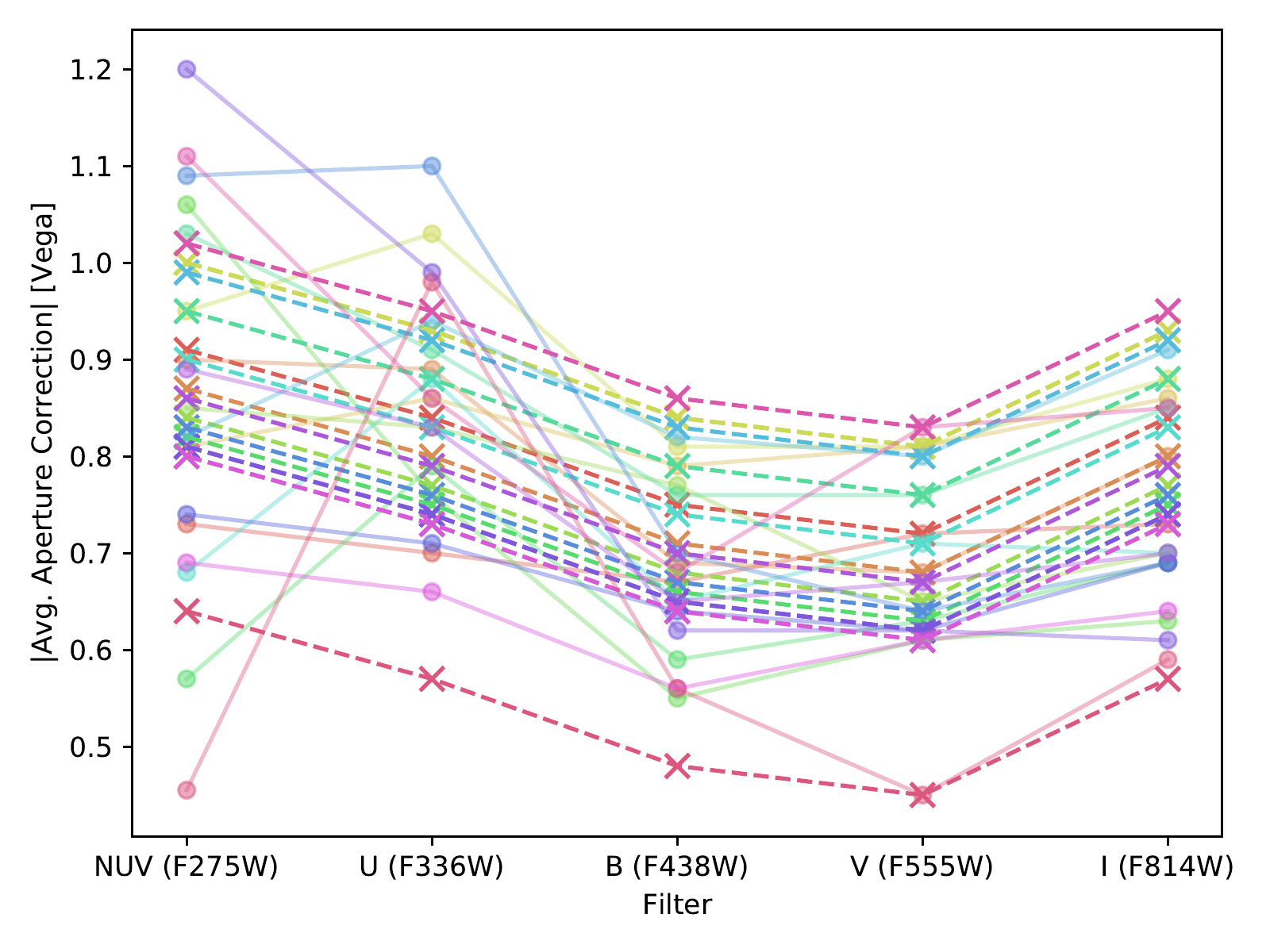}
    \caption{\textit{Left panel --} The average aperture correction values derived for each galaxy in our sample, using the methodology described in Section~\ref{subsec:comp_ap_corr}, for all five PHANGS--HST bands. The distance to each galaxy is shown in the legend, together with the color of the line for the galaxy aperture correction values. Note the pronounced variance in the correction computed in F275W. \textit{Right panel --} The fixed offset applied to the V-band average aperture correction of individual fields (dashed lines and markers), plotted over the results from the left panel (solid lines and markers, transparency applied). Plot colors same as the left panel.}
    \label{fig:5band_apcorr_allfields}
\end{figure*}

We demonstrate the 5-band average aperture corrections we would obtain with and without employing the constant offset approach in Figure~\ref{fig:5band_apcorr_allfields}. The left panel shows the 5-band aperture corrections as obtained via the technique described in Section~\ref{subsec:comp_ap_corr}, for all the galaxies in our sample. The right panel shows the corrections after applying the constant offsets to the V-band correction, as described in this section. The figure demonstrates the wide range in corrections we obtain especially in NUV, as a result of the behavior we demonstrated in Figure~\ref{fig:5band_gc}.  

\begin{figure}
    \centering
    \includegraphics[width=.49\textwidth]{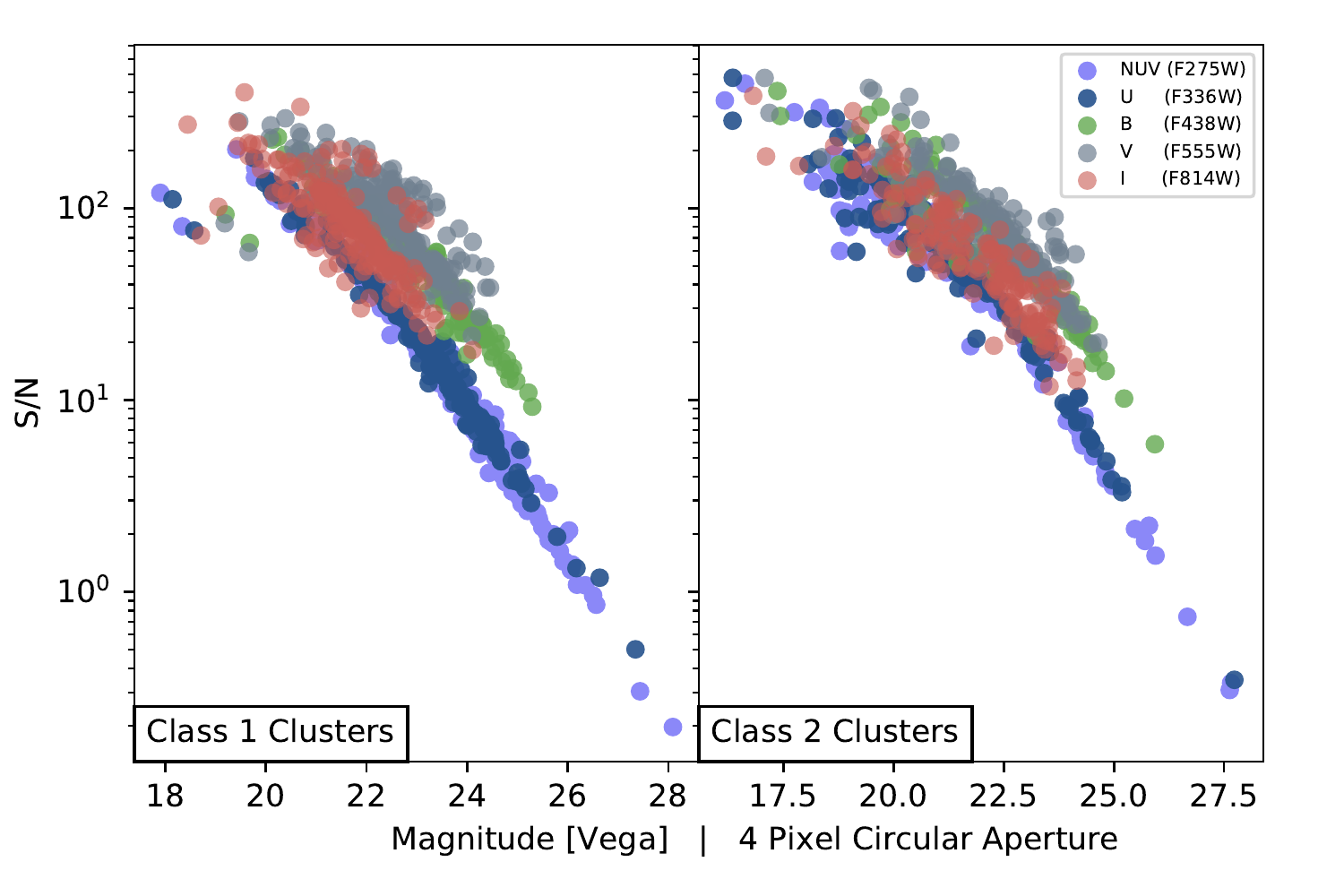}
    \caption{The S/N ratio versus the Vega magnitude of class~1 (left panel) and class~2 (right panel) clusters, in all five PHANGS--HST bands. Both have been derived for a circular aperture with a radius of 4~pixels. Aperture corrections have not been applied to the magnitude values. The data are color-coded by the filter where the measurements are made, as shown in the legend in the right panel. The $y$-axis is displayed in log scale.}
    \label{fig:5band-s/n}
\end{figure}

We further investigate the behavior of class~1 and~2 bright, isolated clusters in all five bands in Figure~\ref{fig:5band-s/n}. The figure shows the S/N versus magnitude distribution of these two morphology types in all five bands. We see that there is a significant shift towards the faint and low S/N part of the plot in bluer bands, especially for class~1 clusters, demonstrating the difficulty in obtaining robust aperture corrections in all five bands.

These results are not surprising given that star formation occurs in a clustered manner, and so young clusters will be found in crowded regions. They are rarely sufficiently isolated to enable a robust measure of the extended radial profile of a cluster, and this represents a basic limitation on the accuracy of aperture corrections.  As discussed above and illustrated in the UB-VI color--color (Figure~\ref{fig:ub-vi}) and mass-age diagrams, the majority of class~1 objects are not among the youngest of our clusters, ergo unlikely to have significant emission in NUV from their red dominant stellar populations. 

\subsection{Aperture Correction Results}
\label{subsec:apcorr_results}

For the \ngal galaxies examined here, we find that corrections for V-band photometry performed in a circular aperture  with a radius of 4 pixels range from $-0.45$ to $-0.83$~mags, with a median of $-0.67$~mag.  By construction the magnitude of the correction will be about a factor of two since the size of the aperture was chosen to roughly correspond to the half-light radius.

\begin{figure*}
    \centering
    \includegraphics[scale=0.6]{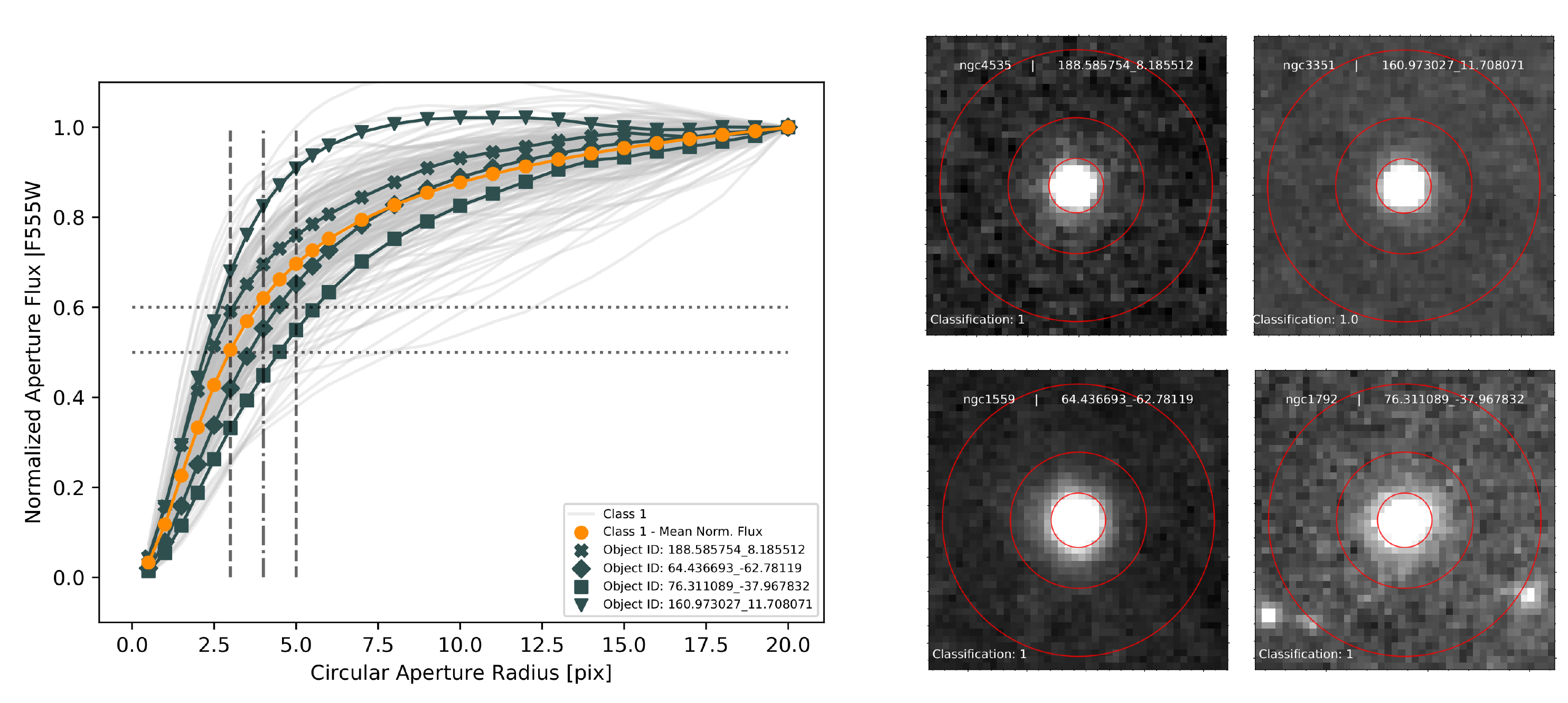}
    \caption{20~pixel growth curve profiles of representative class~1 clusters from our sample, together with their postage stamps. The figure on the left displays the growth curves of the entire set of class~1 clusters that satisfy the viability conditions listed in Section~\ref{subsec:quality_cuts}, in light gray. We display the mean growth curve of this set in orange. Highlighted in dark gray are growth curves belonging to select class~1 clusters whose postage stamps are displayed on the right. We indicate which object the curve belongs to by marking individual pixel sizes with markers of different shape. Object ID's for the markers are given in the legend, as well as on the top part of the postage stamps. The red circles in the postage stamps have radii of 5, 10, and 20~pixels, starting from the innermost circle.}
    \label{fig:cat1_profiles}
\end{figure*}

\begin{figure*}
    \centering
    \includegraphics[scale=0.65]{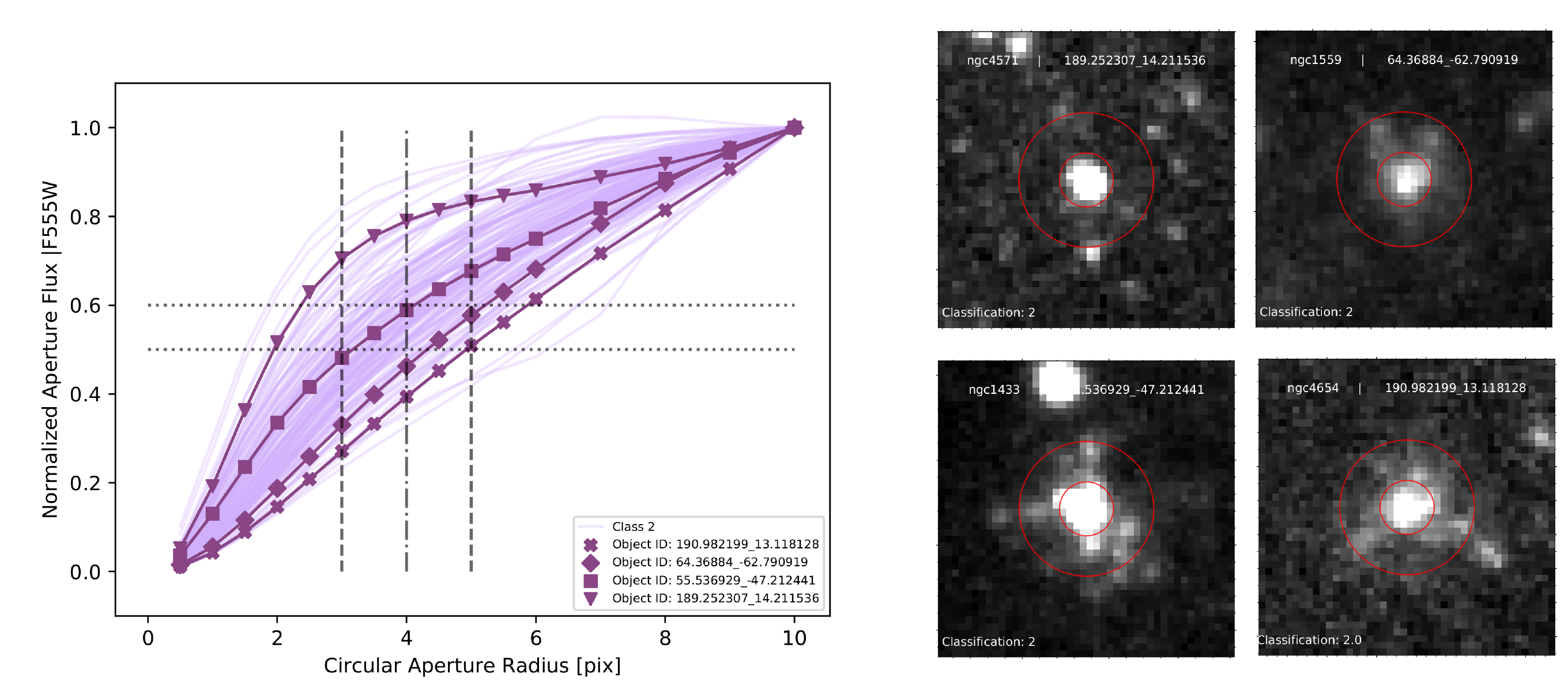}
    \caption{V-band growth curve profiles up to 10~pixels of all viable class~2 clusters from our sample, shown in light purple. Plotted on top of the entire sample are four select clusters, with their postage stamps displayed next to the plot. Different markers are used to denote the clusters for which we display the object ID's both in the legend, and at the top of the postage stamp. The red circles in the postage stamps have radii of 5~and~10 pixels, starting from the innermost circle.}
    \label{fig:cat2_profiles}
\end{figure*}

In Figures~\ref{fig:cat1_profiles} and~\ref{fig:cat2_profiles}, we show the growth profile of every class~1 and~2 cluster from across \ngal galaxies that satisfy the viability conditions described in Section~\ref{subsec:quality_cuts}. These figures demonstrate the success of our sample construction in obtaining clean V-band growth curves up to 20~pixels for class~1, and up to 10~pixels for class~2 objects.

\subsection{Comparison with Previous Results}
\label{subsec:apcorr_comparison}

We can also compare our aperture corrections to those derived by the LEGUS program \citep{legus} for the 3~fields in common; NGC~1433, NGC~1566, and NGC~3351. LEGUS uses a different technique to compute their aperture corrections, and not the three-step technique we introduce in this paper. The first step for their averaged aperture correction approach (i.e., a single correction is computed for all clusters in a galaxy; we do not compare to the CI-based corrections here, see \citealt{Adamo17} and \citealt{Cook19} for details) is similar to ours in that it entails the construction of a bright and isolated cluster sample. They employ a varying aperture radius, fixed at the smallest integer radius that contains at least 50\% of the cluster flux, as derived using this sample for a given galaxy. The sky annulus for this computation is located at a radius of 7~pixels and is 1~pixel wide. Then, the average aperture correction is computed as the difference between the magnitude of the source within a 20~pixel circular aperture (sky median computed between pixels 21 and~22), and the magnitude of the source using the varying aperture size above. Both programs have employed a 4~pixel circular aperture for the three overlapping fields, allowing for a one-to-one comparison of the corrections computed utilizing different techniques.

\begin{figure}
    \centering
    \includegraphics[width=.48\textwidth]{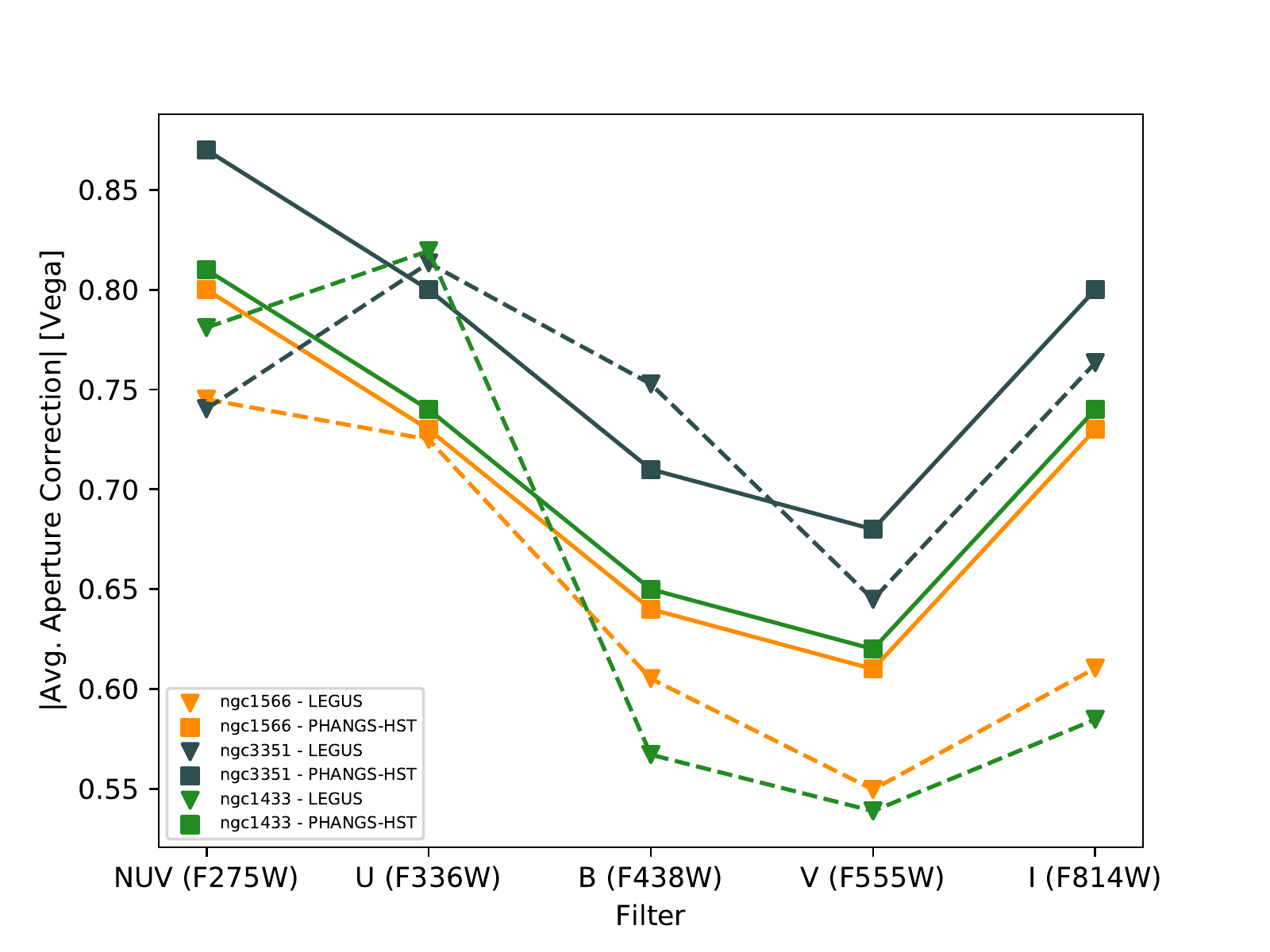}
    \caption{The 5-band average aperture corrections comparing galaxies overlapping between LEGUS and PHANGS--HST. The full lines and square markers represent the PHANGS--HST results, whereas the dashed lines and triangle markers represent LEGUS results. Same colors represent the same galaxy. Both programs employ a circular aperture 4~pixel in radius for the set of galaxies demonstrated here. The PHANGS--HST corrections for NGC~1566 have been offset by $0.01$~mag in all bands, to distinguish its corrections from the identical values we find for NGC~1433.}
    \label{fig:phangs_legus_comp}
\end{figure}

We show the 5-band average aperture correction comparison in Figure~\ref{fig:phangs_legus_comp}. This figure shows that the two programs find V-band corrections that differ by less than $0.1$~mag, with the PHANGS correction being higher. There are no other systematic differences we can establish. Maximum separation occurs in the I-band corrections for NGC~1433 with a difference of $0.16$~mag. The constant offset we apply to our V-band correction is steeper for this case than the mild difference between the LEGUS V~and I-band corrections. This is not the case for NGC~3351, where a similar increase in V~and I-band corrections are found by both programs. On the blue end, we find that the LEGUS NUV corrections for NGC~1433 and NGC~3351 are lower than their U-band corrections, contrary to the expectation from the PSF. We see a similar behavior in the U~and NUV-computed corrections for many galaxies in Figure~\ref{fig:5band_apcorr_allfields}, a behavior we bypass by adopting the corrections obtained via the application of constant offsets to the V-band correction. For NGC~1433, there is also a significant increase from the LEGUS B-band correction to the U-band, much steeper than the constant offset expectation, the consequences of which we discuss below. 

\begin{figure*}
    \centering
    \includegraphics[width=.48\textwidth]{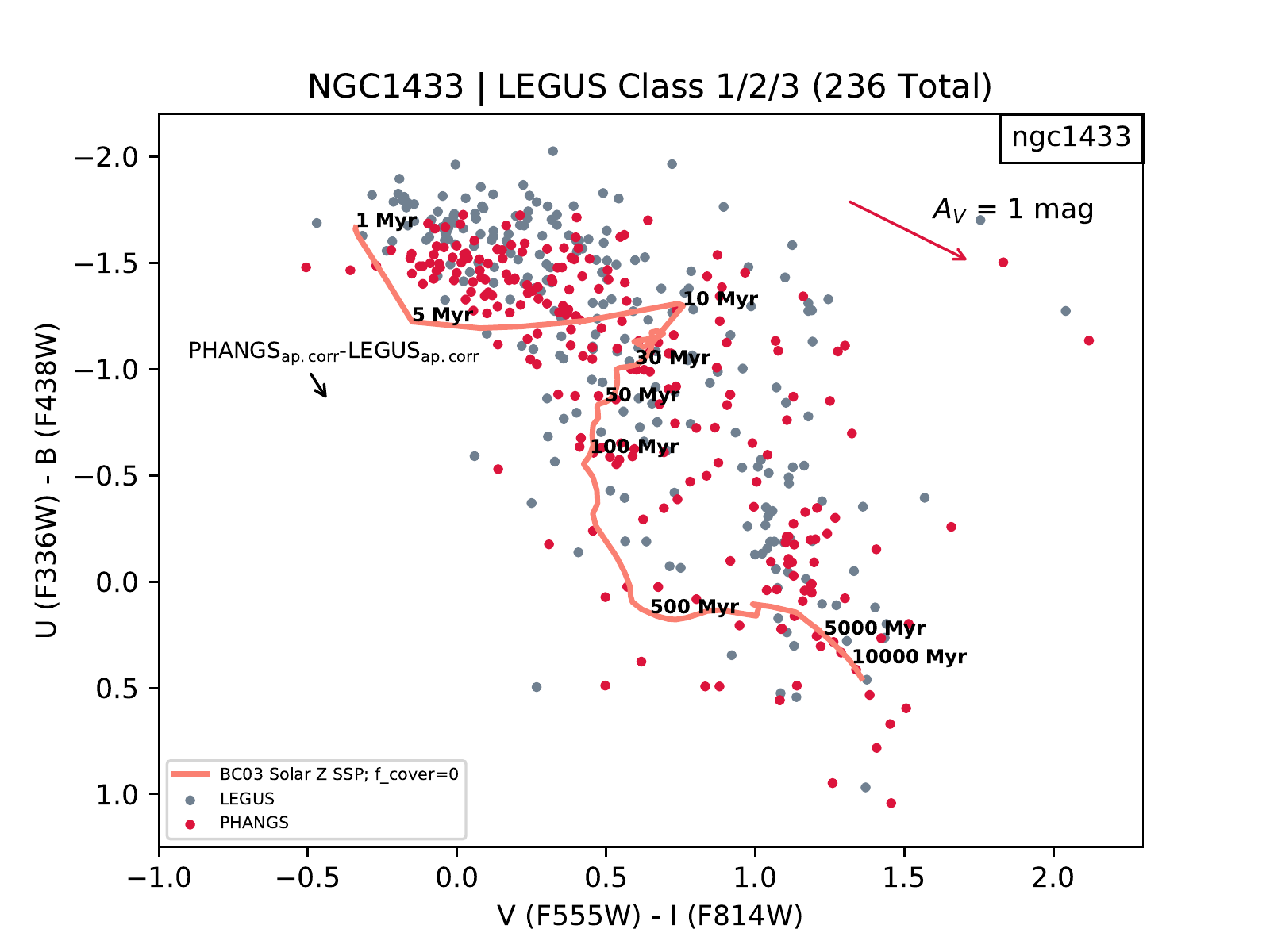}
    \includegraphics[width=.48\textwidth]{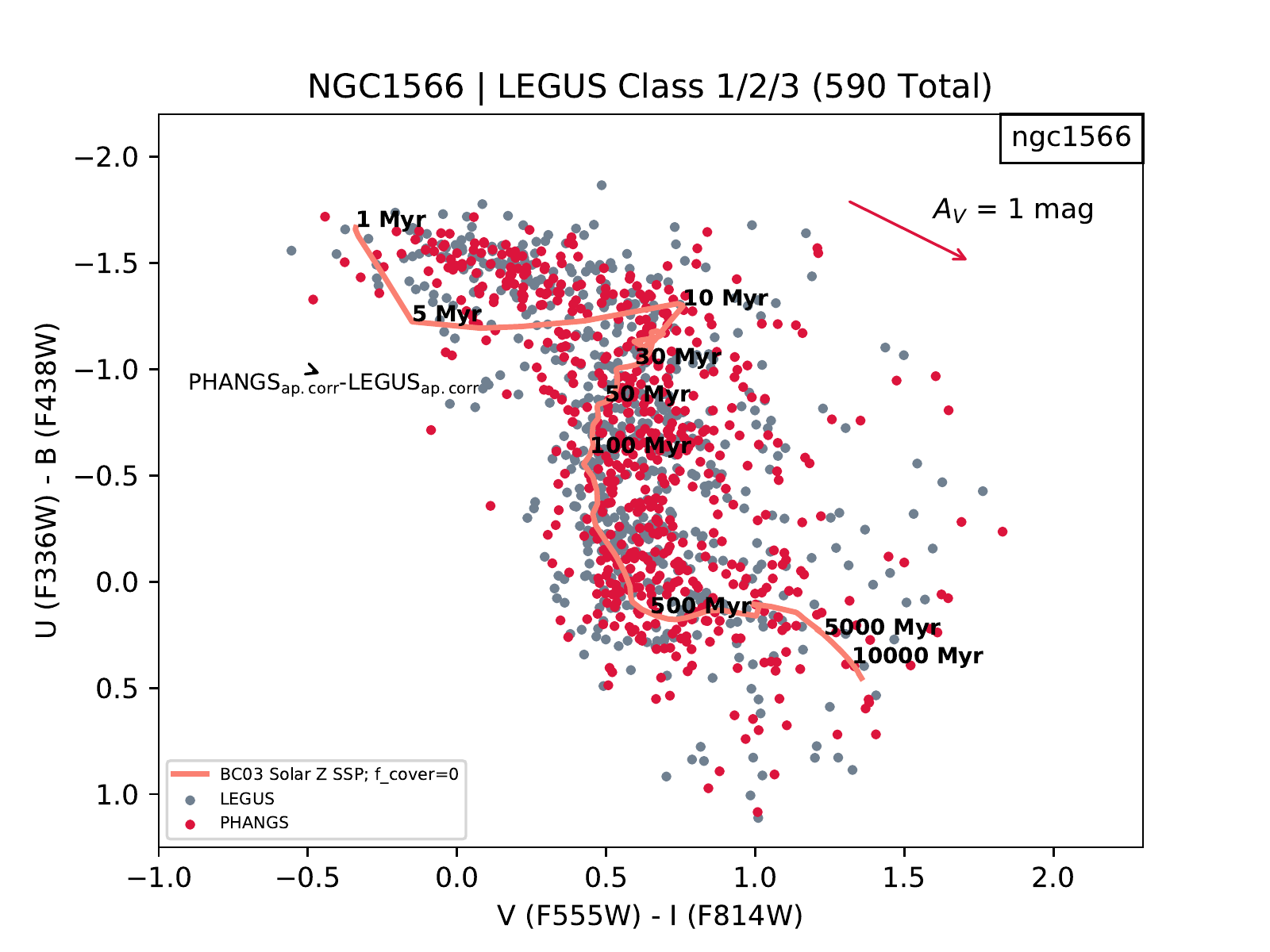}
    \caption{The UB-VI distributions of LEGUS and PHANGS--HST for two galaxies analyzed independently by each program, NGC~1433 (left panel), and NGC~1566 (right panel). The colors are computed for the total (aperture corrected) Vega magnitudes using a circular aperture 4~pixels in radius by both programs. The red markers denote PHANGS--HST data, and gray markers denote LEGUS data. The BC03 track depicting the UB-VI evolution of an SSP is shown as the pink curve. The reddening vector corresponding to $A_{V} = 1$~mag is shown as the red arrow. Both panels also include a vector depicting the shift that occurs on this space solely due to the differences in aperture corrections employed by each program (shown in Figure~\ref{fig:phangs_legus_comp}). The vector is obtained through subtracting the LEGUS corrections from those of PHANGS--HST in each band. We find that this vector is negligible in NGC~1566, which is not the case for NGC~1433. It depicts a significant shift towards the SSP track for especially the young clusters for the PHANGS--HST data (ages less than 10~Myr). The figure also shows that many LEGUS clusters are above the bluest U-B color value of the SSP track in NGC~1433.}
    \label{fig:phangs_legus_ubvi}
\end{figure*}

We further investigate the differences in PHANGS--HST and LEGUS total photometry in Figure~\ref{fig:phangs_legus_ubvi}. This figure demonstrates the distributions in the UB-VI space for PHANGS and LEGUS data, for the class~1, 2, and~3 clusters that are a match between the programs. Colors computed independently for both programs have been aperture corrected by the corrections shown in Figure~\ref{fig:phangs_legus_comp}. We display a vector mapping the difference in the UB-VI values of each program solely due to differences in aperture corrections in each panel. Any shifts in the colors of clusters not accounted for by this vector is a result of the differences in photometry performed by the programs. Following the comparison in Figure~\ref{fig:phangs_legus_comp}, this vector is negligible for NGC~1566 , but not for NGC~1433. As a result of the pronounced difference in the U~and B-band corrections, the distributions are offset. We find that the PHANGS--HST data is consistent with the SSP model in the panel with modest reddening. Much higher reddening values are required to make LEGUS colors consistent with the SSP model, especially at the young end. It is also important to note that due to this offset, many LEGUS clusters are bluer in their U-B color than the bluest U-B value of the SSP track. The color distributions of the two programs for NGC~1566 are indistinguishable.

We gauge the effects of the different corrections with the following test. We compare the ages derived using PHANGS--HST photometry corrected by the aperture corrections in this paper, with the ages derived using PHANGS--HST photometry corrected by the LEGUS aperture corrections. The ages obtained through the two approaches are comparable in NGC~3351 and NGC~1566, but there is an overall mean scatter of $0.1$~dex in NGC~1433. This approaches $0.2$~dex for clusters around the ages of $10{-}50$~Myr, where the SSP track curves back on itself, and the type of differences we describe in the previous paragraph can make a significant difference.

\subsection{Does the aperture correction depend on galaxy distance?}
\label{subsec:apcorr_distance}

\begin{figure*}
    \centering
    \includegraphics[width=.48\textwidth]{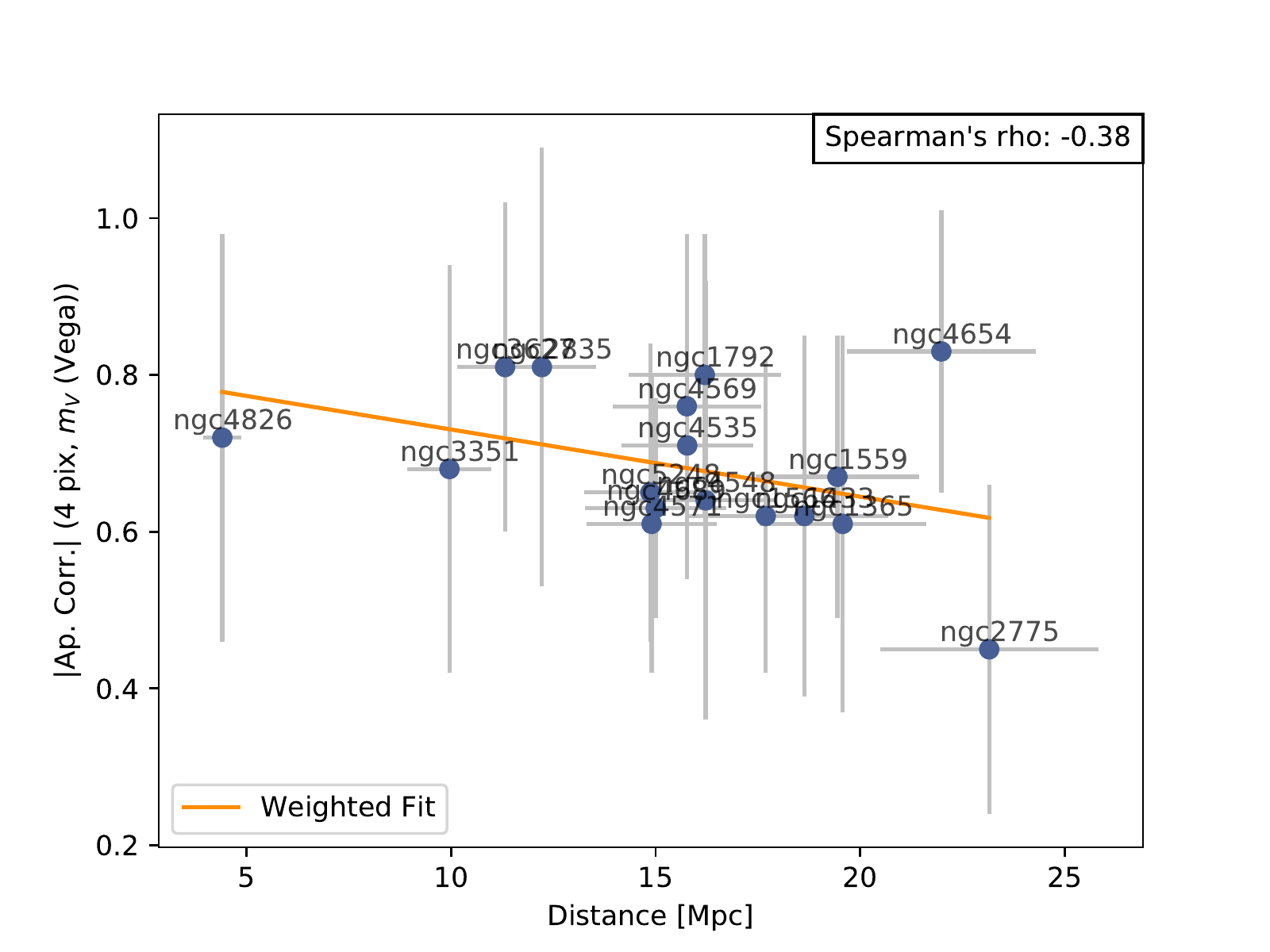}
    \includegraphics[width=.48\textwidth]{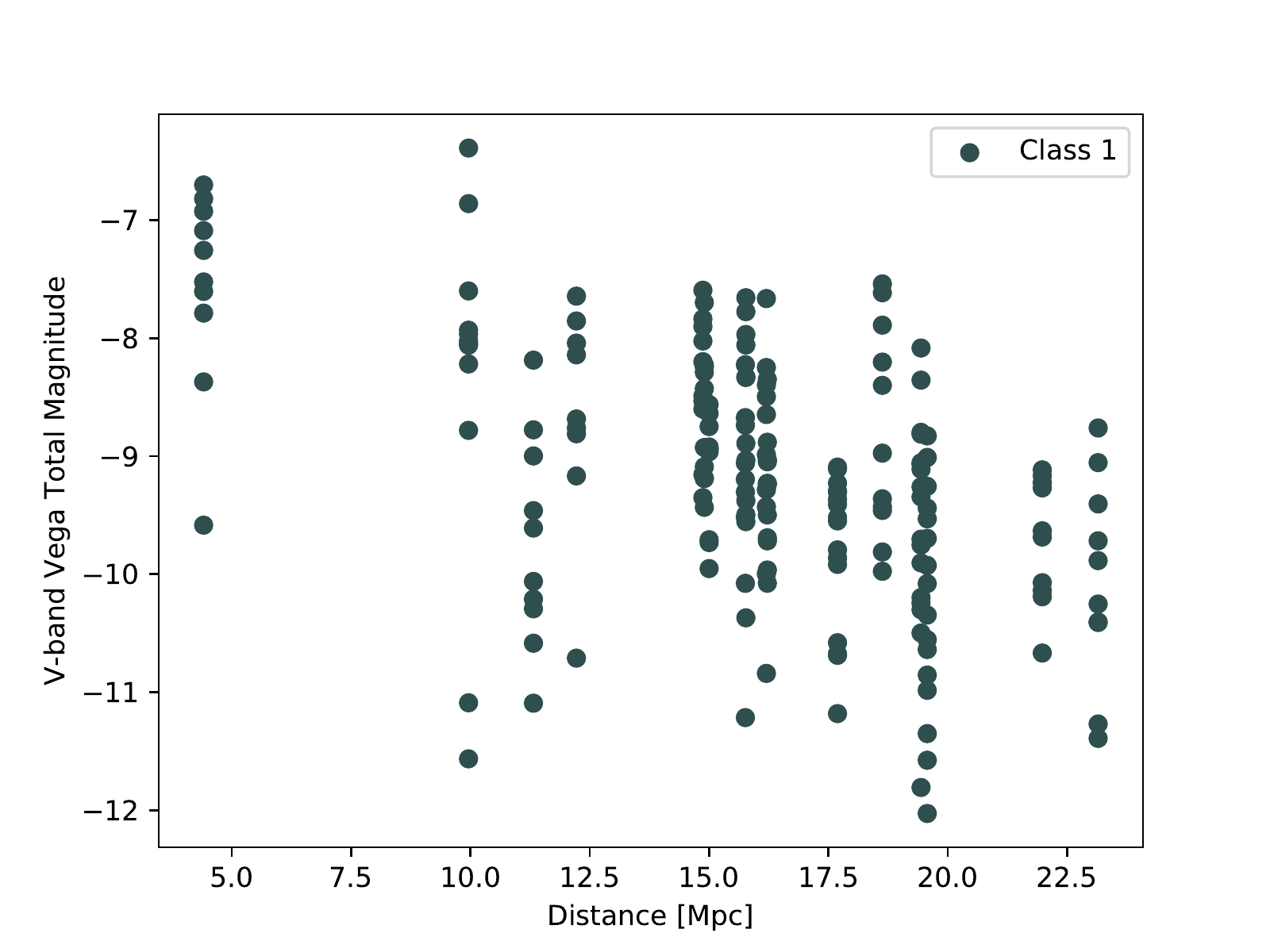}
    \caption{\textit{Left panel --} The absolute value of the average aperture correction of galaxies from our samples versus their distances. The error bars for the average aperture correction values are the standard deviation of the distribution of individual aperture correction clusters. The distances and their uncertainties displayed here are presented in \citet{Anand21}. The orange line is a weighted linear fit to the data. We also provide the Spearman's rank correlation coefficient, $\rho$, for this distribution on the top right of the panel. \textit{Right panel --} The absolute V-band magnitude of class~1 clusters from our sample.}
    \label{fig:ap.corr_vs_distance}
\end{figure*}

Finally, we show the average aperture correction versus distance in the left panel of Figure~\ref{fig:ap.corr_vs_distance}. For identical clusters we would expect the aperture correction to decrease with increasing distance, since a larger fraction of the light in the cluster falls within the 4~pixel measurement radius, i.e., a constant angular size on the sky encompasses a larger linear radius for more distant galaxies. As expected, Figure~\ref{fig:ap.corr_vs_distance} shows a decreasing relation between average aperture correction and distance for our sample, demonstrated both as a weighted fit to the data, and using a Spearman's rank correlation test. The Spearman rank correlation coefficient we find for our distribution is $\rho = -0.38$, consistent with a monotonically decreasing relation, albeit not a strong one.

The right panel of Figure~\ref{fig:ap.corr_vs_distance} shows the reason we do not find a steeper decline. For the more distant galaxies only the brightest, and hence largest clusters are above our magnitude cutoff. Hence the clusters in the nearby galaxies are not identical to those in the more distant galaxies, modifying the expected steep relation by making it flatter. As future work, we plan to revisit the computation of aperture corrections using a more homogeneously selected sample. We expect this to result in corrections that are typically ~0.1  mag different than the current estimates. 

\section{Quantitative Measures of Morphology}
\label{sec:morphology}
In this section we describe the methodology used in our exploration of the application of quantitative measures of morphology for star clusters. The bright, isolated sample of star clusters we analyze in this paper provides an ideal set to test such measures of morphology, due to high S/N, and the low contamination of their cluster light from neighboring objects enforced by the criterion built in to their selection. 

The first measure we analyzed is a non-parametric measure of morphology from \cite{Lotz04}, the $M_{20}$ parameter, which has been extensively applied to galaxies \citep{Lotz08, Peth16}. $M_{20}$, or the normalized  second order moment of the 20\% brightest pixels, is a parameter that measures how spatially separated the brightest regions of the given object are. When the spatial separation, or the spatial variance, of the pixels ranking among the 20\% brightest of pixels are large, the object has a large $M_{20}$ measure. If the spatial separation of the 20\% brightest pixels is low, then so is the $M_{20}$ value of the object. \cite{Lotz04} mainly used $M_{20}$, together with the Gini coefficient, $G$, in an astronomical setting (how egalitarian the distribution of flux is for the given object), to identify merging galaxies. In this paper, we will solely focus on $M_{20}$, with the investigation of other parameters, including~$G$, planned as future work.

We briefly summarize the derivation of $M_{20}$ here, and refer to \cite{Lotz04} for further details. One quantity needed to compute $M_{20}$ is the total second-order moment of object flux, $M_{\rm tot}$. $M_{\rm tot}$ is computed as a sum of the pixel flux multiplied by the squared distance of the pixel to the object center, for each pixel present in the segmentation map:

\begin{equation} \label{eq:Mtot}
M_{\rm tot} = \sum_{i}^{n} M_{i} = \sum_{i}^{n} f_{i}[(x_{i} - x_{c})^{2} + (y_{i} - y_{c})^{2}]. 
\end{equation}

Here $i$ denotes the current pixel, $n$ is the total number of pixels in the segmentation map, and ($x_{c}$, $y_{c}$) are the coordinates of the center of the light source. The center of the source is designated as the coordinates where $M_{\rm tot}$ is at its minimum. $M_{\rm tot}$ provides the normalization factor used when calculating $M_{20}$. The computation of $M_{20}$ involves a sum similar to that in Equation \ref{eq:Mtot}, but performed over only the 20\% brightest pixels:

\begin{equation} \label{eq:M20}
M_{20} \equiv \mathrm{log10} \left( \frac{\sum_{i} M_{i}}{M_{\rm tot}} \right),~\mathrm{while~} \sum f_{i} < 0.2f_{\rm tot}. 
\end{equation}

In this equation $f_{i}$ is the flux value of individual pixels that reside in the segmentation map, and $f_{\rm tot}$ is the sum of the flux from all such pixels. Values of $M_{20}$ are negative as it is the logarithm of a quantity in the range (0,1).  

Being a non-parametric measure, $M_{20}$ does not assume any functional form for the light profile of the sources, making it a parameter well-suited for analyzing the wide morphological range of star clusters. Furthermore, as only the brightest 20\% percent of the pixels determine its value, the measure is less likely to be affected by the varied background within which the clusters reside, making it an appealing parameter to start with. Here we test the behavior of the clusters in our specialized sample in this parameter. Our procedure is as follows. First, we make $11 \times 11$ pixel cutouts centered on our clusters using the V-band image. This size is chosen to comply with BCW visual identification of especially class~3 clusters, where multiple distinct peaks within a 5~pixel radius is required. Next, we run \textit{statmorph}, detailed in \cite{Rodriguez-Gomez19} on these cutouts, to measure $M_{20}$. Together with the cutouts \textit{statmorph} requires their corresponding segmentation maps. Our tests find that the value of $M_{20}$ is robust against the specifics of the segmentation maps provided, where we have provided segmentation maps ranging from those derived by requiring at least 5 contiguous pixels between segments, and those that include every single pixel in the cutout. We also provide \textit{statmorph} with the PHANGS--HST weight maps for use during the derivation of these parameters. The weight maps provide the standard deviation of the flux at each pixel value. 

\begin{figure}
    \centering
    \includegraphics[scale=0.6]{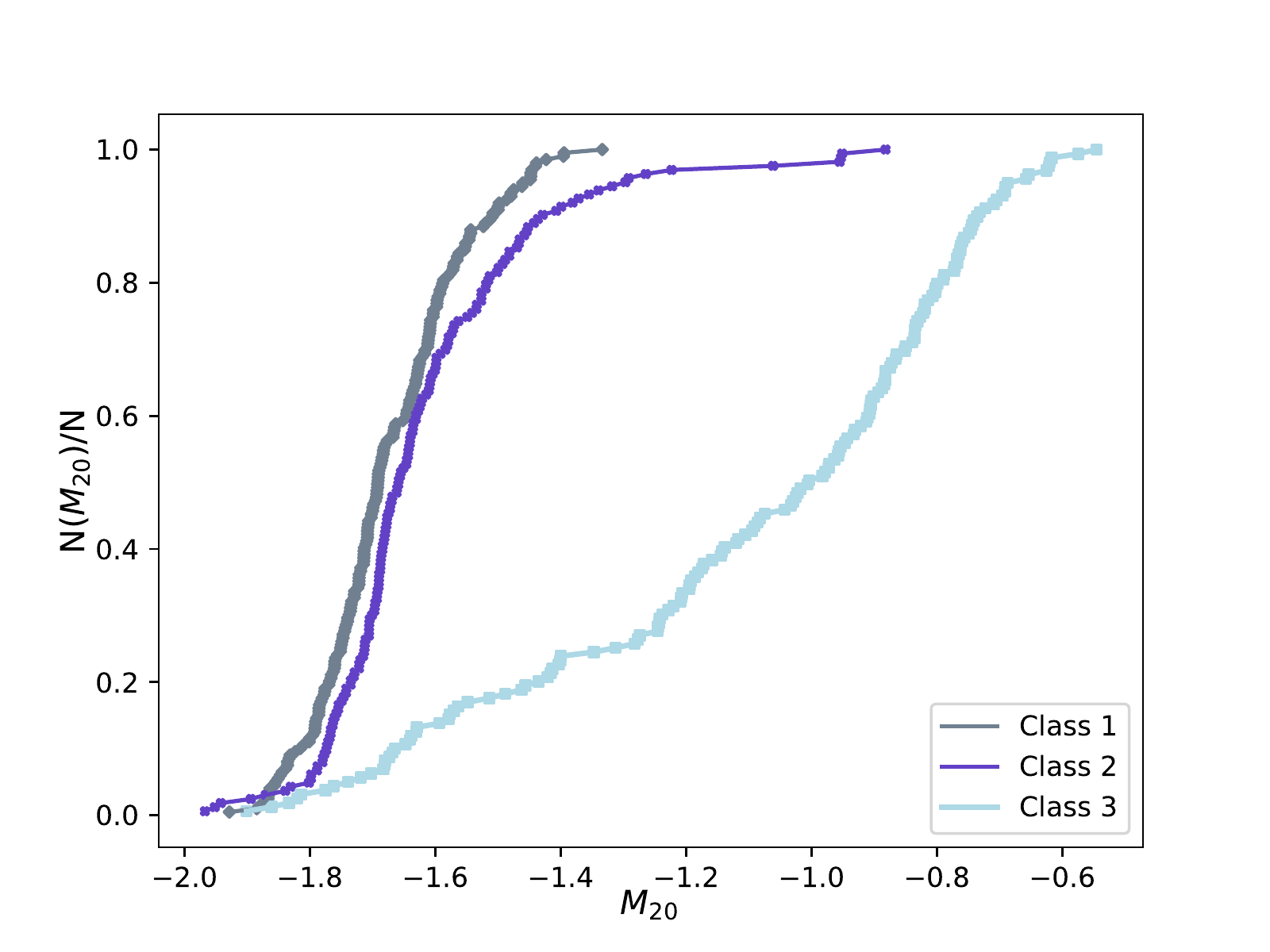}
    \caption{The normalized cumulative histogram depicting the distribution of class~1, 2, and~3 clusters in the $M_{20}$ parameter. Gray, purple, and blue curves and markers represent classes~1, 2, and~3, respectively. We find that class~1 and~2 clusters show similar distributions in their $M_{20}$ values and that class~3 clusters show a distribution dramatically different from them.}
    \label{fig:m20_chist}
\end{figure}

We show our results in Figure~\ref{fig:m20_chist}, as a normalized cumulative histogram for each cluster class. The figure shows that the class~1 and~2 clusters show a similar $M_{20}$ distribution, one dramatically different from the distribution for class~3 clusters. This is confirmed by the two sample Kolmogorov--Smirnov (KS) test. When we compare the distribution of the combined set of class~1 and~2 clusters together with the class~3 clusters, we find the two sample KS-test $p$-value to be $\mathrm{10^{-16}}$, underscoring the exceedingly small probability that these two sets are drawn from the same parent population in their $M_{20}$ distributions. $M_{20}$ is hence able to capture the distinct light profile of class~3 clusters, when measured in our $11 \times 11$ pixel cut-outs. The figure shows that for the majority of the class~3 objects the top 20\% brightest pixels are split among peaks far from the center of the structure, causing an increase in $M_{20}$. This is not the case for the centrally peaked class~1 and class~2 objects as they decidedly reside at the low end of the distribution, indicating that the brightest pixels are concentrated close to the center of the object.

An important inquiry is whether the trend here is driven by the distance to the host galaxy of our clusters. We investigate this with two different tests, both of which find that distance does not significantly affect our results.  In the first test we split our sample into two distance bins: the clusters residing in galaxies within 16 Mpc, and those that reside in galaxies further away. Then we individually compare the $M_{20}$ distributions of class 1, 2, 3 clusters within these bins via two sample KS tests. We find that the distributions of individual classes in the two distance bins do not differ in a statistically meaningful way. Next, we compare the nearest and farthest end of our sample, choosing the clusters in galaxies within 12 Mpc, and farther away than 18 Mpc. The same comparison as the first test finds that distance does not have an effect even when the $M_{20}$ distributions of individual classes at the extreme ends of our distance range are compared. This conclusion is best displayed by the two sample KS tests of our class 3 clusters, as they are the clusters in our sample having the least isolation hence the most susceptibility to distance effects. We find the two sample KS test comparing class 3 clusters in the distance bins of the first test to be $p$ = 0.7, and in the bins of the second test to be $p$ = 0.9, demonstrating that distance is highly unlikely to be the driver of the trend in Figure~\ref{fig:m20_chist}. As a final remark, we note that both tests split the sample into roughly equal number of galaxies and clusters.

\subsection{The $M_{20}$ Measure of Full Cluster Samples}
\label{subsec:m20_full}

\begin{figure*}
    \centering
    \includegraphics[width=.48\textwidth]{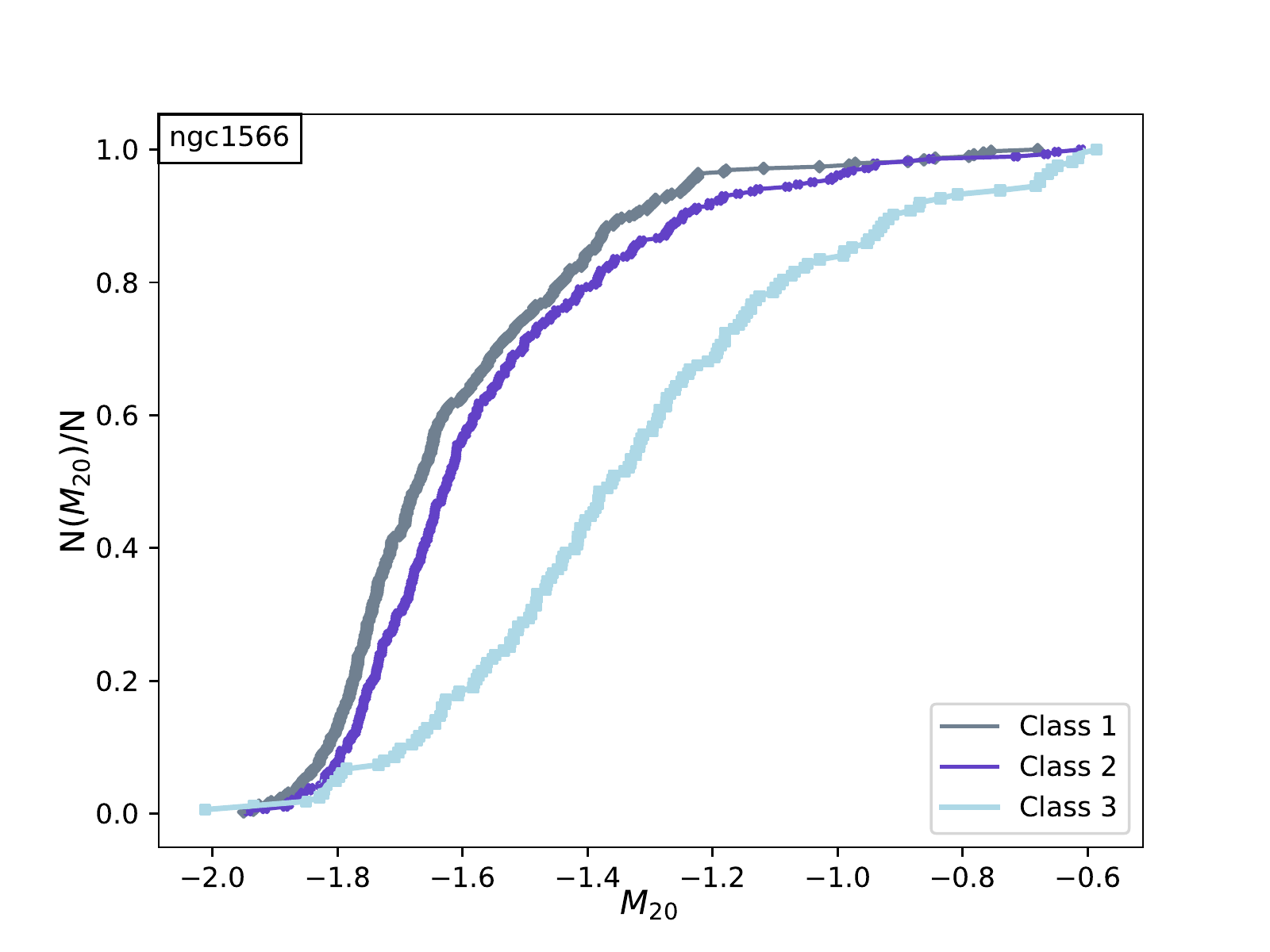}
    \includegraphics[width=.48\textwidth]{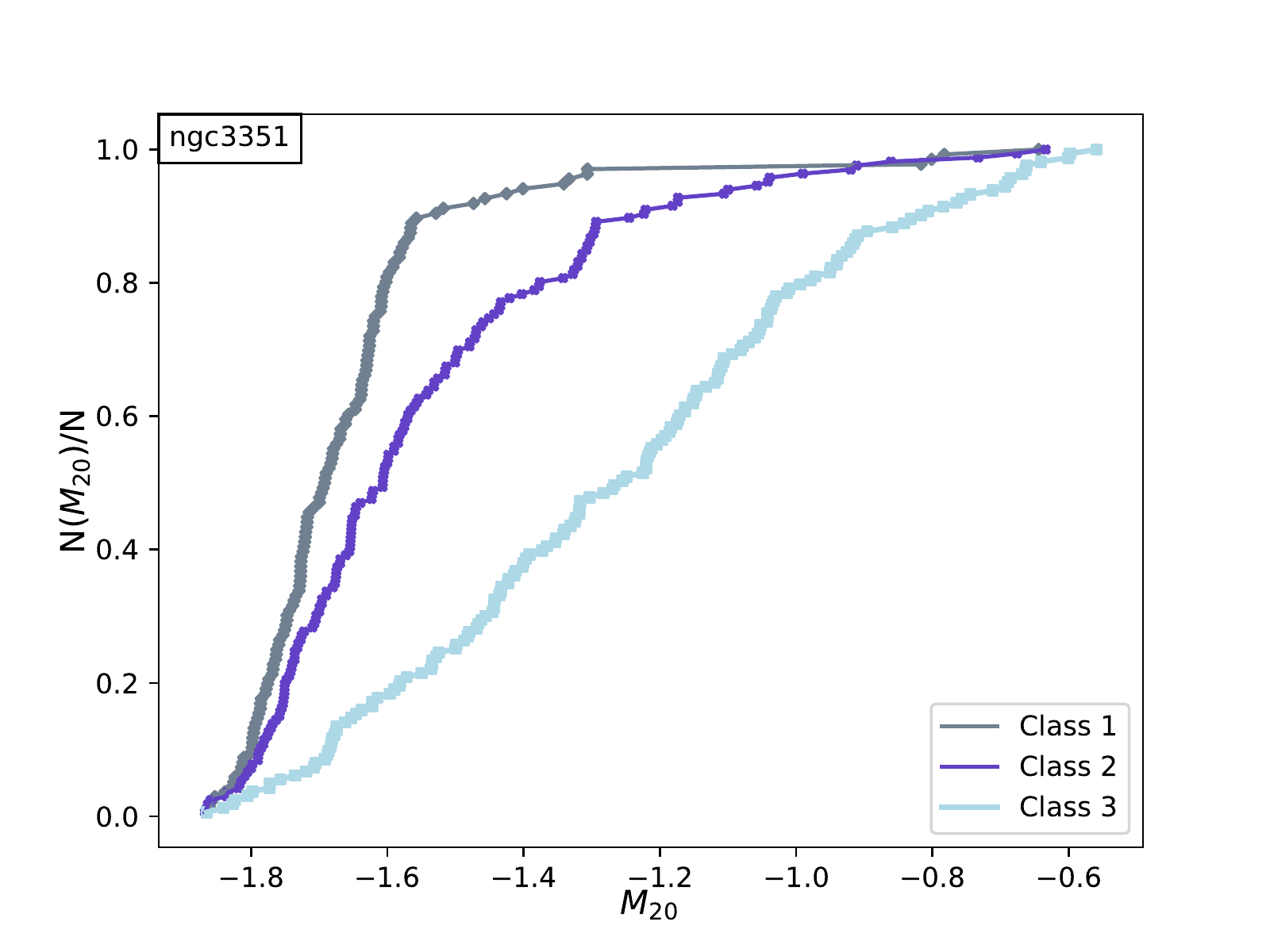}
    \caption{The normalized cumulative histograms showing the $M_{20}$ distribution of full cluster samples. The left panel shows the distribution of class~1, 2, and~3 clusters in NGC~1566, and the right panel in NGC~3351.}
    \label{fig:full_sample_m20}
\end{figure*}

We have presented the $M_{20}$ distribution of our specialized bright, isolated sample. Though this sample serves as an invaluable starting point, testing with a more extensive cluster sample is crucial to gauge the efficacy of this parameter. To this effect, we use the PHANGS--HST cluster catalogs of the overall populations of two of the galaxies from our sample, NGC~1566 and NGC~3351. These samples are formed by visually classifying the likely cluster candidates up to a magnitude limit. The V-band magnitude limit is $m_{V} = 24.5$~mag for NGC~1566 and $m_{V} = 24.0$~mag for NGC~3351. We show our findings in Figure~\ref{fig:full_sample_m20}. This figure shows that $M_{20}$ is sensitive to the distinct light profiles of clusters and stellar associations even when applied to the full cluster catalogs. Even though the maximal separation is smaller than in Figure~\ref{fig:m20_chist}, the multi-peaked class~3's from these more extensive samples still have a significantly different distribution in $M_{20}$ than the centrally peaked clusters. We discuss the implications further in Section~\ref{sec:conclusions}.

\subsection{The UB-VI-$M_{20}$ Space}

\begin{figure*}
    \centering
    \includegraphics[scale=0.95]{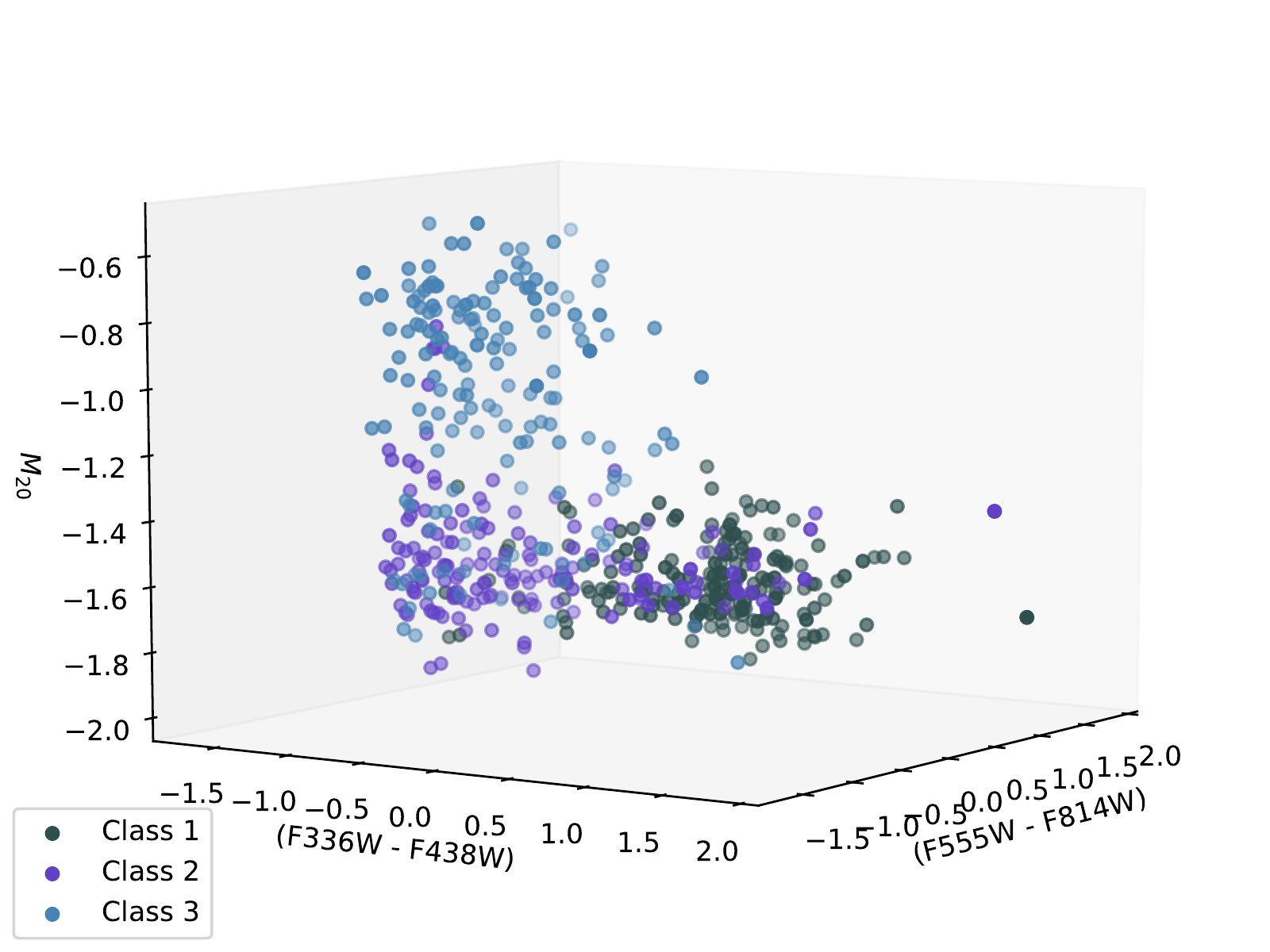}
    \caption{The UB-VI-$M_{20}$ diagram of the bright and isolated cluster sample we analyze in this paper.}
    \label{fig:ub-vi-m20}
\end{figure*}

\begin{figure*}
    \centering
    \includegraphics[scale=0.45]{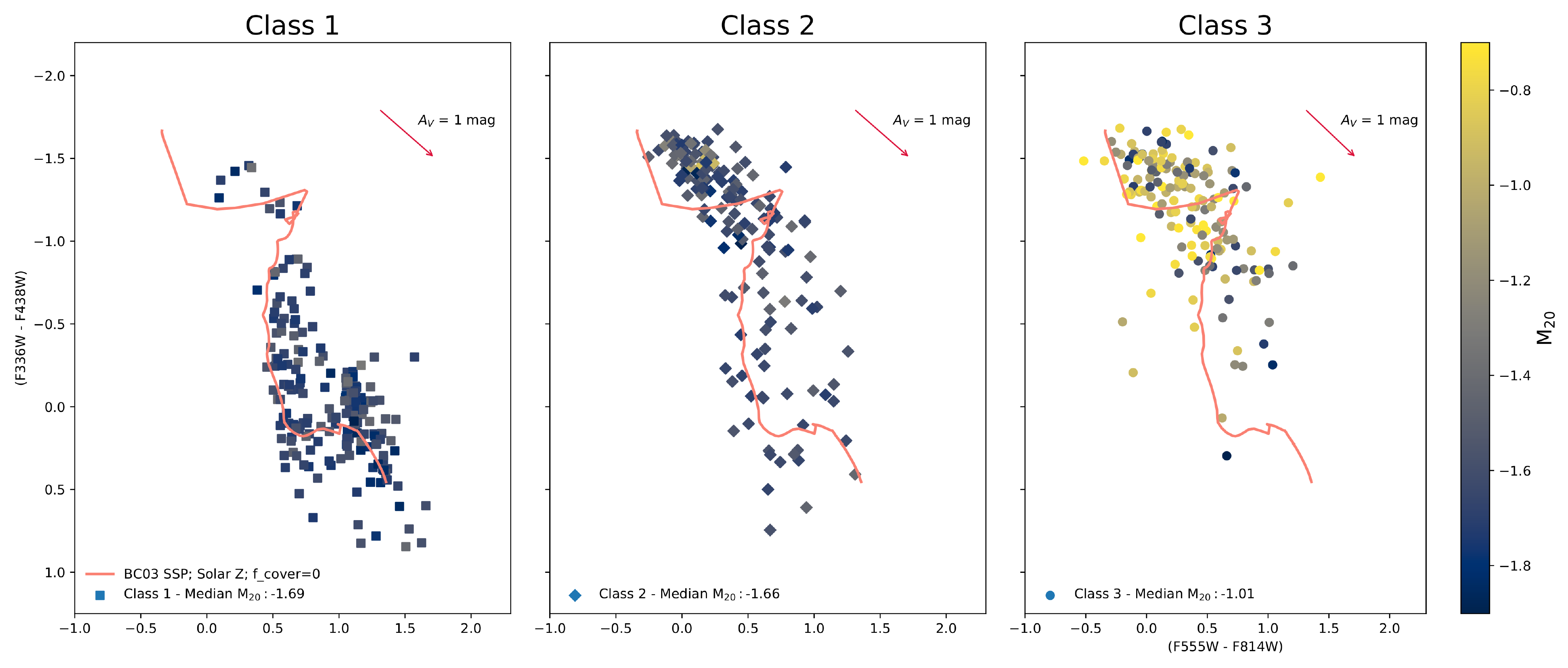}
    \caption{The UB-VI diagram color-coded by $M_{20}$. The figure is presented in three panels, one panel for each of the three class of clusters (left panel shows class~1, middle class~2, right panel shows class~3). We denote the median $M_{20}$ in the legend for each morphology type.}
    \label{fig:ub-vi-color-coded-m20}
\end{figure*}

We have investigated UB-VI color--color diagrams, and $M_{20}$ distributions of our bright and isolated clusters separately. These analyses have found that different morphology types preferentially occupied different sections in these parameters. Now we combine the UB-VI color--color diagram with $M_{20}$, and present the results as a 3D figure in Figure~\ref{fig:ub-vi-m20}, and as the UB-VI diagram color--coded by $M_{20}$ in Figure~\ref{fig:ub-vi-color-coded-m20}. This joint space shows that the degeneracy of class~2 and class~3 objects on the UB-VI diagram is broken quite beautifully by the inclusion of $M_{20}$. This three-dimensional space separates the three morphology types with modest overlap, at least for the specialized sample we analyze in this paper. We discuss this finding in more detail in Section~\ref{sec:conclusions}.

The UB-VI-$M_{20}$ space provides insight into the critical role selection plays in star cluster analysis, especially at ages prior to 50 Myr. Figure~\ref{fig:ub-vi-color-coded-m20} shows this range to be dominated by class 2 and 3 structures. These structures have in past studies been split by reference to whether they are in or are likely to remain in a bound cluster state, at times based on qualitative judgements of morphology such as visual roundness \citep[e.g.][]{Krumholz19}. Class 2 structures, owing to their centrally concentrated morphology, which is generally taken to be a sign of relaxation, are considered to have a higher likelihood of remaining bound. Multi-peaked class 3 associations on the other hand are considered to be likely unbound, on their way to dissolve into the field population \citep[see e.g.][]{bressert10, gieles11}. Aggregation of both populations therefore introduces the risk of making inferences with disparate sub-populations. For example, Figure~\ref{fig:ub-vi-color-coded-m20} demonstrates that inclusion of all structures between the ages of 10-50 Myr results in an overestimation of the bound cluster fraction, in alignment with the discussion in \cite{kruijssen16}. As discussed above, Figures~\ref{fig:ub-vi-m20} and ~\ref{fig:ub-vi-color-coded-m20} demonstrate that $M_{20}$ has distinguishing power between these two classes. A selection based on $M_{20}$ within this age range has the potential to eliminate those clusters unlikely to remain bound, resulting in a more robust measure of the fraction of stars formed in bound clusters and thereby a better comparison with theoretical frameworks \citep[e.g.][]{kruijssen12}. Selection based on $M_{20}$ can also provide a framework for selecting samples for the systematic analysis of star clusters. This can help address the discrepancy in cluster age functions measured when using inclusive and exclusive catalogs, as detailed in \citet{Krumholz19}.

We conclude this subsection with a final remark about the distributions of the three classes in their $M_{20}$ measurements. As demonstrated and discussed above, the distributions of the centrally peaked class 1 and 2 clusters and the multiply peaked class 3 objects are distinct in this parameter. However, there is still considerable overlap, especially at the high $M_{20}$ values. We note that this overlap is to be expected, due to both the vast range of clusters and associations that form this sample, and the evolutionary dynamics between the archetypes within the classification system. Some examples of  both of these can be observed in Figure~\ref{fig:ub-vi-color-coded-m20}. There are young class 1 clusters towards the bluer portion of the UB-VI space, but many more that have evolved to old ages resulting in a distribution skewed towards red colors. This is especially prominent here due to the selection philosophy of our sample favoring isolated class 1's. The distribution of class 2 clusters peak at the blue edge of the diagram, yet there are members of this type at old ages. In Section~\ref{sec:conclusions} we discuss whether the range in class 3 objects may also correspond to a likelihood estimate of evolution as a bound system, though such investigations will require further analysis in future papers.

\section{Color Evolution: Comparison of Observations with Synthetic Stellar Populations}
\label{sec:stellarpop_models}

Throughout the paper we used the UB-VI color-color track of a synthetic simple stellar population (SSP) as a reference and frame of comparison for the colors of our clusters. The underlying assumption is that the set of stars formed in a cluster or association follows an initial mass function (IMF; in our case, the \citealt{chabrier03} IMF) and are all born at the same time. Broadly speaking, this assumption is a plausible one for star clusters \citep{wofford16}; indeed, SSPs have been used for the SED fitting of cluster populations in nearby galaxies for over a decade \citep[e.g.,][]{chandar10b, Adamo17}, including for our initial set of PHANGS-HST cluster catalogs \citep{turner21}. We chose the particular track displayed to have solar metallicity, consistent with expected metallicities for our galaxies which have stellar masses above $\mathrm{10^{10}~M_{\odot}}$, as also confirmed by studies of the nebular gas metallicity based on PHANGS-MUSE integral field spectroscopy \citep[][]{kreckel19, kreckel20}

In this section we investigate how both the UB-VI and the NUVB-VI colors of a variety of stellar population models other than the SSP evolve through their lifetimes, and compare these with the colors of the clusters and associations in our bright, relatively isolated sample. We test the effects of metallicity, nebular emission, and star formation history by generating a set of BC03 models using CIGALE. These tests are motivated by results that challenge the validity of using SSP models in the systematic analysis of star clusters \citep[e.g.][for multiple populations in globular clusters]{piotto15}, and inconsistencies in the age estimates of some of our clusters discussed by \cite{Whitmore20} and \cite{turner21}.

\begin{figure*}
    \centering
    \includegraphics[scale=0.6]{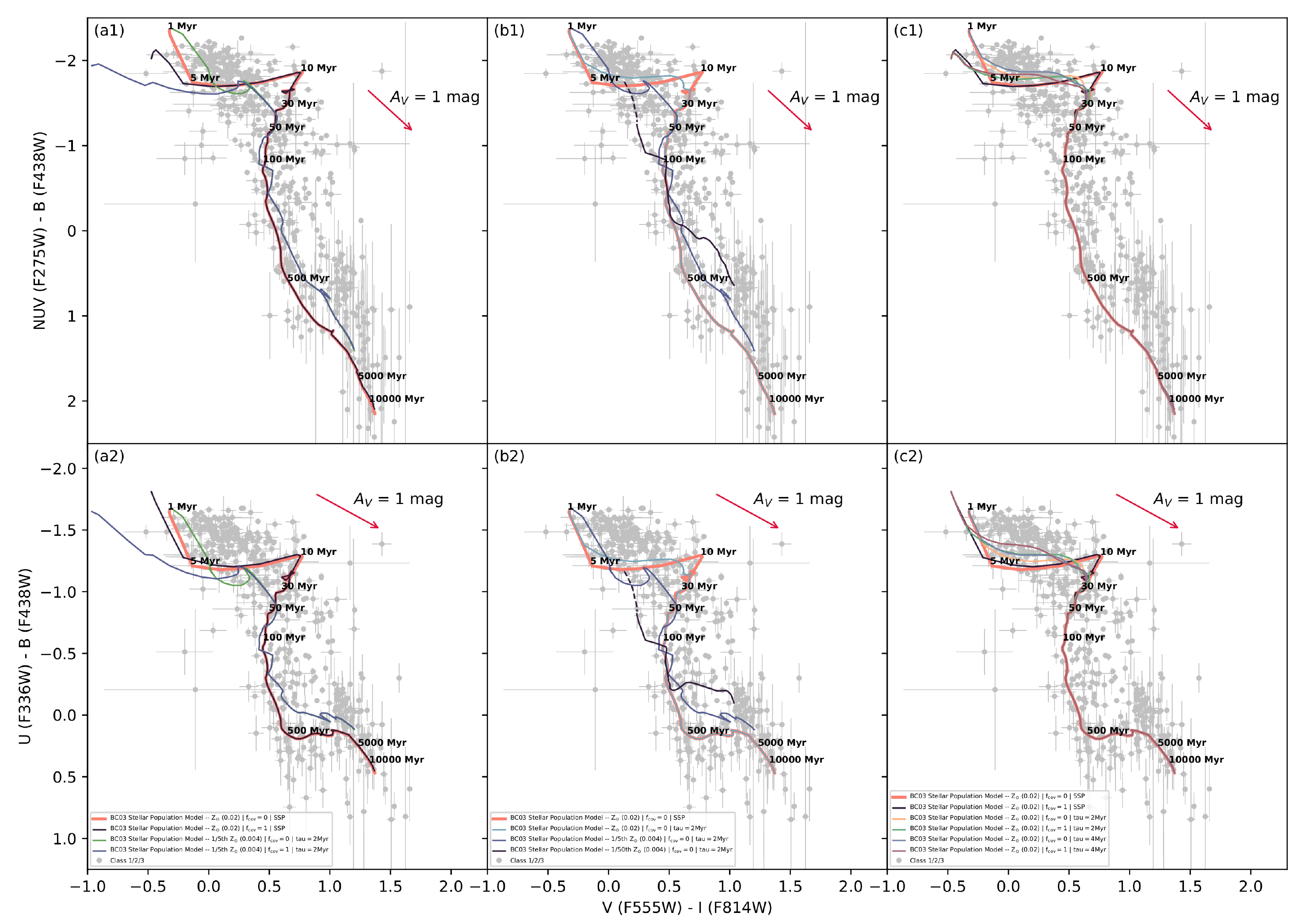}
    \caption{The NUVB-VI (top panels) and UB-VI (bottom panels) evolutionary tracks of various synthetic stellar population models overplotted on the class 1/2/3 structures we analyze in this paper. The data for our clusters and assocations are shown with their 1$\sigma$ uncertainties as the gray markers. The reddening vector corresponding to $A_{V} = 1$~mag is shown as the red arrow. The fiducial solar metallicity and $\mathrm{f_{cov}}$ = 0 SSP model is shown in salmon color in every panel, as it has been throughout the paper.  We denote the specifications of the individual models in the legends of individual panels, and provide details in Section \ref{sec:stellarpop_models}. The figure is presented in three pairs of joint NUVB-VI and UB-VI panels, demonstrating the following: panels (a1) and (a2) demonstrate the combined effects of metallicity and nebular emission, panels (b1) and (b2) demonstrate the effects of metallicity at constant $\mathrm{f_{cov}}$, and panels (c1) and (c2) demonstrate the effects of star formation history at constant metallicity on the colors of stellar population model tracks. All model tracks end at 13.7 Gyr. Apart from the one fiftieth solar metallicity track all tracks start at 1 Myr. This track starts at 50 Myr as its metallicity is too low to provide a plausible frame of comparison for our youngest structures.}
    \label{fig:tracks}
\end{figure*}

Our testing grid is formed through the combination of the following parameters. For metallicities, we test populations with solar metallicity, and those with one fifth or one fiftieth solar abundance. For nebular emission, CIGALE uses nebular templates from \cite{inoue11}, generated using CLOUDY \citep{ferland98, ferland13}. We investigate the effect of nebular emission by varying the escape fraction, or the fraction of photons that do not ionize the gas surrounding the stellar population. We test escape fractions of zero (or covering fraction, i.e. $\mathrm{f_{cov}}$, of one) where all the Lyman-continuum photons ionize the gas, and one ($\mathrm{f_{cov}}$ of zero), where none do. Finally, we compute the model tracks for composite stellar populations (CSP's) by evolving populations formed via an exponential star formation rate, $SFR(t) \sim \exp(-t/\tau)$. We set the $\tau$ value to 2 Myr, and 4 Myr. The 2 Myr model especially goes through multiple e-foldings within 10 Myr, making it a suitable comparison to clustered star formation, where feedback mechanisms destroy the star formation sequence within roughly these timescales. Models with $\tau$ values less than 2 Myr do not show a significant difference, and therefore are not displayed in figures for visual clarity. The resultant model tracks are shown in Figure~\ref{fig:tracks}, plotted together with the colors of our class 1/2/3 objects.

First, we note that the models with lower metallicity and low escape fraction become much bluer than other models at young ages (Figure~\ref{fig:tracks}, panels (a1) and (a2)). This happens because for low metallicity gas, the V-band is affected by very strong emission lines such as [O III]~$\lambda5007\Angstrom$. We do not have data points in the vicinity of this track, or much bluer than the SSP track, as expected since our galaxies have solar metallicities. Examples where this blue region is populated can be seen in \cite{Whitmore20} (e.g., Figures 5 and 6), who studied clusters in NGC~4449 which exhibit sub-solar nebular metallicies.

Examination of the models with sub-solar metallicities also confirms that at old ages, they are more consistent with the observed colors of some of the reddest objects (the clump at V-I$\sim$1.1), which are likely globular clusters. We demonstrate this in panels (b1) and (b2) of Figure~\ref{fig:tracks}. The figure shows a concentration of clusters (almost all class 1, as can be seen in Figure~\ref{fig:ub-vi}) close in proximity to where the lower metallicity tracks end, which correspond to an age of 13.7 Gyr.  The ages for these objects are thus underestimated as a result of SED fitting which use solar metallicity templates ($\sim$1-5 Gyr), as discussed in \cite{Whitmore20} and \cite{turner21}. This result is unsurprising, given that globular clusters were formed at high redshift, when their host galaxies have not had many star formation cycles necessary to undergo significant metal enrichment, and are known to have sub-solar metallicities \citep[e.g.][]{yuan13, horta21}. This sub-population is positioned between the one-fifth and the one-fiftieth solar abundance metallicity tracks, which indicates that their metallicity values are on average within this range. For the NUVB-VI diagram at this age range, we find that some data are offset further to bluer NUV-B. However, there are comparatively large uncertainty in the NUV photometry of these clusters, owing to their old ages and red colors. Most points are comparable to the tracks within their 1$\sigma$ errors. Importantly, there is also the possibility that the model tracks do not represent the low metallicity and very old cluster populations well. That there are clusters more consistent with the solar metallicity track is potentially in alignment with the bimodality of observed colors and metallicities in globular clusters (see e.g. \citealt{zinn85}, or for simulations \citealt{renaud17}). More detailed exploration of issues with the model tracks is outside the bounds of this paper, however, so we leave further investigations to future research. We also note that the large uncertainty in the NUV photometry of some of our clusters would have caused unreliable estimates of the aperture correction values for this band, had we not employed the constant offset approach described in Section~\ref{subsec:other_bands}.

Another point of interest is the behavior of the turn in the tracks at ages of $\sim$10 Myr, demonstrated in panels (c1) and (c2) of Figure~\ref{fig:tracks}. The SSP model is the one that veers to the reddest in V-I color, due to light from red supergiant stars.The turn towards bluer colors at $\sim$10 Myr occurs when red supergiants still dominate the light, but their overall contribution begins to decrease. The CSP model tracks show a less abrupt turn, and do not traverse as far red in V-I color as the SSP. This is due to the higher levels of blue light from the prolonged formation of stars in the CSP's. The CSP's with larger $\tau$ values move further away from the SSP turn, as they have more extended star formation. The observed data points appear to be more consistent with the CSP models with very short e-folding times in the region of the turn.

Examination of the young clusters in our sample shows that they lie above the SSP track in both panels by an average U-B color of approximately 0.3 mags. Some tracks traverse closer to the observed UB-VI colors than the SSP model, especially the CSP models with solar metallicity. Especially the bluest clusters in U-B color are more easily recovered using the solar metallicity plus complete covering CSP tracks with an application of reddening. CSP models with star formation active over short e-folding times ($\sim$2 Myr) may be physically more realistic than the instantaneous formation of stars employed by the SSP model, given observations of very nearby young clusters which have age spreads of a few Myr \citep[e.g., NGC2070,][]{cignoni15}. However, the distribution of our data likely reflects the effects of dust more than the particulars of the model. Our data can be recovered by applying the suitable reddening vector to most of the models we test, except for the low metallicity models with complete covering by gas. A cursory analysis utilizing the intensity of the CO emission from PHANGS-ALMA \citep{leroy21} data is in support of this conclusion. For clusters between the ages of 1 to 3 Myr, we find that the degree of offset of the observed colors from the track correlates with the amount of CO emission at the location of the cluster, which indicates the offset is due to reddening. A more thorough analysis of this will be presented in a future paper, which incorporates the CO maps as a prior for the reddening in our SED modeling.  Given the issues with dust, it is unlikely we can clearly distinguish between any of the models over the SSP model purely based on the distribution in Figure~\ref{fig:tracks}.

\section{Summary, Discussion, and Future Work}
\label{sec:conclusions}

In this paper, we analyze sets of relatively bright, isolated star clusters of three different morphological types across \ngal different galaxies in the PHANGS--HST sample, with distances ranging from 4 to 23~Mpc. The three morphological types consist of two centrally peaked types: class~1's that are symmetric around their center and class~2's that are asymmetric; and one multiply-peaked type,  that is class~3. It has been standard practice to visually identify such samples in each galaxy as one of the initial steps for generating star cluster catalogs for the purposes of deriving aperture corrections. Here, we further use same sample of objects to investigate a quantitative measure of morphology, $M_{20}$, applied for the first time in the context of star cluster morphology. 

In this paper we introduce a three-step procedure we utilize to compute an average aperture correction for each galaxy. Briefly, the three-step procedure aims the recover the total correction to be applied to the star clusters of a target galaxy as a sum of three individual corrections, derived using class~1 and~2 clusters. The specifications of each step are calibrated to recover the light left out by the 4~pixel circular aperture radius, and 7-to-8 pixel sky definition of the PHANGS--HST data products. This technique is used to derive the V-band aperture corrections, and the corrections in bands other than the V-band are computed using constant offsets applied to this value. These offsets are computed using the PSF specifications of each HST band other than the V-band in our imaging. We find this approach to provide an improvement over previous work deriving individual corrections in each band. Deriving reliable aperture corrections in our set of bands ranging from the NUV to the I-band is a nontrivial task. This requires finding clusters with robust light profiles for at least a part of this range, with minimal light contamination so as not to influence our goal of finding realistic corrections to individual clusters. Selecting clusters with reliable NUV light profiles almost invariably means they are in crowded regions, however, where their light profile is affected by neighboring sources even within moderate size apertures. We therefore chose to derive robust corrections in the V-band, and correct for the other bands via the application of constant offsets to this value. We then compare the aperture corrections we derive in this paper with those derived by LEGUS \citep[][]{legus}, for the three overlapping galaxies. For one of these galaxies, NGC~1433, we find that the PHANGS--HST photometry is closer in representation by an SSP model on the UB-VI space. The shift in the LEGUS data is unlikely to be accounted for by reddening, and is likely a result of the difficulty in deriving aperture corrections in bluer bands. This affects the ages of especially the clusters with ages of 5-50 Myr, where the SSP track turns around.

The particular selection of the specialized sample we construct also provides an ideal framework to launch an analysis into quantitative measurement of the light profiles of star clusters, as such an analysis undoubtedly benefits from bright sources in relative isolation. Given the challenges and long gestational periods required to produce complete cluster catalogs, these samples provide an expedited first look at the properties of clusters across many galaxies, keeping in mind that they will not be fully representative of the overall population.

For our initial look into the utility of quantitative measurements of morphologies as applied to star clusters, we focus on one measure, namely the non-parametric measure $M_{20}$. $M_{20}$ is the normalized second order moment of the brightest 20\% pixels of a given source. This parameter was originally introduced for the study of galaxies in \cite{Lotz04}, with the primary focus being the identification of galaxy mergers, as it is especially sensitive to telltale merger signatures such as double nuclei. We chose to start this investigation with $M_{20}$ specifically due both to its ability to differentiate such singly and multiply peaked distributions, and that its use of the 20\% brightest pixels makes it less susceptible to the variety of backgrounds in which clusters reside. Our analysis finds that the $M_{20}$ parameter can distinguish between the centrally concentrated clusters (i.e., class~1 and~2), and multi-peaked associations (i.e., class~3). Higher values of $M_{20}$ indicate larger spatial separation of flux peaks relative to the structure center, and we indeed find that the higher end of the $M_{20}$ distribution of our clusters is dominated by class~3 objects. We find this result to hold true when the $M_{20}$ distribution of full cluster catalogs are investigated, and that distance to the host galaxy does not play a role in this trend. This indicates that $M_{20}$ retains its ability to distinguish between the single and the multi-peaked light profiles even in the absence of strict isolation requirements. Moreover, we find that coupling $M_{20}$ to the UB-VI color--color diagram results in a parameter space where all three categories of objects occupy relatively distinct loci, at least for the bright, isolated sources selected for the analysis we detail in this paper.

Our investigation in this paper reveals a promising venue to address a crucial component of star cluster evolution, concerning bound clusters and unbound associations. This is portrayed most dramatically by the class 1 clusters in our sample formed by identifying structures across 17 galaxies. Our findings show that these clusters occupy both the oldest range in age, dominantly occupying the tail end of the UB-VI space, and the lowest in $M_{20}$ (their median $M_{20}$ is -1.7). Even without kinematic information, it is extremely likely that these are bound clusters, having existed for roughly 100 Myr at the time of our observation. This implies that the old age and low $M_{20}$ part of the UB-VI-$M_{20}$ space is likely occupied by bound clusters. On the other hand, the young end of the age range is strictly populated by class 2 and 3 structures, two classes that are distinct in their $M_{20}$ distribution. It might be possible to chart broad evolutionary sequences in the UB-VI-$M_{20}$ space for these objects. For example, the class 2 objects that remain bound  will evolve along the SSP track in the UB-VI space, as their $M_{20}$ distributions are already highly similar to the older class 1's. The ones that will not remain bound may first move towards higher $M_{20}$ values and occupy a space mainly populated by class 3 objects, in case they attain a multi-peaked profile prior to dissolving into the field stellar population. As for class 3 objects that may eventually remain bound, their $M_{20}$ values may move to lower values as they age towards redder colors, potentially occupying the same UB-VI-$M_{20}$ neighborhood as class 2 objects and finally that of class 1 objects. The wide range in $M_{20}$ spanned by class 3 objects may potentially carry information about their bounded state. If this were the case, this would be invaluable in the systematic analysis of star cluster samples, providing a more complete picture of both young and old structures. We plan to investigate these in upcoming work.

In the full cluster catalogs of the two galaxies we analyze in this paper, we already find a shift of potential interest. As in the case of our bright and isolated sample, the $M_{20}$ distributions of class~1 and~2 clusters in NGC~1566 are similar. This is not the case in NGC~3351, however, where the distributions are distinct. This difference is likely driven by an aspect of the population of class~1 clusters from NGC~3351, as the class~2 distribution for this galaxy is comparable to that of NGC~1566. This warrants further investigation, but our early interpretation as to why we observe this difference is the overabundance of old class~1 clusters in NGC~3351. This potentially points to $M_{20}$ providing further insight into the star cluster populations of galaxies. 

We utilize the observations from our specialized sample as a medium of comparison for synthetic stellar population models. We compare the observed colors of our sample with those of models having varying metallicity, nebular emission, and star formation history. One of these models is the one we use as a reference in color-color figures throughout the paper: an SSP, a population of stars formed at the same time in a single instantaeous burst, with solar metallicity. This is also the model currently employed in the SED fitting of our sample. One sub-population of interest within our sample is globular clusters of ages above 1~Gyr. Our investigation reveals that the color-color distribution of this sub-population is more consistent with models of sub-solar metallicities, as would be expected. These objects are positioned close to the end of the lower metallicity tracks with ages above 10~Gyr, but are positioned between the ages of 0.5~--~5~Gyr relative to the SSP track. This indicates that the common practice of SED fitting with solar metallicity SSP models underestimates the ages of these structures. Another important insight comes from comparing the model tracks with young clusters at ages below 10~Myr. We find that even though the models show significant differences, one cannot be clearly favored over the others purely based on their consistency with observations due to the effects of dust. The data are consistent with multiple models if reasonable amounts of reddening are applied, though the SSP models appear to be disfavored due to the absence of clusters at the 10 Myr turn-around, associated with appearance of red supergiants.

One final plan for future research is to include other quantitative measures of morphology to the analysis started in this paper. There are a set of promising parameters we can use, such as different measures of asymmetry (e.g measuring asymmetry for the inner and outer profiles), the Gini coefficient (a measure of how equally the light is distributed among the pixels of a source) which forms the other half of the space \cite{Lotz04} uses for galaxy merger detection, and many others. The next step is to use this multi-dimensional parameter space of morphology measures and seek trends of cluster morphology with their physical properties, such as mass and age, and their local and global environment. Our findings in this paper alone indicate that the addition of one morphology metric to the commonly used UB-VI space helps break a degeneracy between commonly used morphological types. This multi-dimensional space is expected to provide a concrete framework to help understand the formation and evolution of star clusters. These parameters can also provide a framework for simulations to track the evolution of star clusters and associations, undoubtedly serving as an important step forward in our understanding of the star formation paradigm.

\section*{Acknowledgements}

Based on observations made with the NASA/ESA Hubble Space Telescope, obtained from the data archive at the Space Telescope Science Institute. STScI is operated by the Association of Universities for Research in Astronomy, Inc. under NASA contract NAS 5-26555.  Support for Program number 15654 was provided through a grant from the STScI under NASA contract NAS5-26555.

This research has made use of the NASA/IPAC Extragalactic Database (NED) which is operated by the Jet Propulsion Laboratory, California Institute of Technology, under contract with NASA. 

JMDK gratefully acknowledges funding from the Deutsche Forschungsgemeinschaft (DFG, German Research Foundation) through an Emmy Noether Research Group (grant number KR4801/1-1) and the DFG Sachbeihilfe (grant number KR4801/2-1), as well as from the European Research Council (ERC) under the European Union's Horizon 2020 research and innovation programme via the ERC Starting Grant MUSTANG (grant agreement number 714907).

SCOG and RSK acknowledge support from the DFG via SFB 881 “The Milky Way System” (sub-projects B1, B2 and B8) and from the Heidelberg cluster of excellence EXC 2181-390900948 “STRUCTURES: A unifying approach to emergent phenomena in the physical world, mathematics, and complex data”, funded by the German Excellence Strategy. They also acknowledge funding from the European Research Council via the ERC Synergy Grant ``ECOGAL'' (grant 855130).

TGW acknowledges funding from the European Research Council (ERC) under the European Union’s Horizon 2020 research and innovation programme (grant agreement No. 694343).

ATB would like to acknowledge funding from the European Research Council (ERC) under the European Union’s Horizon 2020 research and innovation programme (grant agreement No.726384/Empire).

\section{Data Availability}

The imaging observations underlying this article can be retrieved from the Mikulski Archive for Space Telescopes at \url{https://archive.stsci.edu/hst/search_retrieve.html} under proposal GO-15654. High level science products, including science ready mosaicked imaging, associated with HST GO-15654 are provided at \url{https://archive.stsci.edu/hlsp/phangs-hst} with digital object identifier \doi{10.17909/t9-r08f-dq31}



\bibliography{/home/sinandeger/Desktop/arXiv_submit/main.bib}

\begin{thebibliography}{}
\makeatletter
\relax
\def\mn@urlcharsother{\let\do\@makeother \do\$\do\&\do\#\do\^\do\_\do\%\do\~}
\def\mn@doi{\begingroup\mn@urlcharsother \@ifnextchar [ {\mn@doi@}
  {\mn@doi@[]}}
\def\mn@doi@[#1]#2{\def\@tempa{#1}\ifx\@tempa\@empty \href
  {http://dx.doi.org/#2} {doi:#2}\else \href {http://dx.doi.org/#2} {#1}\fi
  \endgroup}
\def\mn@eprint#1#2{\mn@eprint@#1:#2::\@nil}
\def\mn@eprint@arXiv#1{\href {http://arxiv.org/abs/#1} {{\tt arXiv:#1}}}
\def\mn@eprint@dblp#1{\href {http://dblp.uni-trier.de/rec/bibtex/#1.xml}
  {dblp:#1}}
\def\mn@eprint@#1:#2:#3:#4\@nil{\def\@tempa {#1}\def\@tempb {#2}\def\@tempc
  {#3}\ifx \@tempc \@empty \let \@tempc \@tempb \let \@tempb \@tempa \fi \ifx
  \@tempb \@empty \def\@tempb {arXiv}\fi \@ifundefined
  {mn@eprint@\@tempb}{\@tempb:\@tempc}{\expandafter \expandafter \csname
  mn@eprint@\@tempb\endcsname \expandafter{\@tempc}}}

\bibitem[\protect\citeauthoryear{{Adamo} et~al.,}{{Adamo}
  et~al.}{2017}]{Adamo17}
{Adamo} A.,  et~al., 2017, \mn@doi [\apj] {10.3847/1538-4357/aa7132}, \href
  {https://ui.adsabs.harvard.edu/abs/2017ApJ...841..131A} {841, 131}

\bibitem[\protect\citeauthoryear{{Anand} et~al.,}{{Anand}
  et~al.}{2021}]{Anand21}
{Anand} G.~S.,  et~al., 2021, \mn@doi [\mnras] {10.1093/mnras/staa3668}, \href
  {https://ui.adsabs.harvard.edu/abs/2021MNRAS.501.3621A} {501, 3621}

\bibitem[\protect\citeauthoryear{{Boquien}, {Burgarella}, {Roehlly}, {Buat},
  {Ciesla}, {Corre}, {Inoue}  \& {Salas}}{{Boquien} et~al.}{2019}]{Boquien19}
{Boquien} M.,  {Burgarella} D.,  {Roehlly} Y.,  {Buat} V.,  {Ciesla} L.,
  {Corre} D.,  {Inoue} A.~K.,   {Salas} H.,  2019, \mn@doi [\aap]
  {10.1051/0004-6361/201834156}, \href
  {https://ui.adsabs.harvard.edu/abs/2019A&A...622A.103B} {622, A103}

\bibitem[\protect\citeauthoryear{{Bressert} et~al.,}{{Bressert}
  et~al.}{2010}]{bressert10}
{Bressert} E.,  et~al., 2010, \mn@doi [\mnras]
  {10.1111/j.1745-3933.2010.00946.x}, \href
  {https://ui.adsabs.harvard.edu/abs/2010MNRAS.409L..54B} {409, L54}

\bibitem[\protect\citeauthoryear{{Bruzual} \& {Charlot}}{{Bruzual} \&
  {Charlot}}{2003}]{BC03}
{Bruzual} G.,  {Charlot} S.,  2003, \mn@doi [\mnras]
  {10.1046/j.1365-8711.2003.06897.x}, \href
  {https://ui.adsabs.harvard.edu/abs/2003MNRAS.344.1000B} {344, 1000}

\bibitem[\protect\citeauthoryear{{Calzetti} et~al.,}{{Calzetti}
  et~al.}{2015}]{legus}
{Calzetti} D.,  et~al., 2015, \mn@doi [\aj] {10.1088/0004-6256/149/2/51}, \href
  {http://adsabs.harvard.edu/abs/2015AJ....149...51C} {149, 51}

\bibitem[\protect\citeauthoryear{{Chabrier}}{{Chabrier}}{2003}]{chabrier03}
{Chabrier} G.,  2003, \mn@doi [\pasp] {10.1086/376392}, \href
  {https://ui.adsabs.harvard.edu/abs/2003PASP..115..763C} {115, 763}

\bibitem[\protect\citeauthoryear{{Chandar} et~al.,}{{Chandar}
  et~al.}{2010}]{chandar10b}
{Chandar} R.,  et~al., 2010, \mn@doi [\apj] {10.1088/0004-637X/719/1/966},
  \href {http://adsabs.harvard.edu/abs/2010ApJ...719..966C} {719, 966}

\bibitem[\protect\citeauthoryear{{Chevallard} \& {Charlot}}{{Chevallard} \&
  {Charlot}}{2016}]{chevallard16}
{Chevallard} J.,  {Charlot} S.,  2016, \mn@doi [\mnras]
  {10.1093/mnras/stw1756}, \href
  {https://ui.adsabs.harvard.edu/abs/2016MNRAS.462.1415C} {462, 1415}

\bibitem[\protect\citeauthoryear{{Cignoni} et~al.,}{{Cignoni}
  et~al.}{2015}]{cignoni15}
{Cignoni} M.,  et~al., 2015, \mn@doi [\apj] {10.1088/0004-637X/811/2/76}, \href
  {https://ui.adsabs.harvard.edu/abs/2015ApJ...811...76C} {811, 76}

\bibitem[\protect\citeauthoryear{{Cook} et~al.,}{{Cook} et~al.}{2019}]{Cook19}
{Cook} D.~O.,  et~al., 2019, \mn@doi [\mnras] {10.1093/mnras/stz331}, \href
  {https://ui.adsabs.harvard.edu/abs/2019MNRAS.484.4897C} {484, 4897}

\bibitem[\protect\citeauthoryear{{Fall}, {Chandar}  \& {Whitmore}}{{Fall}
  et~al.}{2009}]{Fall09}
{Fall} S.~M.,  {Chandar} R.,   {Whitmore} B.~C.,  2009, \mn@doi [\apj]
  {10.1088/0004-637X/704/1/453}, \href
  {https://ui.adsabs.harvard.edu/abs/2009ApJ...704..453F} {704, 453}

\bibitem[\protect\citeauthoryear{{Fall}, {Krumholz}  \& {Matzner}}{{Fall}
  et~al.}{2010}]{Fall10}
{Fall} S.~M.,  {Krumholz} M.~R.,   {Matzner} C.~D.,  2010, \mn@doi [\apjl]
  {10.1088/2041-8205/710/2/L142}, \href
  {https://ui.adsabs.harvard.edu/abs/2010ApJ...710L.142F} {710, L142}

\bibitem[\protect\citeauthoryear{{Ferland}, {Korista}, {Verner}, {Ferguson},
  {Kingdon}  \& {Verner}}{{Ferland} et~al.}{1998}]{ferland98}
{Ferland} G.~J.,  {Korista} K.~T.,  {Verner} D.~A.,  {Ferguson} J.~W.,
  {Kingdon} J.~B.,   {Verner} E.~M.,  1998, \mn@doi [\pasp] {10.1086/316190},
  \href {https://ui.adsabs.harvard.edu/abs/1998PASP..110..761F} {110, 761}

\bibitem[\protect\citeauthoryear{{Ferland} et~al.,}{{Ferland}
  et~al.}{2013}]{ferland13}
{Ferland} G.~J.,  et~al., 2013, \rmxaa, \href
  {https://ui.adsabs.harvard.edu/abs/2013RMxAA..49..137F} {49, 137}

\bibitem[\protect\citeauthoryear{{Franzetti}, {Scodeggio}, {Garilli}, {Fumana}
  \& {Paioro}}{{Franzetti} et~al.}{2008}]{franzetti08}
{Franzetti} P.,  {Scodeggio} M.,  {Garilli} B.,  {Fumana} M.,   {Paioro} L.,
  2008, in {Argyle} R.~W.,  {Bunclark} P.~S.,   {Lewis} J.~R.,  eds,
  Astronomical Society of the Pacific Conference Series Vol. 394, Astronomical
  Data Analysis Software and Systems XVII. p.~642 (\mn@eprint {arXiv}
  {0801.2518})

\bibitem[\protect\citeauthoryear{{Fukui} et~al.,}{{Fukui}
  et~al.}{2014}]{fukui14}
{Fukui} Y.,  et~al., 2014, \mn@doi [\apj] {10.1088/0004-637X/780/1/36}, \href
  {https://ui.adsabs.harvard.edu/abs/2014ApJ...780...36F} {780, 36}

\bibitem[\protect\citeauthoryear{{Gallazzi}, {Charlot}, {Brinchmann}, {White}
  \& {Tremonti}}{{Gallazzi} et~al.}{2005}]{gallazzi05}
{Gallazzi} A.,  {Charlot} S.,  {Brinchmann} J.,  {White} S. D.~M.,   {Tremonti}
  C.~A.,  2005, \mn@doi [\mnras] {10.1111/j.1365-2966.2005.09321.x}, \href
  {https://ui.adsabs.harvard.edu/abs/2005MNRAS.362...41G} {362, 41}

\bibitem[\protect\citeauthoryear{{Gieles} \& {Portegies Zwart}}{{Gieles} \&
  {Portegies Zwart}}{2011}]{gieles11}
{Gieles} M.,  {Portegies Zwart} S.~F.,  2011, \mn@doi [\mnras]
  {10.1111/j.1745-3933.2010.00967.x}, \href
  {https://ui.adsabs.harvard.edu/abs/2011MNRAS.410L...6G} {410, L6}

\bibitem[\protect\citeauthoryear{{Grasha} et~al.,}{{Grasha}
  et~al.}{2015}]{Grasha15}
{Grasha} K.,  et~al., 2015, \mn@doi [\apj] {10.1088/0004-637X/815/2/93}, \href
  {https://ui.adsabs.harvard.edu/abs/2015ApJ...815...93G} {815, 93}

\bibitem[\protect\citeauthoryear{{Grasha} et~al.,}{{Grasha}
  et~al.}{2019}]{Grasha19}
{Grasha} K.,  et~al., 2019, \mn@doi [\mnras] {10.1093/mnras/sty3424}, \href
  {https://ui.adsabs.harvard.edu/abs/2019MNRAS.483.4707G} {483, 4707}

\bibitem[\protect\citeauthoryear{{Han} \& {Han}}{{Han} \& {Han}}{2012}]{han12}
{Han} Y.,  {Han} Z.,  2012, \mn@doi [\apj] {10.1088/0004-637X/749/2/123}, \href
  {https://ui.adsabs.harvard.edu/abs/2012ApJ...749..123H} {749, 123}

\bibitem[\protect\citeauthoryear{{Hannon} et~al.,}{{Hannon}
  et~al.}{2019}]{hannon19}
{Hannon} S.,  et~al., 2019, \mn@doi [\mnras] {10.1093/mnras/stz2820}, \href
  {https://ui.adsabs.harvard.edu/abs/2019MNRAS.490.4648H} {490, 4648}

\bibitem[\protect\citeauthoryear{{He}, {Zhang}, {Ren}  \& {Sun}}{{He}
  et~al.}{2015}]{resnet}
{He} K.,  {Zhang} X.,  {Ren} S.,   {Sun} J.,  2015, arXiv e-prints, \href
  {https://ui.adsabs.harvard.edu/abs/2015arXiv151203385H} {p. arXiv:1512.03385}

\bibitem[\protect\citeauthoryear{{Holtzman} et~al.,}{{Holtzman}
  et~al.}{1992}]{Holtzman92}
{Holtzman} J.~A.,  et~al., 1992, \mn@doi [\aj] {10.1086/116094}, \href
  {https://ui.adsabs.harvard.edu/abs/1992AJ....103..691H} {103, 691}

\bibitem[\protect\citeauthoryear{{Horta}, {Hughes}, {Pfeffer}, {Bastian},
  {Kruijssen}, {Reina-Campos}  \& {Crain}}{{Horta} et~al.}{2021}]{horta21}
{Horta} D.,  {Hughes} M.~E.,  {Pfeffer} J.~L.,  {Bastian} N.,  {Kruijssen}
  J.~M.~D.,  {Reina-Campos} M.,   {Crain} R.~A.,  2021, \mn@doi [\mnras]
  {10.1093/mnras/staa3522}, \href
  {https://ui.adsabs.harvard.edu/abs/2021MNRAS.500.4768H} {500, 4768}

\bibitem[\protect\citeauthoryear{{Inoue}}{{Inoue}}{2011}]{inoue11}
{Inoue} A.~K.,  2011, \mn@doi [\mnras] {10.1111/j.1365-2966.2011.18906.x},
  \href {https://ui.adsabs.harvard.edu/abs/2011MNRAS.415.2920I} {415, 2920}

\bibitem[\protect\citeauthoryear{{Koleva}, {Prugniel}, {Ocvirk}, {Le Borgne}
  \& {Soubiran}}{{Koleva} et~al.}{2008}]{koleva08}
{Koleva} M.,  {Prugniel} P.,  {Ocvirk} P.,  {Le Borgne} D.,   {Soubiran} C.,
  2008, \mn@doi [\mnras] {10.1111/j.1365-2966.2008.12908.x}, \href
  {https://ui.adsabs.harvard.edu/abs/2008MNRAS.385.1998K} {385, 1998}

\bibitem[\protect\citeauthoryear{{Kreckel} et~al.,}{{Kreckel}
  et~al.}{2019}]{kreckel19}
{Kreckel} K.,  et~al., 2019, \mn@doi [\apj] {10.3847/1538-4357/ab5115}, \href
  {https://ui.adsabs.harvard.edu/abs/2019ApJ...887...80K} {887, 80}

\bibitem[\protect\citeauthoryear{{Kreckel} et~al.,}{{Kreckel}
  et~al.}{2020}]{kreckel20}
{Kreckel} K.,  et~al., 2020, \mn@doi [\mnras] {10.1093/mnras/staa2743}, \href
  {https://ui.adsabs.harvard.edu/abs/2020MNRAS.499..193K} {499, 193}

\bibitem[\protect\citeauthoryear{{Kruijssen}}{{Kruijssen}}{2012}]{kruijssen12}
{Kruijssen} J.~M.~D.,  2012, \mn@doi [\mnras]
  {10.1111/j.1365-2966.2012.21923.x}, \href
  {https://ui.adsabs.harvard.edu/abs/2012MNRAS.426.3008K} {426, 3008}

\bibitem[\protect\citeauthoryear{{Kruijssen} \& {Bastian}}{{Kruijssen} \&
  {Bastian}}{2016}]{kruijssen16}
{Kruijssen} J.~M.~D.,  {Bastian} N.,  2016, \mn@doi [\mnras]
  {10.1093/mnrasl/slv182}, \href
  {https://ui.adsabs.harvard.edu/abs/2016MNRAS.457L..24K} {457, L24}

\bibitem[\protect\citeauthoryear{{Krumholz}, {McKee}  \&
  {Bland-Hawthorn}}{{Krumholz} et~al.}{2019}]{Krumholz19}
{Krumholz} M.~R.,  {McKee} C.~F.,   {Bland-Hawthorn} J.,  2019, \mn@doi [\araa]
  {10.1146/annurev-astro-091918-104430}, \href
  {https://ui.adsabs.harvard.edu/abs/2019ARA&A..57..227K} {57, 227}

\bibitem[\protect\citeauthoryear{{Kudryavtseva} et~al.,}{{Kudryavtseva}
  et~al.}{2012}]{kudryavtseva12}
{Kudryavtseva} N.,  et~al., 2012, \mn@doi [\apjl]
  {10.1088/2041-8205/750/2/L44}, \href
  {https://ui.adsabs.harvard.edu/abs/2012ApJ...750L..44K} {750, L44}

\bibitem[\protect\citeauthoryear{{Kuncarayakti}, {Galbany}, {Anderson},
  {Kr{\"u}hler}  \& {Hamuy}}{{Kuncarayakti} et~al.}{2016}]{kuncarayakti16}
{Kuncarayakti} H.,  {Galbany} L.,  {Anderson} J.~P.,  {Kr{\"u}hler} T.,
  {Hamuy} M.,  2016, \mn@doi [\aap] {10.1051/0004-6361/201628813}, \href
  {https://ui.adsabs.harvard.edu/abs/2016A&A...593A..78K} {593, A78}

\bibitem[\protect\citeauthoryear{{Lada} \& {Lada}}{{Lada} \&
  {Lada}}{2003}]{Lada03}
{Lada} C.~J.,  {Lada} E.~A.,  2003, \mn@doi [\araa]
  {10.1146/annurev.astro.41.011802.094844}, \href
  {https://ui.adsabs.harvard.edu/abs/2003ARA&A..41...57L} {41, 57}

\bibitem[\protect\citeauthoryear{{Larson} et~al.}{{Larson}
  et~al.}{2021}]{larson21}
{Larson} K.,  et~al., 2021, MNRAS, in preparation

\bibitem[\protect\citeauthoryear{{Lee} et~al.,}{{Lee} et~al.}{2021}]{lee21}
{Lee} J.~C.,  et~al., 2021, ApJS, submitted, \href
  {https://ui.adsabs.harvard.edu/abs/2021arXiv210102855L} {p. arXiv:2101.02855}

\bibitem[\protect\citeauthoryear{{Leroy} et~al.,}{{Leroy}
  et~al.}{2021}]{leroy21}
{Leroy} A.~K.,  et~al., 2021, \mn@doi [\apjs] {10.3847/1538-4365/abec80}, \href
  {https://ui.adsabs.harvard.edu/abs/2021ApJS..255...19L} {255, 19}

\bibitem[\protect\citeauthoryear{{Lotz}, {Primack}  \& {Madau}}{{Lotz}
  et~al.}{2004}]{Lotz04}
{Lotz} J.~M.,  {Primack} J.,   {Madau} P.,  2004, \mn@doi [\aj]
  {10.1086/421849}, \href {http://adsabs.harvard.edu/abs/2004AJ....128..163L}
  {128, 163}

\bibitem[\protect\citeauthoryear{{Lotz} et~al.,}{{Lotz} et~al.}{2008}]{Lotz08}
{Lotz} J.~M.,  et~al., 2008, \mn@doi [\apj] {10.1086/523659}, \href
  {https://ui.adsabs.harvard.edu/abs/2008ApJ...672..177L} {672, 177}

\bibitem[\protect\citeauthoryear{{Matzner}}{{Matzner}}{2002}]{Matzner02}
{Matzner} C.~D.,  2002, \mn@doi [\apj] {10.1086/338030}, \href
  {https://ui.adsabs.harvard.edu/abs/2002ApJ...566..302M} {566, 302}

\bibitem[\protect\citeauthoryear{{Messa} et~al.,}{{Messa}
  et~al.}{2018}]{Messa18}
{Messa} M.,  et~al., 2018, \mn@doi [\mnras] {10.1093/mnras/stx2403}, \href
  {https://ui.adsabs.harvard.edu/abs/2018MNRAS.473..996M} {473, 996}

\bibitem[\protect\citeauthoryear{{Meurer}}{{Meurer}}{1995}]{Meurer95}
{Meurer} G.~R.,  1995, \mn@doi [\nat] {10.1038/375742a0}, \href
  {https://ui.adsabs.harvard.edu/abs/1995Natur.375..742M} {375, 742}

\bibitem[\protect\citeauthoryear{{Moustakas} et~al.,}{{Moustakas}
  et~al.}{2013}]{moustakas13}
{Moustakas} J.,  et~al., 2013, \mn@doi [\apj] {10.1088/0004-637X/767/1/50},
  \href {https://ui.adsabs.harvard.edu/abs/2013ApJ...767...50M} {767, 50}

\bibitem[\protect\citeauthoryear{{P{\'e}rez}, {Messa}, {Calzetti}, {Maji},
  {Jung}, {Adamo}  \& {Sirressi}}{{P{\'e}rez} et~al.}{2021}]{Perez20}
{P{\'e}rez} G.,  {Messa} M.,  {Calzetti} D.,  {Maji} S.,  {Jung} D.~E.,
  {Adamo} A.,   {Sirressi} M.,  2021, \mn@doi [\apj]
  {10.3847/1538-4357/abceba}, \href
  {https://ui.adsabs.harvard.edu/abs/2021ApJ...907..100P} {907, 100}

\bibitem[\protect\citeauthoryear{{Peth} et~al.,}{{Peth} et~al.}{2016}]{Peth16}
{Peth} M.~A.,  et~al., 2016, \mn@doi [\mnras] {10.1093/mnras/stw252}, \href
  {https://ui.adsabs.harvard.edu/abs/2016MNRAS.458..963P} {458, 963}

\bibitem[\protect\citeauthoryear{{Piotto} et~al.,}{{Piotto}
  et~al.}{2015}]{piotto15}
{Piotto} G.,  et~al., 2015, \mn@doi [\aj] {10.1088/0004-6256/149/3/91}, \href
  {https://ui.adsabs.harvard.edu/abs/2015AJ....149...91P} {149, 91}

\bibitem[\protect\citeauthoryear{{Portegies Zwart}, {McMillan}  \&
  {Gieles}}{{Portegies Zwart} et~al.}{2010}]{pz10}
{Portegies Zwart} S.~F.,  {McMillan} S. L.~W.,   {Gieles} M.,  2010, \mn@doi
  [\araa] {10.1146/annurev-astro-081309-130834}, \href
  {https://ui.adsabs.harvard.edu/abs/2010ARA&A..48..431P} {48, 431}

\bibitem[\protect\citeauthoryear{{Renaud}, {Agertz}  \& {Gieles}}{{Renaud}
  et~al.}{2017}]{renaud17}
{Renaud} F.,  {Agertz} O.,   {Gieles} M.,  2017, \mn@doi [\mnras]
  {10.1093/mnras/stw2969}, \href
  {https://ui.adsabs.harvard.edu/abs/2017MNRAS.465.3622R} {465, 3622}

\bibitem[\protect\citeauthoryear{Robitaille}{Robitaille}{2019}]{aplpy2019}
Robitaille T.,  2019, {APLpy v2.0: The Astronomical Plotting Library in
  Python}, \mn@doi{10.5281/zenodo.2567476}, \url
  {https://doi.org/10.5281/zenodo.2567476}

\bibitem[\protect\citeauthoryear{{Robitaille} \& {Bressert}}{{Robitaille} \&
  {Bressert}}{2012}]{aplpy2012}
{Robitaille} T.,  {Bressert} E.,  2012, {APLpy: Astronomical Plotting Library
  in Python}, Astrophysics Source Code Library (\mn@eprint {ascl} {1208.017})

\bibitem[\protect\citeauthoryear{{Rodriguez-Gomez} et~al.,}{{Rodriguez-Gomez}
  et~al.}{2019}]{Rodriguez-Gomez19}
{Rodriguez-Gomez} V.,  et~al., 2019, \mn@doi [\mnras] {10.1093/mnras/sty3345},
  \href {https://ui.adsabs.harvard.edu/abs/2019MNRAS.483.4140R} {483, 4140}

\bibitem[\protect\citeauthoryear{{Ryon} et~al.,}{{Ryon} et~al.}{2017}]{Ryon17}
{Ryon} J.~E.,  et~al., 2017, \mn@doi [\apj] {10.3847/1538-4357/aa719e}, \href
  {https://ui.adsabs.harvard.edu/abs/2017ApJ...841...92R} {841, 92}

\bibitem[\protect\citeauthoryear{{S{\'e}rsic}}{{S{\'e}rsic}}{1963}]{Sersic63}
{S{\'e}rsic} J.~L.,  1963, Boletin de la Asociacion Argentina de Astronomia La
  Plata Argentina, \href
  {https://ui.adsabs.harvard.edu/abs/1963BAAA....6...41S} {6, 41}

\bibitem[\protect\citeauthoryear{{Simonyan} \& {Zisserman}}{{Simonyan} \&
  {Zisserman}}{2014}]{vgg}
{Simonyan} K.,  {Zisserman} A.,  2014, arXiv e-prints, \href
  {https://ui.adsabs.harvard.edu/abs/2014arXiv1409.1556S} {p. arXiv:1409.1556}

\bibitem[\protect\citeauthoryear{{Thilker} et~al.}{{Thilker}
  et~al.}{2021}]{thilker21}
{Thilker} D.,  et~al., 2021, MNRAS, in preparation

\bibitem[\protect\citeauthoryear{{Turner} et~al.,}{{Turner}
  et~al.}{2021}]{turner21}
{Turner} J.~A.,  et~al., 2021, \mn@doi [\mnras] {10.1093/mnras/stab055}, \href
  {https://ui.adsabs.harvard.edu/abs/2021arXiv210102134T} {502, 1366}

\bibitem[\protect\citeauthoryear{{Wei} et~al.,}{{Wei} et~al.}{2020}]{wei20}
{Wei} W.,  et~al., 2020, \mn@doi [\mnras] {10.1093/mnras/staa325}, \href
  {https://ui.adsabs.harvard.edu/abs/2020MNRAS.493.3178W} {493, 3178}

\bibitem[\protect\citeauthoryear{{Whitmore}, {Sparks}, {Lucas}, {Macchetto}  \&
  {Biretta}}{{Whitmore} et~al.}{1995}]{Whitmore95}
{Whitmore} B.~C.,  {Sparks} W.~B.,  {Lucas} R.~A.,  {Macchetto} F.~D.,
  {Biretta} J.~A.,  1995, \mn@doi [\apjl] {10.1086/309788}, \href
  {https://ui.adsabs.harvard.edu/abs/1995ApJ...454L..73W} {454, L73}

\bibitem[\protect\citeauthoryear{{Whitmore}, {Zhang}, {Leitherer}, {Fall},
  {Schweizer}  \& {Miller}}{{Whitmore} et~al.}{1999}]{Whitmore99}
{Whitmore} B.~C.,  {Zhang} Q.,  {Leitherer} C.,  {Fall} S.~M.,  {Schweizer} F.,
    {Miller} B.~W.,  1999, \mn@doi [\aj] {10.1086/301041}, \href
  {https://ui.adsabs.harvard.edu/abs/1999AJ....118.1551W} {118, 1551}

\bibitem[\protect\citeauthoryear{{Whitmore} et~al.,}{{Whitmore}
  et~al.}{2016}]{Whitmore16}
{Whitmore} B.~C.,  et~al., 2016, \mn@doi [\aj] {10.3847/0004-6256/151/6/134},
  \href {https://ui.adsabs.harvard.edu/abs/2016AJ....151..134W} {151, 134}

\bibitem[\protect\citeauthoryear{{Whitmore} et~al.,}{{Whitmore}
  et~al.}{2020}]{Whitmore20}
{Whitmore} B.~C.,  et~al., 2020, \mn@doi [\apj] {10.3847/1538-4357/ab59e5},
  \href {https://ui.adsabs.harvard.edu/abs/2020ApJ...889..154W} {889, 154}

\bibitem[\protect\citeauthoryear{{Whitmore} et~al.}{{Whitmore}
  et~al.}{2021}]{whitmore21}
{Whitmore} B.,  et~al., 2021, MNRAS, in preparation

\bibitem[\protect\citeauthoryear{{Wofford} et~al.,}{{Wofford}
  et~al.}{2016}]{wofford16}
{Wofford} A.,  et~al., 2016, \mn@doi [\mnras] {10.1093/mnras/stw150}, \href
  {https://ui.adsabs.harvard.edu/abs/2016MNRAS.457.4296W} {457, 4296}

\bibitem[\protect\citeauthoryear{{Yuan}, {Kewley}  \& {Richard}}{{Yuan}
  et~al.}{2013}]{yuan13}
{Yuan} T.~T.,  {Kewley} L.~J.,   {Richard} J.,  2013, \mn@doi [\apj]
  {10.1088/0004-637X/763/1/9}, \href
  {https://ui.adsabs.harvard.edu/abs/2013ApJ...763....9Y} {763, 9}

\bibitem[\protect\citeauthoryear{{Zinn}}{{Zinn}}{1985}]{zinn85}
{Zinn} R.,  1985, \mn@doi [\apj] {10.1086/163249}, \href
  {https://ui.adsabs.harvard.edu/abs/1985ApJ...293..424Z} {293, 424}

\bibitem[\protect\citeauthoryear{{da Cunha}, {Charlot}  \& {Elbaz}}{{da Cunha}
  et~al.}{2008}]{dacunha08}
{da Cunha} E.,  {Charlot} S.,   {Elbaz} D.,  2008, \mn@doi [\mnras]
  {10.1111/j.1365-2966.2008.13535.x}, \href
  {http://adsabs.harvard.edu/abs/2008MNRAS.388.1595D} {388, 1595}

\bibitem[\protect\citeauthoryear{{de Vaucouleurs}}{{de
  Vaucouleurs}}{1948}]{deV48}
{de Vaucouleurs} G.,  1948, Annales d'Astrophysique, \href
  {https://ui.adsabs.harvard.edu/abs/1948AnAp...11..247D} {11, 247}

\makeatother
\end{thebibliography}



\appendix
\section{Our Sample}

We provide the sample of bright, isolated objects we analyzed for this paper in this appendix, in Table~\ref{tab:sample_prop}. We only display the first 10 entries of the table in the paper, and the full table is available online. The table has the following columns: Name of the host galaxy, object coordinates (RA, Dec, X, Y), 5-band 4~pixel circular aperture photometry and their errors in Vega magnitudes, $\chi^{2}$ minimum age, mass, E(B-V) and their errors, CI, $M_{20}$, and the visual morphology class.

\clearpage
\onecolumn
\begin{landscape}
\begin{scriptsize}
\begin{longtable}{|c|c|c|c|c|c|c|c|c|c|c|c|c|c|c|c|c|}
\hline
  \multicolumn{1}{c|}{Galaxy} &
  \multicolumn{1}{c|}{RA} &
  \multicolumn{1}{c|}{Dec} &
  \multicolumn{1}{c|}{X} &
  \multicolumn{1}{c|}{Y} &
  \multicolumn{1}{c|}{$m_{NUV} \pm \delta m_{NUV}$} &
  \multicolumn{1}{c|}{$m_{U} \pm \delta m_{U}$} &
  \multicolumn{1}{c|}{$m_{B} \pm \delta m_{B}$} &
  \multicolumn{1}{c|}{$m_{V} \pm \delta m_{V}$} &
  \multicolumn{1}{c|}{$m_{I} \pm \delta m_{I}$} &
  \multicolumn{1}{c|}{Age$\pm \delta$ Age [Myr]} &
  \multicolumn{1}{c|}{Mass$\pm \delta$ Mass [$\mathrm{M_{\odot}}$]} &
  \multicolumn{1}{c|}{E(B-V)$\pm \delta$ E(B-V)} &
  \multicolumn{1}{c|}{$M_{20}$} &
  \multicolumn{1}{c|}{CI} &
  \multicolumn{1}{c|}{Class} \\
\hline

  NGC 4826 & 194.190374 & 21.695519 & 3108.5 & 4889.5 & 23.88$\pm$0.3 & 22.5$\pm$0.07 & 22.13$\pm$0.02 & 21.4$\pm$0.01 & 20.15$\pm$0.01 & 1.0e+03$\pm$4.5e+02 & 4.4e+04$\pm$3.0e+03 & 0.23$\pm$0.17 & -1.78 & 1.56 & 1\\
  NGC 4826 & 194.170181 & 21.694304 & 4813.3 & 4779.14 & 23.29$\pm$0.19 & 22.12$\pm$0.06 & 22.05$\pm$0.02 & 21.52$\pm$0.02 & 20.76$\pm$0.02 & 8.9e+02$\pm$2.2e+02 & 1.7e+04$\pm$1.6e+03 & 0.0$\pm$0.05 & -1.68 & 1.9 & 1\\
  NGC 4826 & 194.161467 & 21.693118 & 5549.02 & 4671.48 & 22.93$\pm$0.14 & 22.26$\pm$0.06 & 21.98$\pm$0.02 & 21.73$\pm$0.02 & 21.08$\pm$0.03 & 5.4e+02$\pm$1.6e+02 & 9.9e+03$\pm$3.0e+02 & 0.0$\pm$0.1 & -1.5 & 2.08 & 1\\
  NGC 4826 & 194.173786 & 21.684177 & 4508.95 & 3858.93 & 20.45$\pm$0.02 & 20.34$\pm$0.02 & 20.85$\pm$0.01 & 20.7$\pm$0.01 & 20.03$\pm$0.02 & 1.1e+02$\pm$3.7e+01 & 1.5e+04$\pm$2.4e+02 & 0.12$\pm$0.05 & -1.74 & 1.77 & 1\\
  NGC 4826 & 194.162992 & 21.685389 & 5420.34 & 3969.14 & 21.88$\pm$0.06 & 21.22$\pm$0.03 & 21.21$\pm$0.01 & 20.62$\pm$0.01 & 19.52$\pm$0.01 & 1.7e+02$\pm$4.2e+01 & 5.4e+04$\pm$3.8e+03 & 0.47$\pm$0.08 & -1.72 & 1.88 & 1\\
  NGC 4826 & 194.160047 & 21.685459 & 5669.0 & 3975.56 & 22.18$\pm$0.07 & 21.58$\pm$0.04 & 21.7$\pm$0.02 & 21.13$\pm$0.01 & 20.08$\pm$0.01 & 1.2e+02$\pm$5.1e+01 & 2.8e+04$\pm$3.0e+03 & 0.48$\pm$0.04 & -1.71 & 1.58 & 1\\
  NGC 4826 & 194.170105 & 21.683517 & 4819.76 & 3799.01 & 18.88$\pm$0.01 & 19.09$\pm$0.01 & 20.36$\pm$0.01 & 20.43$\pm$0.01 & 20.34$\pm$0.02 & 4.0e+00$\pm$3.0e+00 & 2.0e+03$\pm$5.9e+03 & 0.2$\pm$0.14 & -1.84 & 1.29 & 1\\
  NGC 4826 & 194.18784 & 21.671258 & 3322.31 & 2685.12 & 20.78$\pm$0.03 & 20.25$\pm$0.01 & 20.42$\pm$0.01 & 19.85$\pm$0.01 & 18.76$\pm$0.01 & 5.0e+00$\pm$8.3e+01 & 2.9e+04$\pm$6.2e+04 & 0.91$\pm$0.38 & -1.71 & 1.65 & 1\\
  NGC 4826 & 194.179797 & 21.678429 & 4001.42 & 3336.66 & 22.35$\pm$0.09 & 21.43$\pm$0.04 & 21.55$\pm$0.02 & 20.96$\pm$0.01 & 19.84$\pm$0.02 & 1.7e+02$\pm$1.1e+01 & 4.0e+04$\pm$4.7e+03 & 0.47$\pm$0.05 & -1.63 & 1.65 & 1\\
  NGC 4826 & 194.18759 & 21.670516 & 3343.4 & 2617.64 & 21.11$\pm$0.03 & 21.11$\pm$0.03 & 21.42$\pm$0.01 & 21.3$\pm$0.01 & 20.42$\pm$0.02 & 9.5e+01$\pm$3.2e+01 & 1.0e+04$\pm$1.5e+03 & 0.2$\pm$0.02 & -1.62 & 1.78 & 1\\
  
\hline\caption{\label{tab:sample_prop} The bright, isolated specialized cluster sample we use in this paper. The table has the following columns from left to right: Name of the host galaxy, object coordinates (RA, Dec, X, Y), 5-band 4~pixel circular aperture photometry and their errors in Vega magnitudes, $\chi^{2}$ minimum age, mass, E(B-V) and their errors, CI, $M_{20}$, and the visual morphology class.}
\end{longtable}
\end{scriptsize}
\end{landscape}
\clearpage
\twocolumn

\bsp	
\label{lastpage}
\end{document}